\documentclass[11pt]{article}
\usepackage[margin=1in,paperheight=12in,paperwidth=9in]{geometry}
\usepackage[utf8]{inputenc}
\usepackage{mathtools}
\usepackage{amssymb} 
\usepackage{float}
\usepackage[shortlabels]{enumitem}
\usepackage{amsthm}
\usepackage{tikz}
\usepackage{amsopn}
\usepackage{bbm}
\usepackage{multirow}
\usepackage{url}

\usepackage{natbib}
 \bibpunct[, ]{(}{)}{,}{a}{}{,}%
\DeclarePairedDelimiterX\Basics[1](){ #1}
\usepackage{color}
\usepackage{subfigure}
\usepackage{float}
\usepackage{pifont}
\usepackage{algorithm,algpseudocode}

\usepackage[T1]{fontenc}
\usepackage{etoolbox}
\usepackage{float}
\usepackage{lscape}

\usepackage{authblk}
\renewcommand{\qedsymbol}{$\blacksquare$}
\usepackage{booktabs}

\usepackage{amsmath}

\DeclarePairedDelimiter\floor{\lfloor}{\rfloor}
\newcommand{\one}{\mathbbm{1}}
\newcommand{\dd}{\mathrm{d}}

\newcommand{\TGAN}{\textsc{Tail-GAN}}

\usepackage{pgfplots}
\usetikzlibrary{arrows.meta}

\usepgfplotslibrary{groupplots}
\usepgfplotslibrary{dateplot}
\newcommand{\Gbar}{\overline{G}}
\newcommand{\Dbar}{\overline{D}}
\newcommand{\ba}{\begin{eqnarray}}
\newcommand{\ea}{\end{eqnarray}}

\newtheorem{theorem}{Theorem}[section]

\newtheorem{assumption}[theorem]{Assumption}

\newtheorem{proposition}[theorem]{Proposition}
\newtheorem{remark}[theorem]{Remark}
\newtheorem{example}[theorem]{Example}
\usepackage{xcolor}

\usepackage[colorlinks=true,linkcolor=red,filecolor=green,citecolor=blue]{hyperref}
\setcitestyle{square}

\newcounter{noteMCctr} 

\newcounter{noteCZctr} 

\newcommand{\bb}{\hspace{-1mm} $\bullet$ }

\usepackage{placeins}

\pgfplotsset{compat=1.17}

\title{\textbf{{\TGAN}: \\Learning to Simulate Tail Risk Scenarios}\footnote{An earlier version of this paper circulated under the title ``{\TGAN}: Nonparametric Scenario Generation for Tail Risk Estimation''. First draft: March 2022.}}

\author[1]{Rama Cont}
\author[2]{Mihai Cucuringu}
\author[3]{Renyuan Xu}
\author[4]{Chao Zhang}

\affil[1]{\small Mathematical Institute, University of Oxford, UK.}
\affil[2]{\small Department of Mathematics, University of California Los Angeles, US.}
\affil[3]{\small Department of Finance and Risk Engineering, New York University, US.}
\affil[4]{\small FinTech Thrust, Hong Kong University of Science and Technology (Guangzhou), CN.}

\date{This Version: May 11, 2025}

\begin{document}

\maketitle
\begin{abstract}
The estimation of loss distributions for dynamic portfolios requires the simulation of scenarios representing realistic joint dynamics of their components. We propose a novel data-driven approach for simulating realistic, high-dimensional multi-asset scenarios, focusing on accurately representing tail risk for a class of static and dynamic trading strategies. We exploit the joint elicitability property of Value-at-Risk (VaR) and Expected Shortfall (ES) to design a Generative Adversarial Network (GAN)  that learns to simulate price scenarios preserving these tail risk features. We demonstrate the performance of our algorithm on synthetic and market data sets through detailed numerical experiments. In contrast to previously proposed data-driven scenario generators, our proposed method correctly captures tail risk for a broad class of trading strategies and demonstrates strong generalization capabilities.  In addition, combining our method with principal component analysis of the input data enhances its scalability to large-dimensional multi-asset time series, setting our framework apart from the univariate settings commonly considered in the literature.
\end{abstract}


 {
   \hypersetup{linkcolor=blue}
   \tableofcontents
 }

%%%%%%%%%%%%%%%%%%%%%%%%%%%%%%%%%%%%%%%%%%%%%%%%%%%%%%%%%%%%%%%%%%%%%%%%%%%%%%%%
%%%%%%%%%%%%%%%%%%%%%%%%%%%%%%%% Introduction %%%%%%%%%%%%%%%%%%%%%%%%%%%%%%%%%%
%%%%%%%%%%%%%%%%%%%%%%%%%%%%%%%%%%%%%%%%%%%%%%%%%%%%%%%%%%%%%%%%%%%%%%%%%%%%%%%%
\section{Data-driven simulation of financial scenarios}
Scenario simulation is extensively used in finance for   evaluating the loss distribution of portfolios and trading strategies, often with a focus on the estimation of risk measures such as Value-at-Risk and Expected Shortfall (\citet{glasserman2003}). The estimation of such risk measures for static and dynamic portfolios involves the simulation of scenarios representing realistic joint dynamics of  their components. This requires both a realistic representation of the temporal dynamics  of individual assets  (\textit{temporal dependence}), as well as an adequate representation  of their co-movements  (\textit{cross-asset dependence}). 
The use of scenario simulation for risk estimation has increased in light of the Basel Committee's Fundamental Review of the Trading Book (FRTB), an international standard that regulates the amount of capital banks ought to hold against market risk exposures (see \citet{FRTB_Basel}). FRTB particularly revisits and emphasizes the use of Value-at-Risk vs Expected Shortfall as a measure of risk under stress, thus ensuring that banks appropriately capture tail risk events. 
%In addition, FRTB requires banks to develop clear methodologies to specify how various extreme scenarios are simulated,  and how the stress scenario risk measures are constructed using these scenarios. This suite of capital rules has taken effect in January 2022 to strengthen the financial system, with an eye towards capturing tail risk events that came to light during the 2007-2008 financial crisis.

A common approach in scenario simulation is to use parametric models in the literature. The specification and estimation of such parametric models pose challenges in situations in which one is interested in heterogeneous portfolios or intraday dynamics. As a result of these issues, and along with the scalability constraints inherent in nonlinear models, many applications in finance have focused on Gaussian factor models for scenario generation, even though they fail to capture many stylized features of market data.

%\subsection{Data-driven scenario generation}

Over the past decade, with the evolution of deep learning techniques, {\it generative models} based on machine learning techniques 
ave emerged as an efficient alternative to parametric models for simulating patterns extracted from complex, high-dimensional datasets.
In particular, Generative Adversarial Networks (GANs) (\citet{goodfellow2014generative}) have been successfully used for  generating images (\citet{goodfellow2014generative,
radford2015unsupervised}), audio (\citet{oord2016wavenet}) and text (\citet{fedus2018maskgan, zhang2017adversarial}), as well as the simulation of financial market scenarios \cite{takahashi2019modeling,wiese2020quant,yoon2019time,vuletic2023fin,volgan}.

Most generative models for financial time series have been based on a Conditional-GAN (CGAN) architecture \citep{mirza2014conditional,fu2019time,koshiyama2020generative,ni2020conditional,li2020generating,vuletic2023fin,volgan,wiese2020quant}, with the notable exception of \citet{buehler2020generating}, who employed a Variational Autoencoder (VAE). Furthermore, except for \citet{volgan,yoon2019time}, these studies primarily focus on generating univariate time series.

When scenarios are intended for use in risk management applications, a relevant criterion is whether they correctly capture the risk of portfolios or commonly used trading strategies. 
The frameworks discussed above typically use divergence measures such as cross-entropy (\citet{goodfellow2014generative, chen2016infogan}) or   Wasserstein distance (\citet{arjovsky2017wasserstein})  to measure the similarity of the generated data to the input data.  
Such global divergence measures are in fact not exactly computable but are approximated using the data sample; they are thus dominated by typical sample values, and may fail to account for ``tail'' events, which occur with a small probability in the training sample. As a consequence, such training criteria may lead to poor performance if one is primarily interested in {\it tail properties} of the distribution.
 
A related concern with the use of data-driven market generators for risk management is {\it model validation}.
Unlike generative models for images, which may be validated by visual inspection,  validation of scenario generators requires a quantitative approach and needs to be addressed in a systematic manner. 
In particular, it is not clear whether  scenarios simulated by such black-box market generators  are  realistic enough to be useful for applications in risk management. 
\subsection{Contributions}
To address these challenges, we propose a novel training methodology for scenario generators based on a training objective which directly relates to the performance in targeted use cases. To achieve this goal, we design an objective function for the scenario generator which ensures that the output scenarios accurately represent the tail risk of a broad class of benchmark strategies. 

Generally speaking, the goal of a dynamic scenario generator is to sample from an unknown distribution on a space of  scenarios  $\mathbf{p}=(\mathbf{p}_t, t\in \mathbb{T})\in \Omega $  representing price trajectories for a set of financial assets on discrete time set $\mathbb{T}$. The main use of these scenarios is to compute and analyze the profit and loss (PnL) of various trading strategies along each trajectory. Each --dynamic or static-- trading strategy $\varphi^k$ $(1\leq k \leq K)$ thus defines a map
\begin{eqnarray}
    \Pi^k :  \Omega &\mapsto & \mathbb{R}\nonumber\\
   \mathbf{p}=(\mathbf{p}_t,t\in \mathbb{T}) &\mapsto & \Pi^k(\mathbf{p})=\sum_{\mathbb{T}}\varphi^k(t_i,\mathbf{p}).(\mathbf{p}_{t_{i+1}}-\mathbf{p}_{t_{i}})\qquad \label{eq.pnl}
\end{eqnarray} 
whose value $\Pi^k(\bf{p})$  represents the terminal PnL of the trading strategy $\varphi^k$ in scenario $\bf{p}$.
A probability measure $\mathbb{P}$ on the set $\Omega$ of market scenarios thus induces a {\it one-dimensional} distribution for the (scalar) random variable $\Pi^k(\bf{p})$. We therefore observe that each trading strategy $\varphi^k$ {\it projects} the (high-dimensional) probability measure $\mathbb{P}$ into a one-dimensional distribution, whose properties are easier to learn from data.

A set  of trading strategies $\{\varphi^k\}_{k=1, \cdots, K}$ thus allows to ``project'' the high-dimensional probability measure $\mathbb{P}$ onto $K$ scalar random variables $\{\Pi^k(\mathbf{p})\}_{k=1, \cdots, K}$. The associated (one-dimensional)   distributions  serve as tractable {\it features} of $\mathbb{P}$, which the algorithm attempts to learn.

Our idea is to start with a set of user-defined benchmark trading strategies $\{\varphi^k\}_{k=1, \cdots, K}$ and to guide the training of the scenario generator to learn the properties of the associated loss distributions. Focusing on a finite set of trading strategies leads to a dimension reduction, reducing the task to learning $K$ one-dimensional distributions.

 We then exploit the joint elicitability property of Value-at-Risk (VaR) and Expected Shortfall (ES) to design a training criterion sensitive to tail risk measures of these loss distributions. This results in a GAN architecture which is capable of learning to simulate price scenarios which preserve tail risk characteristics for the set of benchmark trading strategies.
 We study various theoretical properties of the proposed algorithm, and assess its   performance through detailed numerical experiments on synthetic data and market data.

\paragraph{\bf Theoretical contributions.}  On the theoretical side, we establish a universal approximation result {\it in the distributional sense}: given any target loss distribution and any (spectral) risk measure, there exists a generator network of sufficient size whose output distribution approximates the target distribution with a given accuracy as quantified by the risk measure. The proof relies on a semi-discrete optimal transport technique, which may be interesting in its own right.

\paragraph{\bf Empirical performance.} Our extensive numerical experiments, using both synthetic and market data, show that  {\TGAN} provides accurate tail risk estimates and is able to capture key univariate and multivariate statistical properties of financial time series, such as heavy tails, autocorrelation, and cross-asset dependence patterns. 
 We show that, by including dynamic trading strategies in the training set of benchmark portfolios,  
{\TGAN} provides more realistic outputs and a better representation of tail risks than classical GAN methods previously applied to financial time series.
Last but not least, we illustrate that combining {\TGAN} with Principal Component Analysis (PCA) enables the design of scenario generators scalable to a large number of  assets.

\subsection{Related literature}
\label{literature} 

The use of GANs with bespoke loss functions for simulation of financial time series has also been explored in \cite{vuletic2023fin,ni2020conditional,volgan}. However, these methods are not focused on tail scenarios.

The idea of incorporating \textit{quantile} properties into the simulation model has been explored in  \citet{ostrovski2018autoregressive}, which introduced an autoregressive implicit quantile network (AIQN). The goal therein is to train a generator  via supervised learning so that the quantile divergence between the empirical distributions of the training data and the generated data is minimized. However, the quantile divergence adopted in AIQN is an {\it average performance} across all quantiles, which provides no guarantees for the tail risks. In addition, the generator  trained with supervised learning may suffer from accuracy issues and the lack of generalization power (see Section \ref{sec:onlyG} for a detailed discussion).

\citet{bhatia2020exgan} employed GANs conditioned on the  statistics of extreme events  to generate   samples using Extreme Value Theory (EVT).       In contrast, our approach is fully non-parametric and does not rely on the parametrization of tail probabilities.

\subsection{Outline} Section \ref{sec:tails} introduces the concept of a tail-sensitive {\it score function}, which is then utilized in Section \ref{sec:framework} to formulate a training criterion for adversarial training of a generative model, referred to as {\TGAN}. Section \ref{sec:methodologies} outlines the methodology for model validation and comparison, which is subsequently applied in Section \ref{sec:model_validation} to assess the performance of {\TGAN} on synthetic data. Finally, Section \ref{sec:market_simulation} evaluates the performance of {\TGAN} on intraday financial data.

\section{Tail risk measures and score functions}\label{sec:tails}

Our goal is to design a  training approach for learning a distribution which is sensitive  to the tail(s) of the distribution. This can be achieved by using a tail-sensitive loss function. 
To this end, we exploit 
recent results on the {\it elicitability} of risk measures (\citet{AS2014,FZG2015,gneiting2011making}) to design a score function related to tail risk measures used in financial risk management.

We define these tail risk measures and the associated score functions in Section \ref{sec:tail_risk_measure} and discuss some of their properties  in Section \ref{sec:optimization_landscape}.

\subsection{Tail risk measures}\label{sec:tail_risk_measure}

Tail risk refers to the risk of large portfolio losses. \textit{Value at Risk (VaR)} and \textit{Expected Shortfall (ES)} are commonly used statistics for measuring the tail risk of portfolios. 
  
 Consider a random variable $X:\Omega\mapsto \mathbb{R}$, representing the 
PnL of a trading strategy  over time index set $\mathbb{T}$, such as those defined in \eqref{eq.pnl}, where $\Omega$ represents the set of market scenarios and $X(\omega)$ the PnL of the strategy in market scenario $\omega\in \Omega$.
 Given a probability measure $\mathbb{P}$ specified on the set $\Omega$ of market scenarios, we denote by $\mathbb{P}_X\in {\cal P}(\mathbb{R})$ the (one-dimensional) distribution of $X$ under $\mathbb{P}$. 
 
 The Value-at-Risk (VaR) of the portfolio at the terminal time for a confidence level $0<\alpha<1$ is then defined as the $\alpha$-quantile of $\mathbb{P}_X$: 
\begin{eqnarray}
\mbox{VaR}_{\alpha}(X,\mathbb{P}) :=\mathbb{P}_X^{-1}(\alpha)= \inf\{x\in \mathbb{R}:\mathbb{P}_X (x)\ge \alpha\}. \nonumber
\end{eqnarray} 
Expected Shortfall (ES) is an alternative risk measure which is  sensitive to  the tail of the loss distribution:
\begin{eqnarray}\label{ES}
\mbox{ES}_{\alpha}(X,\mathbb{P}) :=\frac{1}{\alpha}\int_0^{\alpha} \mbox{VaR}_{\beta}(X,\mathbb{P}) \dd\beta. \nonumber
\end{eqnarray}
Note that VaR and ES only depend on $\mathbb{P}_X$ and one could also write e.g. $\mbox{VaR}_{\alpha}(\mathbb{P}_X),\mbox{ES}_{\alpha}(\mathbb{P}_X)$.
We will consider such tail risk measures under different probabilistic models, each represented by a probability measure $\mathbb{P}$ on the space $\Omega$ of market scenarios, and the notation above emphasizes the dependence on $\mathbb{P}$.

VaR, ES are examples of {\it statistical} risk measures  which may be represented as a functional  $\rho: \mathcal{P}(\mathbb{{R}})\rightarrow \mathbb{R}$ of the loss distribution (\citet{xuedong2013,kusuoka2001law}). We will use the notation: $\rho(X,\mathbb{P}):=\rho(\mathbb{P}_X)$.

\paragraph{Elicitability and score functions.}  VaR, ES, and more generally, most risk measures considered in financial applications are examples of ``statistical functionals'' i.e. maps \\
$T:\mathcal{F}\mapsto \mathbb{R}$ defined on a set $\mathcal{F}\subset {\cal P}(\mathbb{R})$ of probability distributions.

A statistical functional is {\it elicitable} if there exists a score function whose minimization over a samples yields a consistent estimator in the large sample limit  \citep{gneiting2011making}.
More specifically: a statistical functional $T:\mathcal{F}\mapsto \mathbb{R}^d$ is {\it elicitable}  if there is a score function $S:\mathbb{R}^d\times \mathbb{R}\mapsto \mathbb{R}$ such that 
\begin{eqnarray}
T(\mu) = \mathop{\arg \min}_{x\in \mathbb{R}^d} \int S(x, y) \mu(\dd y), \label{eq.elicitability}
\end{eqnarray}
This means that by taking an IID sample $Y_1,...,Y_n\sim \mu$ from $\mu$ and minimizing 
$$ \hat{x}_n=\mathop{\arg\min_{x\in \mathbb{R}^d}}\sum_{i=1}^n S(x,Y_i)$$
we get a consistent estimator  of $T(\mu)$: $\hat{x}_n\to T(\mu)$.

 $S$ is called a {\it strictly consistent} score for $T$ if the minimizer in \eqref{eq.elicitability} is unique. Examples of elicitable statistical functionals  include the mean $T(\mu)=\int x\, \mu(\dd x)$ with $S(x, y) = (x - y)^2$, and the median $T(\mu) = \inf\{x\in \mathbb{R}:\mu(X \leq x)\ge 0.5\}$ with $S(x, y) = |x - y|$.
It was first shown  by \citet{Weber2006} that ES is not elicitable, whereas VaR$_{\alpha}$  is elicitable whenever the $\alpha$-quantile is unique. However, it turns out that  the {\it pair}  $(\mbox{VaR}_{\alpha}(\mu),\mbox{ES}_{\alpha}(\mu))$ is {\it jointly} elicitable. In particular, %\citet{FZG2015} proves that  
the following result in \citet[Theorem 5.2]{FZ2016} gives a family of strictly consistent score functions  for $(\mbox{VaR}_{\alpha}(\mu),\mbox{ES}_{\alpha}(\mu))$:
\begin{proposition}{\rm \citet[Theorem 5.2]{FZ2016}}\label{thm:elicitability} %Assume $\nu$ is a one-dimensional distribution such that 
Assume $\int |x|\mu({\rm d}x)<\infty$.
If $H_2:\mathbb{R}\to\mathbb{R}$ is strictly convex and $H_1:\mathbb{R}\to\mathbb{R}$ is such that   
\begin{eqnarray}\label{eq:r_function}
v\mapsto R_{\alpha}(v,e):= \frac{1}{\alpha}v H_2^{\prime}(e)+H_1(v),
\end{eqnarray}
is strictly increasing for each $e\in \mathbb{R}$, then the score function 
\begin{eqnarray}\label{eq:score_function}
S_{\alpha}(v,e,x) &=& (\one_{\{x\leq v\}}-\alpha)(H_1(v)-H_1(x))\nonumber\\
&&+\frac{1}{\alpha} H_2^{\prime}(e)\one_{\{x\leq v\}}(v-x)+ H_2^{\prime}(e)(e - v) - H_2(e), 
\end{eqnarray}
is strictly consistent for $({\rm VaR}_{\alpha}(\mu),{\rm ES}_{\alpha}(\mu))$, i.e.
\begin{eqnarray}\label{M_statistics}
({\rm VaR}_{\alpha}(\mu), {\rm ES}_{\alpha}(\mu))=\arg \min_{(v,e)\in \mathbb{R}^2} \int S_{\alpha}(v,e,x) \mu(\dd x).
\end{eqnarray}
\end{proposition}

\subsection{Score functions for tail risk measures}\label{sec:optimization_landscape}

The computation of the estimator \eqref{M_statistics} involves the optimization of
\begin{eqnarray}\label{eq:expected_score}
s_\alpha(v,e):=\int S_{\alpha}(v,e,x)\mu(\dd x), 
\end{eqnarray}
for a given one-dimensional distribution $\mu$. While any choice of $H_1,H_2$ satisfying the conditions of Proposition \ref{thm:elicitability} theoretically leads to consistent estimators in \eqref{M_statistics}, 
different choices of $H_1$ and $H_2$ lead to optimization problems with different landscapes, with some being easier to optimize than others. We use a specific form of the score function, proposed by \citet{AS2014},which has been adopted by practitioners for backtesting purposes:
\vspace{-2mm}
\begin{eqnarray}\label{eq:quant_score}
S_{\alpha}(v,e,x) = \frac{W_{\alpha}}{2}(\one_{\{x\leq v\}}-\alpha) (x^2-v^2) 
+\one_{\{x\leq v\}}e(v-x) + \alpha e \left(\frac{e}{2} - v\right), \,\,{\rm with}\,\,  \frac{{\rm ES}_{\alpha}(\mu)}{{\rm VaR}_{\alpha}(\mu) }\geq W_{\alpha} \ge 1.\quad
\end{eqnarray}
This choice is special case of   \eqref{eq:score_function}, where $H_1$ and $H_2$ are given by
\begin{eqnarray}
H_1(v) =-\frac{W_{\alpha}}{2}v^2,\,\, H_2(e) = \frac{\alpha}{2}e^2, \quad{\rm with}\quad  \frac{{\rm ES}_{\alpha}(\mu)}{{\rm VaR}_{\alpha}(\mu) }\geq W_{\alpha} \ge 1. \nonumber
\end{eqnarray}
Then  \eqref{eq:quant_score} satisfies the conditions in Proposition \ref{thm:elicitability} on  $\{(v,e)\in \mathbb{R}^2\,\,\vert\,\, W_{\alpha}\,v \leq e \le v \le 0\}$.
\begin{figure} [H]
    \centering
\subfigure[$s_\alpha(v,e)$.]{ \label{fig:quant_lanscape}
\includegraphics[width=.33\textwidth, trim=1.5cm 1.0cm 2.0cm 2.0cm,clip]{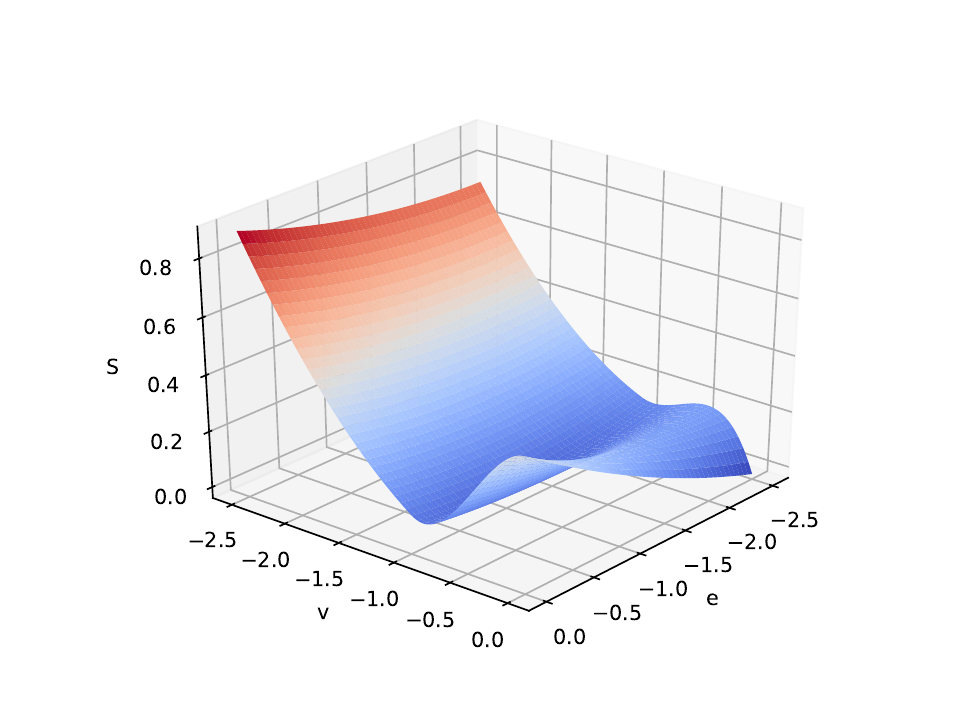}
}
\subfigure[$s_\alpha(v,e)$ as a function of $e$ with given value of $v$.]{ \label{fig:quant_es}
\includegraphics[width=.3\textwidth, trim=2.3cm 0.5cm 3cm 0.5cm,clip]{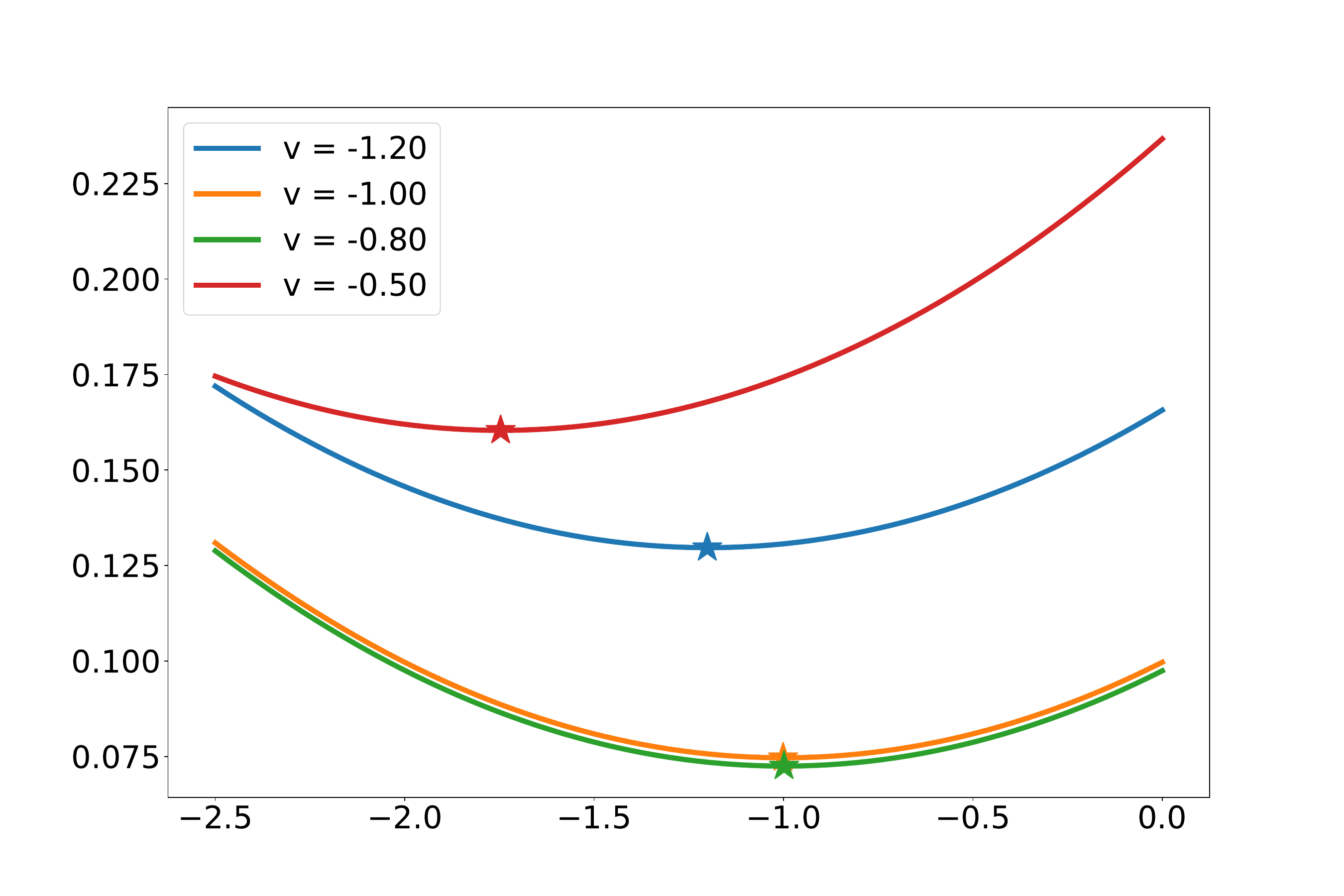}
}
\subfigure[$s_\alpha(v,e)$ as a function of $v$ with given value of $e$.]{ \label{fig:quant_var}
\includegraphics[width=.3\textwidth, trim=2.3cm 0.5cm 3cm 0.5cm,clip]{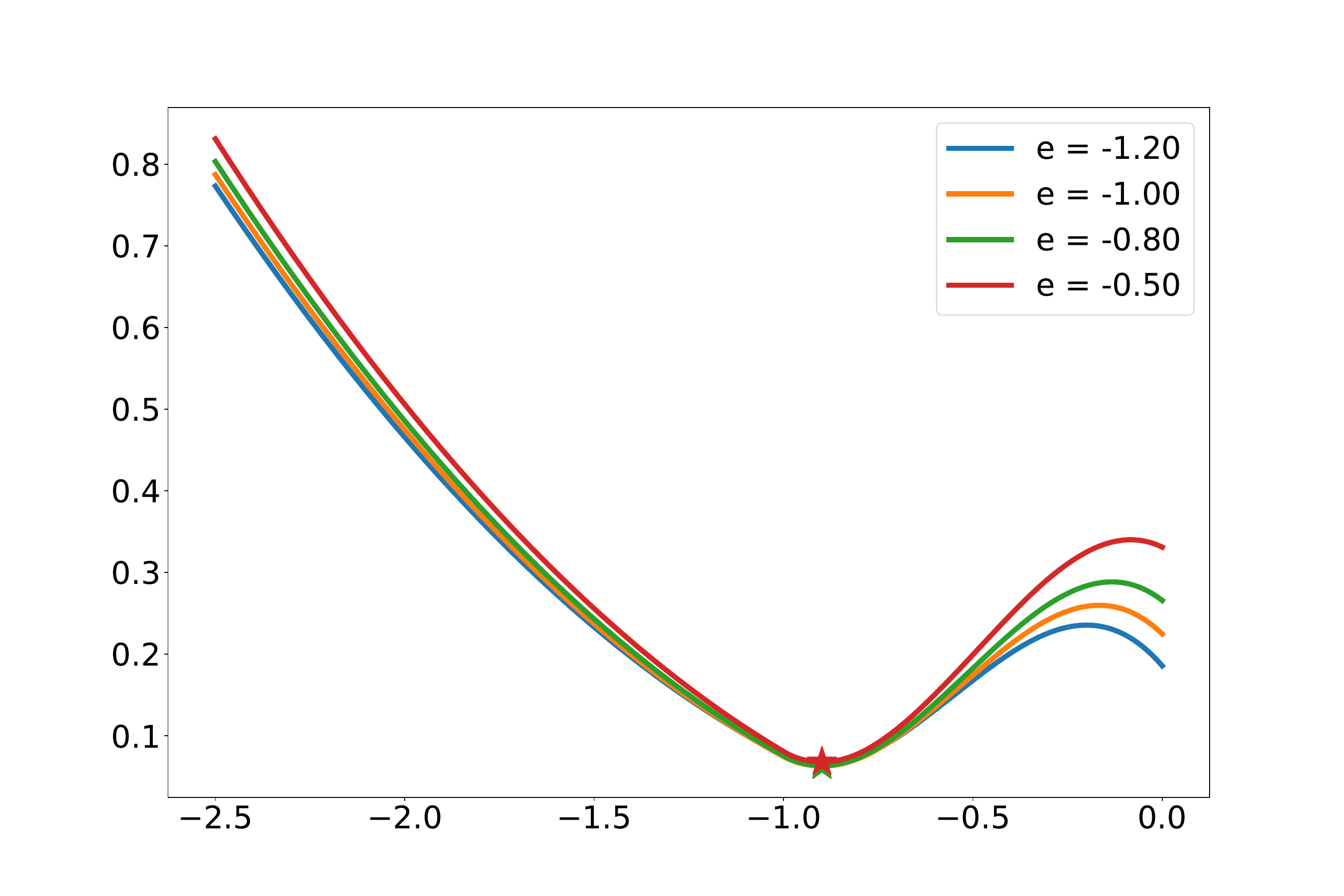}
}
\caption{Landscape of $s_\alpha(v,e)$ based on \eqref{eq:quant_score} with $\alpha=0.05$ for the  uniform distribution on $[-1,1]$.}
\label{fig:quant_illustration}
\end{figure}
 
 The following  proposition shows that  the score function \eqref{eq:quant_score} leads to an optimization problem with desirable properties, as shown in Figure \ref{fig:quant_illustration}:

\begin{proposition}\label{thm:optimization}% Consider a one-dimensional distribution $\nu$ with CDF $F_{\nu}$ such that 
(1) Assume ${\rm VaR}_{\alpha}(\mu)<0$, for $\alpha< 1/2$. Then the score $s_\alpha(v,e)$ based on \eqref{eq:quant_score} is strictly consistent for $({\rm VaR}_{\alpha}(\mu),{\rm ES}_{\alpha}(\mu))$ and its Hessian  is positive semi-definite on the region
$$\mathcal{B} = \left\{(v,e)\,\,\left\vert\,\,v \leq {\rm VaR}_{\alpha}(\mu), \,\,\text{and}\,\,  W_{\alpha} v \leq e \leq v \leq 0\right.\right\}.$$
(2) If there exist 
$\delta_{\alpha}\in (0,1)$, $\varepsilon_{\alpha}\in  \left(0,\frac{1}{2}-\alpha\right)$,  $z_{\alpha}\in \left(0,\frac{1}{2}-\alpha\right)$, and $W_{\alpha}>\frac{1}{\sqrt{\alpha}}$ such that 
\begin{eqnarray}\label{eq:addition_assumption}
\frac{\mu( \dd x)}{\dd x} \ge \delta_{\alpha}  \mbox{ for } x\in  \left[{\rm VaR}_{\alpha}(\mu),{\rm VaR}_{\alpha+\varepsilon_{\alpha}}(\mu)\right]\quad {\rm and } \quad  {\rm ES}_{\alpha}(\mu) \ge W_{\alpha} {\rm Var}_{\alpha}(\mu) + z_{\alpha},
\end{eqnarray}
then the Hessian of $ s_{\alpha}(v,e)$ is positive semi-definite on the region 
$$\widetilde{\mathcal{B}} = \left\{(v,e)\,\,\left\vert\,\,v\leq {\rm VaR}_{\alpha+\beta_{\alpha}}(\mu), \,\,\text{and}\,\,  W_{\alpha} v  + z_{\alpha}\leq  e \leq  v \leq 0\right.\right\}\quad {\rm where}\qquad
\beta_{\alpha}= \min \left\{\varepsilon_{\alpha}, \frac{z_{\alpha}\delta_{\alpha}}{2\,W_{\alpha}}\right\}.$$
\end{proposition}
The proof is given in Appendix \ref{proof1}.

\begin{example}[Example for condition \eqref{eq:addition_assumption}] Condition \eqref{eq:addition_assumption} holds when $X$ has a strictly positive density under measure $\mu$. Take an example where $X$ follows the standard normal distribution.
Denote $f(x) = \frac{1}{\sqrt{2\pi}}\exp\left(-\frac{x^2}{2}\right)$ as the density function,  and $F(y)=\int_{-\infty}^{y}f(x)\dd x$ as the cumulative density function for $X$. Then we have ${\rm  VaR}_{\alpha}(\mu) = F^{-1}(\alpha)$ and ${\rm ES}_{\alpha}(\mu) = -\frac{f(F^{-1}(\alpha))}{\alpha}$. Setting  $\alpha=0.05$ and $\varepsilon_{\alpha}=0.05$, we have ${\rm VaR}_{0.05}(\mu)\approx-1.64$ and ${\rm ES}_{0.05}(\mu)\approx-2.06$ by direct calculation. Then we can set $\delta_{\alpha} = f(F^{-1}(0.05))\approx 0.103$, $W_{\alpha}=5$ and $z_{\alpha}=\frac{1}{4}$. Hence \eqref{eq:addition_assumption} holds for $\beta_{\alpha} = \min\{\varepsilon_{\alpha},\frac{z_{\alpha \delta_{\alpha}}}{2W_{\alpha}}\} \approx 0.0025$.
\end{example}

Proposition \ref{thm:optimization} implies that $s_{\alpha}(v,e)$ has a well-behaved optimization landscape on regions $\mathcal{B}$ and  $\widetilde{\mathcal{B}}$ if the corresponding conditions  are  satisfied. In particular, the minimizer of $s_{\alpha}(v,e)$, i.e., $(\mbox{VaR}_{\alpha}(\mu),\mbox{ES}_{\alpha}(\mu))$, is on the boundary of region $\mathcal{B}$. $\widetilde{\mathcal{B}}$ contains an open ball with center $(\mbox{VaR}_{\alpha}(\mu),\mbox{ES}_{\alpha}(\mu))$.

In summary, $s_{\alpha}(v,e)$ has a positive semi-definite Hessian in a neighborhood of the minimum, which leads to desirable properties for convergence. Other choices for $H_1$ and $H_2$ exist, but  may have undesirable properties (\citet{FZ2016,FZG2015}) as the following example shows.
\begin{example}
 Let $X$ be uniformly distributed on $[-1,1]$ and $\alpha=0.05$.
  When $H_2(x)=\exp(x)$,   
\[
 \mathbb{P}(X \leq v)v-\mathbb{E}( X)+(e-v)\alpha = \frac{1}{4}v^2+\left(\frac{1}{2}-0.05\right)v+{\frac{1}{4}}+0.05e,
\]
which yields  
\[
\frac{\partial^2 s_{\alpha}}{ \partial e^2} = {\frac{\exp(e)}{\alpha}}\left[ \frac{1}{4}(v+0.9)^2+0.0475+\alpha+0.05e\right]. 
\] 
Letting $v=-0.9$, we arrive at $\frac{\partial^2 s_{\alpha}}{\partial e^2}|_{v=-0.9}<0$ for all {$e<-1.95$}, so $s_\alpha$ is not convex.
\end{example}
These results supports the choice of \eqref{eq:quant_score}, as proposed by \citet{AS2014}.

\section{Learning to generate tail scenarios}
%\section{Modelling Framework of Tail Generative Adversarial Network}
\label{sec:framework}

We now introduce the {\TGAN} algorithm, a data-driven algorithm  for simulating multivariate price scenarios which preserves the tail risk features of a set of benchmark trading strategies. 

\subsection{Discriminating probability measure using loss quantiles of trading strategies}
Let $\Omega\subset \mathbb{R_+}^{M\times T}$ be a set of {\it market scenarios}
$\mathbf{p}=(\mathbf{p}_t, t\in \mathbb{T})=(p_{t_i,m})_{i=1,\cdots,T,m=1,\cdots,M} $  representing price trajectories for $M$  financial assets. 
Here $\mathbb{T}=\{t_i=i\Delta, i=1,\cdots,T\}$ represents the (discrete) set of possible trading times over the risk horizon $T_{h}=\Delta \times T$.
Any {\it trading strategy} is then described by a (non-anticipative) map
\begin{eqnarray}
    \varphi :  \mathbb{T}\times \Omega &\mapsto & \mathbb{R}^M\nonumber\\
  (t,\mathbf{p}) &\mapsto & \varphi(t,\mathbf{p})\label{eq.strategy}
\end{eqnarray}
whose value $\varphi(t,\mathbf{p})$ represents the vector of portfolio holdings of the strategy at time $t$ in scenario ${\mathbf{p}}\in \Omega$. The value of such a portfolio at $t$ is $V_t(\varphi)=\mathbf{p}_t.\varphi(t,\mathbf{p})$. 
We consider {\it self-financing} trading strategies, any profit/loss only arises from the accumulation of capital gains, so the total profit over the horizon in scenario $\mathbf{p}\in \Omega$ is given by
$$ 
\Pi(\varphi)=V_{t_T}(\varphi)-V_{0}(\varphi)=\sum_{\mathbb{T}}\varphi(t_i,\mathbf{p}).(\mathbf{p}_{t_{i+1}}-\mathbf{p}_{t_{i}}).$$
Any specification of probability measure $\mathbb{P}$ on the set $\Omega$ of market scenarios then induces a distribution for the (one-dimensional) random variable $\Pi(\varphi)$.

Let $\mathcal{S}(\Omega)$ be the set of all (continuous and bounded) self-financing strategies. 
The starting point of our approach is to note that any probability measure on $\Omega$ is {\it uniquely determined} by the (one-dimensional) distributions of the variables $\{ \Pi(\varphi),\ \varphi\in \mathcal{S}(\Omega)\}$:\\
{\bf Proposition}:
Let $\mathbb{P}^1$ and    $\mathbb{P}^2$ be probability measures on $\Omega$. If for any self-financing trading strategy $\varphi\in \mathcal{S}(\Omega)$, $\Pi(\varphi)$ has the same distribution under $\mathbb{P}^1$ and    $\mathbb{P}^2$, then $\mathbb{P}^1=\mathbb{P}^2$:
$$ \left( \forall \varphi\in {\cal S}(\Omega), \quad {\rm Law}(\Pi(\varphi), \mathbb{P}^1)={\rm Law}(\Pi(\varphi), \mathbb{P}^2)\ \right) \Rightarrow  \mathbb{P}^1=\mathbb{P}^2.$$
Note that for this statement to hold it is not sufficient to consider only static portfolios i.e. buy-and-hold strategies. This would only entail equality for the terminal distributions at $T_h$. It is thus imperative to include dynamic trading strategies.

As a one-dimensional distribution is determined by its quantiles, this means a probability measure $\mathbb{P}$ on $\Omega$ is uniquely determined by knowledge of loss quantiles for all self-financing strategies: denoting by $Q_\alpha(X,\mathbb{P})$ the quantile of a random variable $X$ at level $\alpha\in [0,1]$ under $\mathbb{P}$,
$$ \left( \forall \varphi\in {\cal S}(\Omega), \quad \forall \alpha\in [0,1],\qquad  Q_\alpha(\Pi(\varphi), \mathbb{P}^1)=Q_\alpha(\Pi(\varphi),  \mathbb{P}^2)\ \right) \Rightarrow  \mathbb{P}^1=\mathbb{P}^2.$$
This means that {\bf loss quantiles of self-financing strategies discriminate between probability measures} on $\Omega$.
One can thus learn a probability measure on the high-dimensional space $\Omega\subset \mathbb{R_+}^{M\times T}$ by learning/matching quantiles of the {\it one-dimensional} loss distributions for various trading strategies. These loss quantiles may therefore be considered as {\it features} which, furthermore, have a straightforward financial interpretation.
Moreover,  the quantile levels for $\alpha$ close to 0 or 1 reflect by definition the level of tail risk of a strategy.

We therefore propose a learning approach based on such loss quantiles as features.
We  consider a set $\{\varphi^k\}_{k=1, \cdots, K}$  of self-financing trading strategies, which we call {\it benchmark} strategies. These may be specified by the end-user,  based on the type of trading strategies they are interested in analyzing.
These may be static or dynamic (time- and scenario-dependent) strategies; furthermore this set may be augmented by other trading strategies.
For comparability, each strategy is allocated the same initial capital.
 Each  trading strategy $\varphi^k$ thus defines a map
\begin{eqnarray}
    \Pi^k :  \Omega &\mapsto & \mathbb{R}\nonumber\\
   \bf{p}=(\bf{p}_t,t\in \mathbb{T}) &\mapsto & \Pi^k(\bf{p})=\sum_{\mathbb{T}}\varphi^k(t_i,\bf{p}).(\bf{p}_{t_{i+1}}-\bf{p}_{t_{i}})\qquad \label{eq.pnl}
\end{eqnarray} 
whose value $\Pi^k(\bf{p})$  represents the profit of the  strategy $\varphi^k$ in scenario $\bf{p}$ over the horizon $T_h=\Delta \times T$.
Each trading strategy $\varphi^k$ thus ``projects'' the (high-dimensional) probability measure $\mathbb{P}$ onto a one-dimensional distribution. 
The trading strategies $\{\varphi^k\}_{k=1, \cdots, K}$ thus ``project'' the high-dimensional probability measure $\mathbb{P}$ onto $K$ scalar random variables $\{\Pi^k(\mathbf{p})\}_{k=1, \cdots, K}$. We use the  quantiles  of these variables as tractable {\it features} of $\mathbb{P}$, which the algorithm attempts to learn.

The benchmark strategies considered in this framework include both static portfolios and dynamic trading strategies,  capturing properties of the price scenarios from different perspectives. Static portfolios explore the correlation structure among the assets, while dynamic trading strategies, such as mean-reversion and trend-following strategies probe  temporal properties such as 
mean-reversion or the presence of trends. 

Such benchmark strategies may also be used to assess the adequacy of a scenario generator for risk computations:  a scenario generator is considered to be adequate  if VaR or ES estimates for the benchmark strategies based on its output scenarios meet a predefined accuracy threshold.

This sets the stage for using these loss quantiles as elements of a financially interpretable adversarial training approach for a generative model.

We adopt an adversarial training approach, structured as an iterative max–min game between a generator and a discriminator. The generator simulates price scenarios, while
the discriminator evaluates the quality of the simulated samples using {\it tail risk measures}, namely  VaR and ES. To train the generator and the discriminator, we use a tail-sensitive objective function that leverages the elicitability of VaR and ES, and guarantees the consistency of the estimator.
Building on this foundation, we now provide details on the design of the discriminator and the generator.

\subsection{Training the discriminator: minimizing the score function}\label{sec:discriminator}

Ideally, the discriminator $\Dbar$ takes strategy PnL distributions as inputs, and outputs two values for each of the $K$ strategies, aiming to provide the correct $({\rm VaR}_{\alpha},{\rm ES}_{\alpha})$, by minimizing the score function \eqref{eq:score_function}. However, it is impossible to access the true distribution of $\mathbf{p}$, denoted as $\mathbb{P}_r$, in practice. Therefore we consider a sample-based version of the discriminator, which is easy to train in practice. To this end, we consider PnL samples $\{\mathbf{p}_i\}_{i=1}^n$ with a fixed size $n$ as the input of the discriminator. Mathematically, we write
\begin{eqnarray}\label{eq:sample_discriminator}
{D}^* \in \arg \min_{{D}} \frac{1}{K}\sum_{k=1}^{K} \frac{1}{n} \sum_{i=1}^n%\mathbb{E}_{\mathbf{p}\sim\mathbb{P}_r^{(n)}}
\left[S_{\alpha}\left( \overbrace{ {D}(\underbrace{\Pi^k(\mathbf{p}_j), j \in [n]}_{\text{strategy PnL samples}})}^{\text{VaR and ES prediction  from } D};\mathbf{p}_i\right)\right].
\end{eqnarray}
% Here the expectation  in  \eqref{eq:theoretical_discriminator} is replaced by the empirical mean using finite samples $\{\mathbf{p}_i\}_{i=1}^n$. 
In \eqref{eq:sample_discriminator}, we search the discriminator $D$ over all Lipschitz functions parameterized by the neural network architecture. Specifically, the discriminator adopts a neural network architecture  with $\tilde{L}$ layers, and the input dimension is $\tilde{n}_1 := n$ and the output dimension is $\tilde{n}_{\tilde{L}} := 2$.  Note that
the $\alpha$-VaR of a distribution can be approximated by the $\floor{\alpha n}{}^{th}$ smallest value in a sample of size $n$ from this distribution, which is permutation-invariant to the ordering of the samples. Since the discriminator's goal is to predict the $\alpha$-VaR and $\alpha$-ES, incorporating a sorting function into the architecture design can potentially enhance the stability of the discriminator. We denote this (differentiable) neural sorting function as $\widetilde{\Gamma}$ (\citet{grover2019stochastic}), with details deferred to Appendix \ref{app:neural_sorting}.

In summary, the discriminator  is given by 
\begin{eqnarray}\label{eq:discriminator}
D(\mathbf{x}^{k}; \mathbf{\delta}) = \widetilde{\mathbf{W}}_{\tilde{L}} \cdot \sigma \left( \widetilde{\mathbf{W}}_{\tilde{L}-1} \ldots \sigma (\widetilde{\mathbf{W}}_1{\widetilde{\Gamma}(\mathbf{x}^{k})}+\widetilde{\mathbf{b}}_1) \ldots + \widetilde{\mathbf{b}}_{\tilde{L}-1} \right) +\widetilde{\mathbf{b}}_{\tilde{L}},
\end{eqnarray}
where $\mathbf{\delta}=(\widetilde{\mathbf{W}},\widetilde{\mathbf{b}})$ represent all the parameters in the  neural network. Here we have $\widetilde{\mathbf{W}} = (\widetilde{\mathbf{W}}_1,\widetilde{\mathbf{W}}_2,\ldots,\widetilde{\mathbf{W}}_{\tilde{L}})$ and  $\widetilde{\mathbf{b}} = (\widetilde{\mathbf{b}}_1,\widetilde{\mathbf{b}}_2,\ldots,\widetilde{\mathbf{b}}_{\tilde{L}})$ with $\widetilde{\mathbf{W}}_l \in \mathbb{R}^{n_l \times n_{l-1}}$, $\widetilde{\mathbf{b}}_l \in \mathbb{R}^{n_l \times 1}$ for $l=1,2,\ldots,\tilde{L}$. In the  neural network literature, the $\widetilde{\mathbf{W}}_l$'s are often called the {\it weight} matrices, the $\widetilde{\mathbf{b}}_l$'s are called {\it bias} vectors.
The outputs of the discriminator are two values for each of the $K$ strategies, (hopefully) representing the $\alpha$-VaR and  $\alpha$-ES. The operator $\sigma(\cdot)$ takes a vector of any dimension as input, and applies a function component-wise. $\sigma(\cdot)$ is referred to as the {\it activation function}. Specifically, for any $q\in \mathbb{Z}^+$ and any vector $\mathbf{u}=(u_1,u_2,\ldots,u_q)^{\top} \in \mathbb{R}^{q}$, we have that $\sigma(\mathbf{u}) = (\sigma(u_1),\sigma(u_2),\ldots,\sigma(u_q))^{\top}.$ Several popular choices for the activation function include ReLU with $\sigma(u) = \max(u,0)$, Leaky ReLU with $\sigma(u) = a_1\,\max(u,0) - a_2\, \max(-u,0)$ and $a_1,a_2>0$, and smooth functions such as $\sigma(\cdot) = \tanh(\cdot)$. We sometimes use the abbreviation $D_{\mathbf{\delta}}$ or $D$ instead of $D(\cdot;\mathbf{\delta})$ for notation simplicity.

Accordingly, we define $\mathcal{D}$ as a class of discriminators
\begin{eqnarray}\label{eq:distriminator_class}
\mathcal{D}(\tilde{L},\tilde{n}_1,\ldots,\tilde{n}_{\tilde{L}}) &=& \hspace{-1mm} \Big\{D: \mathbb{R}^{n} \rightarrow \mathbb{R}^{2} \,\,\Big\vert \,\, D \mbox{ takes the form in } \eqref{eq:discriminator} \mbox{ with } \tilde{L} \mbox{ layers and }  \tilde{n}_l \mbox{ as the } \nonumber\\
& \hspace{-2mm} & \hspace{-3mm}  \quad \mbox{ width of each layer}, \|\widetilde{\mathbf{W}}_l\|_{\infty}, \|\widetilde{\mathbf{b}}_l\|_{\infty}<\infty \mbox{ for } l=1,2,\ldots, \tilde{L}  \Big\},
\end{eqnarray}
where $\|\cdot\|_{\infty}$ denotes the max-norm. 

\subsection{Design of the generator and universal approximation}\label{sec:generator}

For the generator, we use a neural network with $L \in \mathbb{Z}^{+}$ layers. Denoting by  $n_l$   the width of the $l$-th layer, the  functional form of the generator is given by 
\begin{eqnarray}\label{eq:generator}
G(\mathbf{z}; \mathbf{\gamma}) = \mathbf{W}_L \cdot \sigma \left( \mathbf{W}_{L-1} \ldots \sigma (\mathbf{W}_1 \mathbf{z}+\mathbf{b}_1)\ldots +\mathbf{b}_{L-1}\right) +\mathbf{b}_L,
\end{eqnarray}
in which $\mathbf{\gamma}:=(\mathbf{W},\mathbf{b})$ represents  the parameters in the  neural network, with $\mathbf{W} = (\mathbf{W}_1,\mathbf{W}_2,\ldots,\mathbf{W}_L)$ and  $\mathbf{b} = (\mathbf{b}_1,\mathbf{b}_2,\ldots,\mathbf{b}_L)$. Here $\mathbf{W}_l \in \mathbb{R}^{n_l \times n_{l-1}}$ and  $\mathbf{b}_l \in \mathbb{R}^{n_l \times 1}$ for $l=1,2,\ldots,L$, where $n_0 = N_{z}$ is the dimension of the input variable.

We define $\mathcal{G} $ as a class of generators that satisfy given regularity conditions:
\begin{eqnarray}\label{eq:generator_class}
\mathcal{G}(L,n_1,n_2,\ldots,n_L) &=& \Big\{G: \mathbb{R}^{N_z} \rightarrow \mathbb{R}^{M \times T} \,\,\Big\vert \,\, G\mbox{ takes the form in } \eqref{eq:generator} \mbox{ with }L\mbox{ layers and }  n_l \mbox{ as the } \nonumber\\
&&\quad \mbox{ width of each layer}, \|\mathbf{W}_l\|_{\infty}, \|\mathbf{b}_l\|_{\infty}<\infty \mbox{ for } l=1,2,\ldots,L \Big\}.
\end{eqnarray}
To ease the notation, we may use the abbreviation $G_{\mathbf{\gamma}}(\cdot)$ or drop the dependency of $G(\cdot\,; \mathbf{\gamma})$ on the neural network parameters $\mathbf{\gamma}$ and conveniently write $G(\cdot)$. We further denote $\mathbb{P}_G$ as the distribution of price series generated by $G$.

{\color{black}
\paragraph{Universal approximation property of the generator.} 
 We first demonstrate the universal approximation power of the generator under the VaR and ES criteria, and then we provide a similar result for more general risk measures that satisfy certain H\"older regularity property. 

\iffalse
{\color{purple}[double check if it can be removed] Given two probability measures $\mu$ and $\nu$ on $\mathbb{R}^d$, recall in \eqref{eq:push_forward_mapping} that a transport map $\Phi$ between $\mu$ and $\nu$ is a measurable map $\Phi : \mathbb{R}^d \rightarrow \mathbb{R}^d$ such that $\nu = \Phi\#\mu $. We denote by $\Gamma(\mu,\nu)$ the set of transport plans between $\mu$ and $\nu$ which consists of all coupling measures $\gamma$ of $\mu$ and $\nu$, i.e., $\gamma ( A \times \mathbb{R}^d ) = \mu( A )$ and $\gamma ( \mathbb{R}^d \times B ) = \gamma ( B ) $ for any measurable $A,B \subset \mathbb{R}^d$ . 
}
\fi
We assume that the portfolio values are Lipschitz-continuous with respect to price paths:
\begin{assumption}[Lipschitz continuity of portfolio values] \label{ass:strategy} For $k=1,2,\cdots,K$,  
\begin{eqnarray}
 \exists \ell_k>0,\qquad  |\Pi^k(p)- \Pi^k(q)|\leq \ell_k \|p-q\|, \quad \forall  p.q \in \Omega. \nonumber
\end{eqnarray}
% In addition, we assume $\Pi^k(0)=0$.  
\end{assumption}

Recall $\mathbb{P}_r$ and $\mathbb{P}_z$ are respectively the target distribution and the  distribution of the input noise.
\begin{assumption}[Noise Distribution and Target Distribution]\label{ass:distribution}  $\mathbb{P}_r$ and $\mathbb{P}_z$ are probability measures on $\Omega$ (i.e.  $N_z =M\times T$)
satisfying the following conditions:
\begin{itemize}
    \item $\mathbb{P}_z\in  \mathcal{P}^2(\Omega)$ has a density. 
    \item $\mathbb{P}_r$ has a bounded moment of order $\beta>1$: $\int \|x\|^\beta \mathbb{P}_r(dx) <\infty$. 
\end{itemize}
\end{assumption}
\begin{assumption}
    The random variables $\Pi^k(X)$  have continuous   densities $f_k$  under $\mathbb{P}_r$ with $f_k({\rm VaR}_\alpha(\Pi^k,\mathbb{P}_r)>0$. \label{ass:pnldensity}
\end{assumption}
%Note that if a $\beta$-th moment is bounded for some $\beta > 2$, Lyapunov's inequality (see pp. 143 in \citet{grimmett2020probability}) implies that the second moment is bounded, and we simply take $ \beta= 2$.

\begin{theorem}[Universal Approximation under VaR and ES Criteria]\label{thm:universial_VaR_ES} Under Assumptions \ref{ass:strategy}-\ref{ass:distribution}-\ref{ass:pnldensity}, for any  $\varepsilon>0$  
\begin{itemize}
    \item there exists a fully connected feed-forward neural network $G_1$, with length $L = \mathcal{O}(\log(\varepsilon^{-2}))$,  width   $N=\mathcal{O}(\varepsilon^{-2\log 2})$  and   ReLU activation, such that 
\begin{eqnarray}
  \Big|{\rm VaR}_\alpha \Big(\Pi^k,\mathbb{P}_{\nabla G_1}\Big) - {\rm VaR}_\alpha \Big(\Pi^k,\mathbb{P}_r \Big)\Big|<\varepsilon; \nonumber
\end{eqnarray}
\item  there exists a fully connected feed-forward neural network $G_2$, with length $L = \mathcal{O}(\log(\varepsilon^{-\frac{\beta}{\beta-1}}))$,  width   $N=\mathcal{O}(\varepsilon^{-\frac{\beta}{\beta-1}\log 2})$  and   ReLU activation, such that 
\begin{eqnarray}
  \Big|{\rm ES}_\alpha\ \Big(\Pi^k,\mathbb{P}_{\nabla G_2} \Big) \Big)- {\rm ES}_\alpha \Big(\Pi^k,\mathbb{P}_r \Big)\Big|<\varepsilon. \nonumber
\end{eqnarray}
\end{itemize}
\end{theorem}
Theorem \ref{thm:universial_VaR_ES} implies that the gradient of a feed-forward neural network with fully connected layers of  equal-width neural network is capable of generating  scenarios which reproduce the tail risk properties (VaR and ES) for the benchmark strategies with arbitrary accuracy. This justifies the use of this   simple network architecture for {\TGAN}. The size of the network, namely the width and the length, depends on the tolerance of the error $\varepsilon$ and further depends  on $\beta$ in the case of ES.  

The proof (given in Appendix \ref{proof2}) consists in using  the theory of (semi-discrete) optimal transport to build a  transport map of the form $\Phi = \nabla\psi$ which pushes   the source distribution $\mathbb{P}_z$ to the empirical distribution $\mathbb{P}_r^{(n)}$. The potential $\psi$ has an explicit form in terms of the maximum of finitely many affine functions. Such  an explicit structure enables the representation of  $\psi$ with a finite deep neural network \citep{lu2020universal}.

We now provide a universal approximation result for more general law-invariant risk measures $\rho: \mathcal{P}(\mathbb{R}) \rightarrow \mathbb{R}$. To start, denote by 
\begin{eqnarray}
    \mathcal{W}_p(\mu,\nu) = \inf_{\xi\in \mathcal{C}(\mu,\nu)} \Big[\mathbb{E}_{(X,Y)\sim \xi}\|X-Y\|^p\Big]^{1/p}, \nonumber
\end{eqnarray}
the Wasserstein distance of order  $p \in [1,\infty)$  between two probability measures $\mu$ and $\nu$ on $\mathbb{R}^d$, 
where $\mathcal{C}(\mu,\nu)$ denotes the collection of all distributions on $\mathbb{R}^d\times \mathbb{R}^d$ with marginal distributions $\mu$ and $\nu$.

\begin{assumption}[H\"older continuity of $\rho$]\label{ass:rho}
\begin{eqnarray}\label{eq:risk_bound1}
  \exists L>0,\quad \exists \kappa\in(0,1], \quad   \Big|\rho(\mu)-\rho(\nu))\Big| \leq L \Big(\,\mathcal{W}_1\Big(\mu,\nu\Big)\,\Big)^{\kappa}, \quad \forall \mu,\nu \in \mathcal{P}(\mathbb{R})
\end{eqnarray}
\end{assumption}
\begin{remark}  The optimized certainty equivalent   (\citet{ben2007old}), spectral risk measures (\citet{acerbi2002spectral}), and utility-based shortfall (\citet{follmer2002convex}) satisfy this assumption.\end{remark}

\begin{theorem}[Universal Approximation in risk metric]\label{thm:universial_general} Under Assumptions \ref{ass:strategy}, \ref{ass:distribution} and \ref{ass:rho},  for any  $\varepsilon$ there exists a fully connected feed-forward neural network $G_3$ under ReLU activation, with length $L$ and  width   $N$  specified below, such that 
\begin{eqnarray}
   \Big|\,\rho\Big(\Pi^k, \mathbb{P}_{\nabla G_3}\Big)-\rho\Big(\Pi^k,\mathbb{P}_r\Big)\,\Big|<\varepsilon. \nonumber
\end{eqnarray}
%{\color{purple}<define $\rho(\mu)$ and $\rho(\Pi^k,\mathbb{P})$ (slight abuse of notation).>}
More specifically,
\begin{enumerate}
    \item[(1)] $L = \mathcal{O}(\log(\varepsilon^{-\frac{\beta}{\kappa(\beta-1)}}))$ and    $N=\mathcal{O}(\varepsilon^{-\frac{\beta}{\kappa(\beta-1)}\log 2})$ when $M= T=1$ and $1<\beta \leq 2$; 
    \item[(2)] $L = \mathcal{O}(\log(\varepsilon^{-\frac{2}{\kappa}}))$ and    $N=\mathcal{O}(\varepsilon^{-\frac{2}{\kappa}\log 2})$  when $M=T=1$ and $\beta \ge 2$;
    \item[(3)]  $L = \mathcal{O}(\log(\varepsilon^{-\frac{M\times T}{\kappa}}))$ and    $N=\mathcal{O}(\varepsilon^{-\frac{M\times T}{\kappa}\log 2})$  when $M\times T \ge 2$ and $\frac{1}{M\times T} + \frac{1}{\beta}<1$;%\frac{\beta}{\kappa(\beta-1)}
     \item[(4)]   $L = \mathcal{O}(\log(\varepsilon^{-\frac{\beta}{\kappa(\beta-1)}}))$ and    $N=\mathcal{O}(\varepsilon^{-\frac{\beta}{\kappa(\beta-1)}\log 2})$ when $M\times T \ge 2$ and $\frac{1}{M\times T} + \frac{1}{\beta}\ge 1$.
\end{enumerate}
\end{theorem}
%\end{remark}
As suggested in Theorem \ref{thm:universial_general}, the depth  of the neural network depends on $\beta$, $M\times T$, and $\kappa$. $M = T =1$ corresponds to the simulation of a single price value of a single asset, which is not much of an interesting case.  When $M\times T \ge 2$ and $\beta > \frac{M\times T}{M\times T-1}$, the complexity of the neural network, characterized by $\frac{M\times T}{\kappa}$, depends on the ratio between the dimension of the price scenario $M\times T$ and the Lipschitz exponent $\kappa$. When $\mathbb{P}_r$ is   heavy-tailed in the sense that $1<\beta < \frac{M\times T}{M\times T-1}$, the complexity of the  network is determined by $\frac{\beta}{(\beta-1)\kappa}$.
The proof of Theorem \ref{thm:universial_general} is deferred to Appendix  \ref{proof3}.

\subsection{Loss function: from bi-level to max-min game}\label{sec:loss_function}
We now design a loss function to jointly train   the generator and the discriminator. The goal of the generator is to generate samples  such that the (optimal) discriminator $D^*$ cannot tell the difference compared to the target distribution. Mathematically,
\begin{eqnarray}\label{eq:generator_supervised0}
G^* \in \arg \min_{G\in \mathcal{G}}S_{\alpha}\Big(D^*\big(\Pi^k(\mathbf{q}_i);i\in[n]\big),\Pi^k(\mathbf{p}_j)\Big), \,\, \textrm{ with } \,\, \mathbf{q}_i\sim \mathbb{P}_G, \mathbf{p}_j\sim \mathbb{P}_r,  i,j=1,2,\ldots,n
\end{eqnarray}
where $D^*$ is the solution to \eqref{eq:distriminator_class}. The coupled optimization problem \eqref{eq:generator_supervised0}, subject to \eqref{eq:distriminator_class}, constitutes a bi-level optimization framework, where the problem for 
$G$ serves as the upper level and the problem for $D$ as the lower level. However, bi-level optimization is often challenging to train in practice. To address this, we propose the following max-min game formulation.

\vspace{0.3cm}
 \fbox{%
  \parbox{16cm}{
\begin{eqnarray} 
\max_{D \in \mathcal{D}}\min_{G\in \mathcal{G}}
\frac{1}{Kn}\, \sum_{k=1}^K\sum_{j=1}^n\Big[
S_{\alpha}\Big(D\big(\Pi^k(\mathbf{q}_i);i\in[n]\big),\Pi^k(\mathbf{p}_j)\Big) - \lambda \,\, S_{\alpha}\Big(D(\Pi^k(\mathbf{p}_i);i\in[n]),\,\,\Pi^k(\mathbf{p}_j)\Big)\Big]
\label{newminimax_distribution_sample} 
\end{eqnarray}}}
\vspace{0.2cm}

\noindent where $\mathbf{p}_i,\mathbf{p}_j\sim \mathbb{P}_r$ and $\mathbf{q}_i\sim \mathbb{P}_G$ ($i,j=1,2,\ldots,n$). 
The discriminator $D$ takes $n$ PnL samples as the input and aims to provide the VaR and ES values of the  sample distribution as the output. The score function $S_{\alpha}$ is defined in \eqref{eq:quant_score}.

\begin{theorem}[Equivalence of the formulations]\label{thm:equivalent_formulation_informal} 
Under mild conditions, the bi-level optimization problem \eqref{eq:generator_supervised0} subject to \eqref{eq:distriminator_class} is equivalent to the max-min game \eqref{newminimax_distribution_sample} for any $\lambda>0$.
\end{theorem}
A detailed statement may be found in Theorem \ref{thm:equivalent_formulation} in Appendix \ref{app:equivalence} along with its proof in Appendix \ref{proof4}.

The max-min structure of \eqref{newminimax_distribution_sample}  encourages the exploration of the generator to simulate scenarios that are not exactly the same as what is observed in the input price scenarios, but are equivalent under the criterion of the score function, hence improving     generalization. We refer the readers  to  Section \ref{sec:onlyG} for a comparison between {\TGAN} and supervised learning methods and a demonstration of the generalization power of {\TGAN}.

\begin{figure}[!htbp]
\centering
\includegraphics[width=0.8\linewidth]{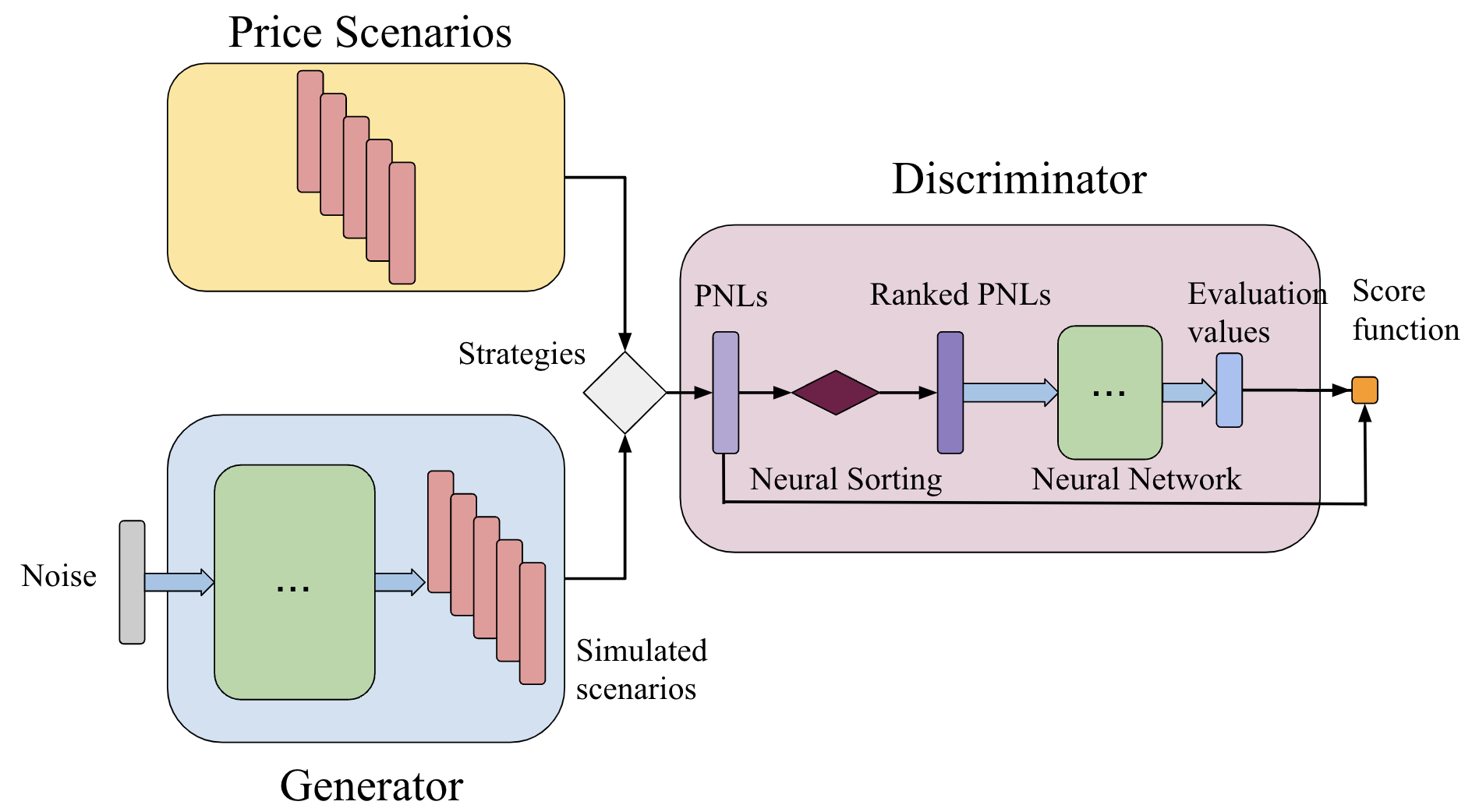}
%\vspace{-3mm}
\caption{Architecture of {\TGAN}. The (blue) thick arrows represent calculations with learnable parameters, and the (black) thin arrows represent calculations with fixed parameters.}
\label{fig:architecture}
\end{figure}

Compared to the binary entropy \citep{goodfellow2014generative} or Wasserstein distance \citep{arjovsky2017wasserstein} used as loss functions in previous GAN algorithms, the objective function \eqref{newminimax_distribution_sample} is more sensitive to  tail risk  and leads to an output which better approximates the $\alpha$-ES and $\alpha$-VaR values. 
 The architecture of {\TGAN} is depicted in Figure \ref{fig:architecture}. 
\FloatBarrier

\section{Numerical experiments: methodology and performance evaluation}\label{sec:methodologies} 
Before we  proceed with the numerical experiments on both synthetic data sets and real financial data sets, we first describe the methodologies, performance evaluation criterion, and several baseline models for comparison. 

\subsection{Methodology}
Algorithm \ref{alg:algorithm1} provides a detailed description of the training procedure of {\TGAN}, which allows us to train the generator  with different benchmark trading strategies.

\begin{algorithm}[!htbp]
  \caption{{\TGAN}.}\label{alg:algorithm1}
  \hspace*{\algorithmicindent} \ \\
  \textbf{Input:}\\
  % Statex 
  \bb Price scenarios $\mathbf{p}_1,\dots,\mathbf{p}_{N_{}}\in\Omega$. \\
  \bb Scenario-based P\&L computation   for trading strategies: $\mathbf{\Pi}=(\Pi^1,\dots,\Pi^K) :\Omega\to \mathbb{R}^K$.\\
  \bb Hyperparameters: learning rate $l_D$ for  discriminator  and $l_G$ for generator; number of training epochs; batch size $N_B$; dual parameter $\lambda$. 
   % \Statex 
\medskip 
  \begin{algorithmic}[1]
    \For{number of epochs} 
        \For{$j = 1 \to \floor{N/N_B}$}
            \State Generate  $N_B$ IID noise samples $\{\mathbf{z}_i, i\in[N_B]\}\sim \mathbb{P}_{\mathbf{z}}$.
            \State Sample a batch $\mathcal{B}_j \subset \{1,\dots,N\}$ of size  $N_B$ from the input data $\{\mathbf{p}_i, i =1, \dots, N\}$.
            \State Compute the loss of the discriminator on the batch $\mathcal{B}_j$
            \begin{equation} 
            \begin{aligned}
                \mathcal{L}_D(\mathbf{\delta}) = \frac{1}{K\,N_B}\,\, \sum_{k=1}^K \sum_{n\in \mathcal{B}_j}  
                &\Big[S_{\alpha}\big(D_{{\mathbf{\delta}}}(\Pi^k(G_{{\mathbf{\gamma}}}({\mathbf{z}_{i}})),i\in [N_B]), \Pi^k(\mathbf{p}_n)\big)\nonumber\\
                &-\lambda S_{\alpha}\big(D_{{\mathbf{\delta}}}(\Pi^k(\mathbf{p}_i), i \in \mathcal{B}_j), \Pi^k(\mathbf{p}_n)\big)\Big]. \nonumber
            \end{aligned}
            \end{equation}
            \State Update the discriminator 
            \begin{equation}
            \mathbf{\delta} \gets \mathbf{\delta} + l_D \nabla\mathcal{L}_D(\mathbf{\delta}). \nonumber
            \end{equation}
            \State Generate  $N_B$ IID noise samples $\{\tilde{\mathbf{z}}_i, i\in[N_B]\}\sim \mathbb{P}_{\mathbf{z}}$.
            \State Compute the loss of the generator
            \begin{equation}
            \begin{aligned}
                \mathcal{L}_G(\mathbf{\gamma})= \frac{1}{K\,N_B}\,\, \sum_{k=1}^K \sum_{n \in \mathcal{B}_j} S_{\alpha}\big(D_{{\delta}}(\Pi^k(G_{{\mathbf{\gamma}}}({\tilde{\mathbf{z}}_{i}})),i\in[N_B]), \Pi^k(\mathbf{p}_n)\big).
                \nonumber
            \end{aligned}
            \end{equation}  
            \State Update the generator 
            \begin{equation}
                \mathbf{\gamma} \gets \mathbf{\gamma} - l_G \nabla\mathcal{L}_G(\mathbf{\gamma}).   \nonumber  
            \end{equation}
        \EndFor
    \EndFor \\   $\mathbf{\gamma}^*= \mathbf{\gamma},\qquad \mathbf{\delta}^*=\mathbf{\delta}$.
 \end{algorithmic}
 
 \noindent \hspace*{\algorithmicindent} \ \\ \textbf{Outputs:}\\ 
 \bb  $\mathbf{\delta}^*$: trained discriminator weights; \quad  $\mathbf{\gamma}^*$: trained generator weights.\\
\bb  Simulated scenarios: $ G_{\mathbf{\gamma}^*}(\mathbf{z}_i)$ where $ \mathbf{z}_i\sim \mathbb{P}_{\mathbf{z}}$ IID.
\end{algorithm}

We train {\TGAN} using a range of \textbf{static and dynamic} trading strategies across multiple assets.
Static trading strategy refers to a buy-and-hold portfolio. Dynamic trading strategies have scenario- and time-dependence portfolios. We consider two types of dynamic strategies: mean-reversion strategies and trend-following strategies. 

We compare {\TGAN} with four benchmark models:
\begin{itemize}
    \item[(1)] {\TGAN-Raw}: {\TGAN} trained (only) with static buy-and-hold strategies on {\it individual asset}.
    \item[(2)] {\TGAN-Static}: {\TGAN} trained (only) with static multi-asset portfolios.
    \item[(3)] Historical Simulation Method (HSM): Using VaR and ES computed from historical data as the prediction  for VaR and ES of future data. %sampling historical data and computing the empirical risk measure estimates. 
    \item[(4)] Wasserstein GAN (WGAN): trained on asset return data with  Wasserstein distance as the loss function. We refer to \citet{arjovsky2017wasserstein} for more  details on WGAN.
\end{itemize}

 {\TGAN-Raw} is trained with the returns of single assets, similarly to previous GAN-based generators such as Quant-GAN (\citet{wiese2020quant})or  \citet{ koshiyama2020generative,ni2020conditional,takahashi2019modeling,li2020generating}). {\TGAN-Static} is trained with the PnLs of multi-asset portfolios which is more flexible than {\TGAN-Raw} by allowing different capital allocations among different assets. This could potentially capture the correlation patterns among different assets. In addition to the static portfolios, {\TGAN} also includes dynamic strategies to capture temporal dependence information in financial time series.

\FloatBarrier

\subsection{Performance evaluation criteria}\label{sec:performance_evaluation_criteria}
We introduce the following criteria to compare  the scenarios simulated using {\TGAN} with other simulation models: (1) tail behavior comparison; (2) structural characterizations such as correlation and auto-correlation; and (3) model verification via (statistical) hypothesis testing  such as the Score-based Test and Coverage Test. The first two evaluation criteria are applied throughout the numerical analysis for both synthetic and real financial data, while the hypothesis tests are only for synthetic data as they require the knowledge of ``oracle'' estimates of the true data generating process.

\paragraph{Tail behavior.} To evaluate the accuracy of  VaR (ES) estimates for the trading strategies  computed under the generated scenarios, we compute, for any strategy $k$ $(1 \leq k \leq K)$,  the relative error of VaR is defined as $\frac{\left|{\rm VaR}_{\alpha}\left( \Pi^k,\mathbb{P}_G^{(\mathfrak{N})} \right) - {\rm VaR}_{\alpha}\left(\Pi^k,\mathbb{P}_{r}\right)\right|}{\left|{\rm VaR}_{\alpha}\left(\Pi^k,\mathbb{P}_{r}\right)\right|}$, where ${\rm VaR}_{\alpha}\left( \Pi^k,\mathbb{P}_{r}\right)$ is the  true $\alpha$-VaR  for strategy $k$  and ${\rm VaR}_{\alpha}\left( \Pi^k,\mathbb{P}_G^{(\mathfrak{N})} \right)$  the empirical estimate using $\mathfrak{N}$ generated samples. Similarly, for the estimates ${\rm ES}_{\alpha}\left( \Pi^k,\mathbb{P}_G^{(\mathfrak{N})} \right)$, we define the relative error as $\frac{\left|{\rm ES}_{\alpha}\left( \Pi^k,\mathbb{P}_G^{(\mathfrak{N})} \right) - {\rm ES}_{\alpha}\left( \Pi^k,\mathbb{P}_{r}\right)\right|}{\left|{\rm ES}_{\alpha}\left( \Pi^k,\mathbb{P}_{r}\right)\right|}.$ 

We then use the following {\it average relative error} for VaR and ES as the overall measure of model performance:
\begin{equation}
    {\rm RE}(\mathfrak{N}) = \frac{1}{2K} \sum_{k=1}^{K} \left( \frac{\left|{\rm VaR}_{\alpha}\left( \Pi^k, \mathbb{P}_G^{(\mathfrak{N})}\right) - {\rm VaR}_{\alpha}\left(\Pi^k, \mathbb{P}_{r}\right)\right|}{\left|{\rm VaR}_{\alpha}\left(\Pi^k,\mathbb{P}_{r}\right)\right|} + \frac{\left|{\rm ES}_{\alpha}\left( \Pi^k,\mathbb{P}_G^{(\mathfrak{N})} \right) - {\rm ES}_{\alpha}\left(\Pi^k, \mathbb{P}_{r}\right)\right|}{\left|{\rm ES}_{\alpha}\left(\Pi^k, \mathbb{P}_{r}\right)\right|}\right). \nonumber
\end{equation}
A useful benchmark  to compare the above relative error with is the {\it sampling error}, when using a finite a sample of same size from the true distribution:
\begin{equation}
    {\rm SE}(\mathfrak{N}) = \frac{1}{2K} \sum_{k=1}^{K} \left( \frac{\left|{\rm VaR}_{\alpha}\left( \Pi^k,\mathbb{P}_r^{(\mathfrak{N})}\right) - {\rm VaR}_{\alpha}\left(\Pi^k, \mathbb{P}_{r}\right)\right|}{\left|{\rm VaR}_{\alpha}\left(\Pi^k, \mathbb{P}_{r}\right)\right|} + \frac{\left|{\rm ES}_{\alpha}\left( \Pi^k,\mathbb{P}_r^{(\mathfrak{N})}\right) - {\rm ES}_{\alpha}\left(\Pi^k, \mathbb{P}_{r}\right)\right|}{\left|{\rm ES}_{\alpha}\left(\Pi^k, \mathbb{P}_{r}\right)\right|}\right),\nonumber
\end{equation}
where ${\rm VaR}_{\alpha}\left( \Pi^k,\mathbb{P}_r^{(\mathfrak{N})}\right)$ and ${\rm ES}_{\alpha}\left( \Pi^k,\mathbb{P}_r^{(\mathfrak{N})}\right)$ are the ``oracle'' estimates for VaR and ES of strategy $k$ using a sample of size $\mathfrak{N}$ from the true probability distribution.

Clearly ${\rm SE}(\mathfrak{N}) $ is a lower bound for accuracy and one cannot do better than having  ${\rm RE}(\mathfrak{N})$  of the same order or magnitude as ${\rm SE}(\mathfrak{N}) $. we note that most studies on generative models for time series  e.g. (\citet{wiese2020quant}) fail to use such benchmarks.

{ We also use \textit{rank-frequency distribution} to visualize the tail behaviors of the simulated data versus the market data. Rank-frequency distribution is  a discrete form of the quantile function, i.e., the inverse cumulative distribution, giving the size of the element at a given rank. By comparing the rank-frequency distribution of the market data and simulated data of different strategies, we gain an understanding of how good the simulated data is in terms of the risk measures of different  strategies. }

\paragraph{Temporal and cross-asset dependence.}{
To test whether  {\TGAN} is capable of capturing structural properties, such as temporal and cross-asset dependence,  we calculate  of the output price scenarios for each generator : (1) the sum of the absolute difference between the \textit{correlation} coefficients of the input price scenario and those of generated price scenario, and (2) the sum of the absolute difference between the
\textit{autocorrelation} coefficients (up to 10 lags) of the input price scenario and those of the generated price scenario.}

\paragraph{Hypothesis testing for synthetic data.} 
Given the benchmark strategies  and a scenario generator, we are interested in examining how accurate risk measures  for benchmark strategies estimated from simulated scenarios  are compared to ``oracle'' estimates given knowledge of the data generating process. Here, the generator may represent {\TGAN}, {\TGAN}-Raw, {\TGAN}-Static, HSM, or WGAN.

We explore two methods, the Score-based Test and the Coverage Test, to verify the relationship between the generator  and the data.  We first introduce the \textit{Score-based Test} to verify the % following 
hypothesis
\begin{equation}
\begin{aligned}
\mathcal{H}_{0}: \qquad  &
\mathbb{E}_{\mathbf{p}\sim\mathbb{P}_r}\left[ S_{\alpha}\big({\rm VaR}_{\alpha}\left( \Pi^k , \mathbb{P}_{G}\right), {\rm ES}_{\alpha}\left( \Pi^k , \mathbb{P}_{G}\right), {\Pi^k(\mathbf{p})}\big)\right]\\
&=\mathbb{E}_{\mathbf{p}\sim\mathbb{P}_r}\left[ S_{\alpha}\big({\rm VaR}_{\alpha}\left( \Pi^k ,\mathbb{P}_{r}\right), {\rm ES}_{\alpha}\left( \Pi^k ,\mathbb{P}_{r}\right), {\Pi^k(\mathbf{p})}\big)\right],\nonumber
\end{aligned}
\end{equation}
where $\mathbb{P}_{G}$ is the distribution of the output time series from the generator $G$. By making use of the joint elicitability property of VaR and ES, \citet{FZG2015} proposed the following test statistic:
$$
T^k=\frac{\bar{S}_{G}^{k}-\bar{S}_{0}^{k}}{\hat{\sigma}^{k}}, \qquad {\rm where}
$$
\begin{eqnarray}
\bar{S}_{G}^{k}&=&\frac{1}{\mathfrak{N}} \sum_{i=1}^{\mathfrak{N}} S_{\alpha}\big({\rm VaR}_{\alpha}\left( \Pi^k, \mathbb{P}_{G}\right), {\rm ES}_{\alpha}\left( \Pi^k, \mathbb{P}_{G}\right), {\Pi^k(\mathbf{p}_i)}\big), \nonumber\\
\bar{S}_{0}^{k}&=&\frac{1}{\mathfrak{N}} \sum_{i=1}^{\mathfrak{N}} S_{\alpha}\big({\rm VaR}_{\alpha}\left( \Pi^k,\mathbb{P}_{r}\right), {\rm ES}_{\alpha}\left( \Pi^k,\mathbb{P}_{r}\right), {\Pi^k(\mathbf{p}_i)}\big), \nonumber\\
\hat{\sigma}^{k} &=& \sqrt{\frac{\hat{\sigma}_{G}^2 + \hat{\sigma}_{0}^2}{\mathfrak{N}}}.\nonumber
\end{eqnarray}
Here $\{\mathbf{p}_i\}_{i=1}^{\mathfrak{N}}$ represents the observations from $\mathbb{P}_r$ and $\{\Pi^k(\mathbf{p}_i)\}_{i=1}^\mathfrak{N}$ represents the PnL observations of strategy $k$ under $\mathbb{P}_r$. $\mathbb{P}_G$ denotes the distribution of generated data from generator  $G$. ${\rm VaR}_{\alpha}\left( \Pi^k,\mathbb{P}_{G}\right)$ and ${\rm ES}_{\alpha}\left( \Pi^k,\mathbb{P}_{G}\right)$ represent the estimates of VaR and ES for PnLs of strategy $k$ evaluated under $\mathbb{P}_{G}$. ${\rm VaR}_{\alpha}\left( \Pi^k,\mathbb{P}_{r}\right)$ and ${\rm ES}_{\alpha}\left( \Pi^k,\mathbb{P}_{r}\right)$ represent the ground-truth estimates of VaR and ES for PnLs of strategy $k$ evaluated under $\mathbb{P}_{r}$. 
Furthermore, 
$\hat{\sigma}_{G}^2$ and $\hat{\sigma}_{0}^2$ are the empirical variances of \\$S_{\alpha}\big({\rm VaR}_{\alpha}\left( \Pi^k,\mathbb{P}_{G}\right), {\rm ES}_{\alpha}\left( \Pi^k,\mathbb{P}_{G}\right), {\Pi^k(\mathbf{p})}\big)$ and $S_{\alpha}\big({\rm VaR}_{\alpha}\left( \Pi^k,\mathbb{P}_{r}\right), {\rm ES}_{\alpha}\left( \Pi^k,\mathbb{P}_{r}\right), {\Pi^k(\mathbf{p})}\big)$, respectively. Under $\mathcal{H}_{0}$, the test statistic $T^{k}$ has expected value equal to zero, and the asymptotic normality of the test statistic $T^k$ can be similarly proved as in \citet{diebold2002comparing}. 

We also examine   the {\it Coverage Test}  (\citet{kupiec1995techniques}) which measures generator performance based on the exceedance rate of estimated quantile levels. 
\citet{kupiec1995techniques} proposed the following test statistic: 
\begin{equation}
{\rm LR} =-2 \ln \left(\frac{(1-\alpha)^{\mathfrak{N}-C^{k}(\mathfrak{N})} \alpha^{C^{k}(\mathfrak{N})}}{\left(1-\frac{C^{k}(\mathfrak{N})}{\mathfrak{N}}\right)^{\mathfrak{N}-C^{k}(\mathfrak{N})}\left(\frac{C^{k}(\mathfrak{N})}{\mathfrak{N}}\right)^{C^{k}(\mathfrak{N})}}\right),\nonumber
\end{equation}
where $C^{k}(\mathfrak{N}) = \sum_{i=1}^{\mathfrak{N}} \one_{\left\{\Pi^k(\mathbf{p}_i) < {\rm VaR}_{\alpha}\left( \Pi^k,\mathbb{P}_{G}\right)\right\}}$ represents the number of exceedances  observed across $\mathfrak{N}$ scenarios.
Under the null hypothesis  
\begin{equation}
    \mathcal{H}_{0}: \mathbb{P}\left(\Pi^k(\mathbf{p}) < {\rm VaR}_{\alpha}\left(\Pi^k,\mathbb{P}_{G}\right)\right)=\alpha. \nonumber
\end{equation}
where $\Pi^k(\mathbf{p})$ represents the PnL of strategy $k$, we have ${\rm LR} \sim \chi_{1}^{2}$.

\paragraph{In-sample test vs out-of-sample test.} Throughout the experiments,  both  in-sample tests  and out-of-sample tests are used to evaluate the trained generators. In particular, the in-sample tests are performed on the training data, whereas the out-of-sample tests are performed on the testing data. For each {\TGAN} variant, the in-sample test uses the same set of strategies as in its loss function. For example, the in-sample test for {\TGAN}-Raw is performed with buy-and-hold strategies on individual assets.  Out-of-sample tests use different trading strategies, not necessarily seen in the training phase.

\section{Numerical experiments with synthetic data}
\label{sec:model_validation}

We test the performance of {\TGAN} on a synthetic data set, for which we can  compare to benchmarks computed using the exact loss distribution.  We divide the entire data set into two disjoint subsets, i.e. the training set and the test set, with no overlap in time periods. The training set  is used to learn the model parameters, and the testing data is used to evaluate the out-of-sample (OOS) performance of different models. In this examination, 50,000 samples are used for training and 10,000 samples are used for performance evaluation.

The main takeaway from our comparison against benchmark simulation models is that the  consistent tail-risk behavior is difficult to attain by {\it only} training on price sequences, without incorporating the dynamic trading strategies in the loss function, as we propose to do in our pipeline. As a consequence, if the user is indeed interested in including dynamic trading strategies in the portfolio, training a generator  on raw asset returns, as suggested by \citet{wiese2020quant}, will be insufficient.

\subsection{Synthetic multi-asset scenarios}  \label{sec:validation_highdimension}
We now test the method on a simulated data set, consisting of five correlated return series with different temporal patterns and tail behaviors. 
More specifically we simulate a 5-dimensional vector series $X(t)$ whose components are respectively given by:
\begin{itemize}
    \item $X_1(t)\sim N(0,1)$ IID
\item $X_2(t)$ is an AR(1) process 
AR(1) with autocorrelation $\phi_1>0$, \item $X_3(t)$ is an AR(1) process  with autocorrelation $\phi_2<0$, \item $X_4(t)$ is a  GARCH(1, 1) process with Student-$t(\nu_1)$ noise and \item $X_5(t)$ is a  GARCH(1, 1) process with Student-$t(\nu_2)$ noise.
\end{itemize}
Details are provided in Appendix \ref{app:para_syn}.
We examine the performance with one quantile value $\alpha=0.05$. The architecture of the network configuration is summarized in Table \ref{tab:configuration} in Appendix \ref{app:configuration}. Experiments with other quantile levels or multiple quantile levels are demonstrated in Section \ref{sec:risk_levels}.

Figure \ref{fig:model_validation_performance_is} reports the convergence of in-sample errors\footnote{The in-sample error of WGAN is not reported in Figure \ref{fig:model_validation_performance_is} because WGAN uses a different training metric.},  and Table \ref{tab:syn_error} summarizes the out-of-sample errors of {\TGAN-Raw}, {\TGAN-Static}, {\TGAN} and WGAN. 

\begin{figure}[t]
    \centering
    \includegraphics[width=.6\textwidth, trim=1.2cm 0cm 3cm  3cm,clip]{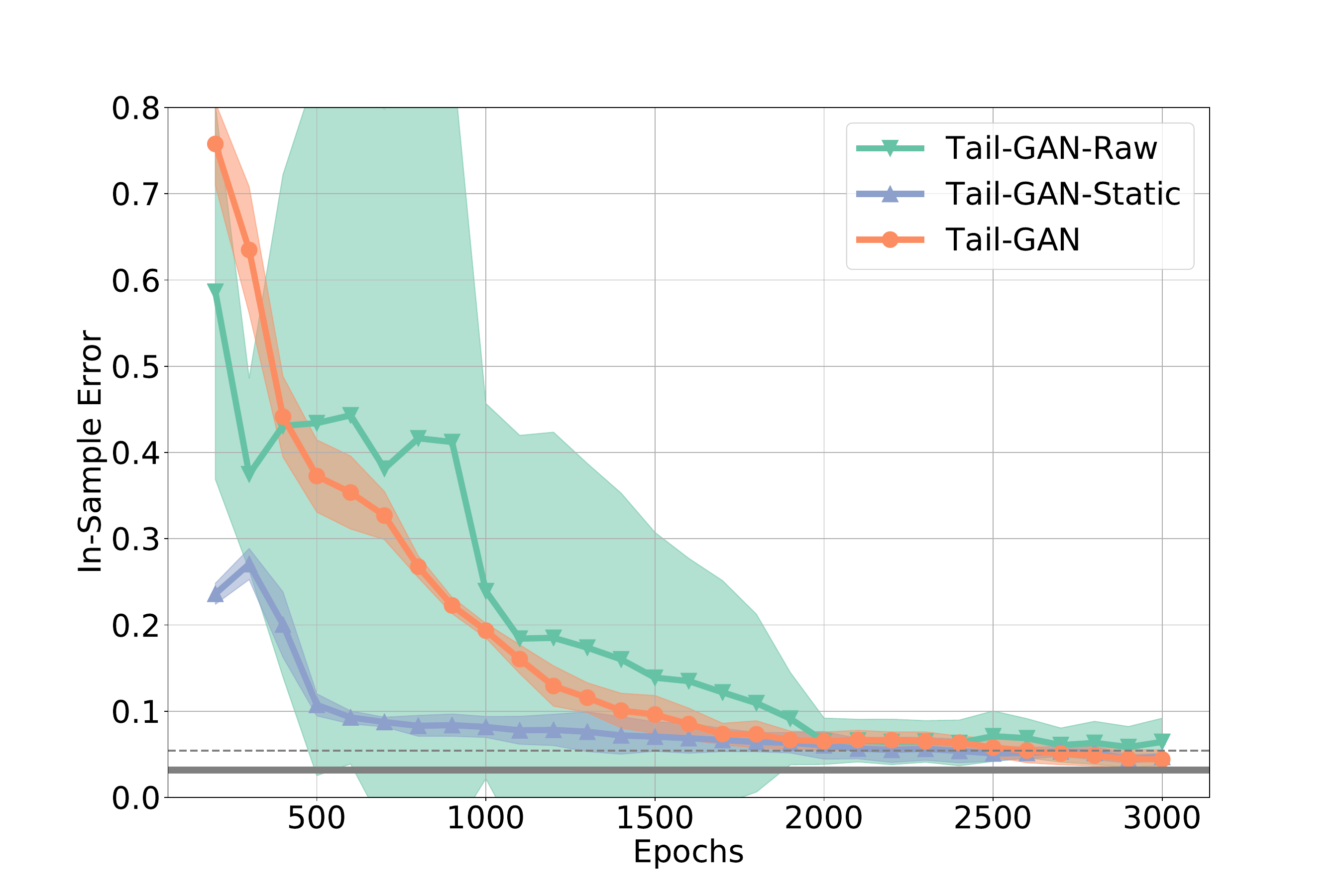}
    \vspace{-3mm}
    \caption{Training performance: relative error ${\rm RE}(1000)$ with 1000 samples. Grey horizontal line: average simulation error {${\rm SE}(1000)$}. Dotted line: average simulation error plus one standard deviation. Each experiment is repeated  five  times with different random seeds. The performance is visualized with mean (solid lines) and standard deviation (shaded areas).}
    \label{fig:model_validation_performance_is}
\end{figure}

\paragraph{Performance accuracy.}
We draw the following observations from Figure \ref{fig:model_validation_performance_is} and Table \ref{tab:syn_error}. 
\begin{itemize}
\itemsep0em
\item For the evaluation criterion ${\rm RE}(1000)$ (see Figure \ref{fig:model_validation_performance_is}), all three generators, {\TGAN-Raw}, {\TGAN-Static} and {\TGAN}, converge within 2000 epochs with errors smaller than 10\%. This implies that all three generators are able to capture the static information contained in the market data.  
\item For the evaluation criterion   ${\rm RE}(1000)$, with both static portfolios and dynamic strategies on out-of-sample tests (see Table \ref{tab:syn_error}), only {\TGAN} converges to an error 4.6\%, whereas the other two generators fail to capture the dynamic information in the market data.  
\item Compared to {\TGAN-Raw} and {\TGAN-Static}, {\TGAN} has the lowest training variance across multiple experiments (see standard deviations in Table \ref{tab:syn_error}). This implies that {\TGAN} has the most stable performance.

\item Compared to {\TGAN}, WGAN yields less competitive out-of-sample performance in terms of generating scenarios that have consistent tail risks of static portfolios and dynamic strategies. Indeed, the objective function of WGAN is not very sensitive to the tail of the distribution, which does not guarantee the accuracy of tail risk measures for   dynamic strategies.
 
\end{itemize}

\begin{table}[H]
    \centering
    \resizebox{0.95\textwidth}{!}{\begin{tabular}{lcccccc}
\toprule
            & SE(1000)   & HSM & {\TGAN-Raw} & {\TGAN-Static} & {\TGAN} & WGAN \\\midrule
Out of sample    &  3.0 & 3.4  & 83.3 & 86.7 & 4.6 & 21.3 \\
    Error (\%)       &  (2.2) & (2.6) & (3.0) & (2.5) & (1.6) & (2.2) \\\bottomrule
\end{tabular}}
    \caption{Mean and standard deviation (in parentheses) of relative errors for out-of-sample tests.
    Each experiment is repeated five times with different random seeds.} 
    \label{tab:syn_error}
\end{table}

{Figure \ref{fig:synthetic_tail_ar5} shows the empirical quantile function of the strategy PnLs evaluated with price scenarios sampled from {\TGAN-Raw}, {\TGAN-Static},  {\TGAN}, and WGAN. The testing strategies are, from left to right (in Figure \ref{fig:synthetic_tail_ar5}), static single-asset portfolio (buy-and-hold strategy),   single-asset mean-reversion strategy  and  single-asset trend-following strategy. 
We only demonstrate here the performance of the AR(1) model, and the results for other assets are provided in Figure \ref{fig:synthetic_tail} in Appendix \ref{app:supply_results}.\footnote{We observe from Figure \ref{fig:synthetic_tail} that for other input scenarios such Gaussian, AR(1), and GARCH(1,1), WGAN is able to generate scenarios that match the tail risk characteristics of benchmark trading strategies.}
We compare the rank-frequency distribution of  PnLs evaluated with input price scenarios (in blue),  three {\TGAN} generators (in orange, red and green, resp.), and WGAN (in purple). Based on the results depicted in Figure \ref{fig:synthetic_tail_ar5}, we conclude that }
\begin{itemize}
\itemsep0em
\item All three variants of {\TGAN} are able to capture the tail properties of the static single-asset portfolio at quantile levels above 1\%, as shown in the first column of Figure \ref{fig:synthetic_tail_ar5}. 
\item For dynamic strategies, only {\TGAN} is able to generate accurate tail statistics, as shown in the second and third columns of Figure \ref{fig:synthetic_tail_ar5}.%
\item {\TGAN-Raw} and {\TGAN-Static} underestimate the risk of the mean-reversion strategy at $\alpha=5\%$ quantile level, and overestimate the risk of the trend-following strategy at $\alpha=5\%$ quantile level, as illustrated in the second and third columns of Figure \ref{fig:synthetic_tail_ar5}.

\item  WGAN fails to generate scenarios that retain consistent risk measures for heavy tailed models such as  GARCH(1,1) with $t(5)$ noise. 

\end{itemize}

\begin{figure}[!htbp]
\centering
\includegraphics[width=1.0\textwidth, trim=5cm 0mm 0cm 0cm,clip]{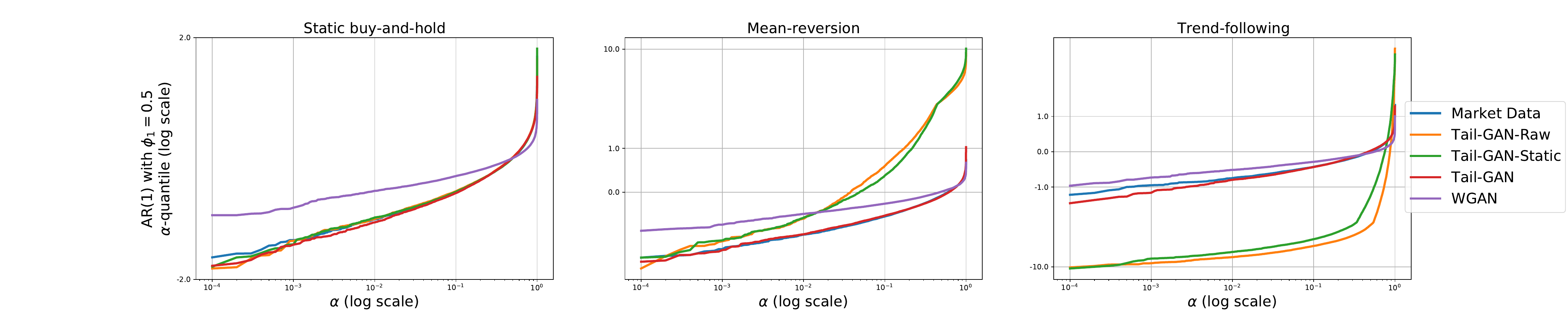}
\caption{Tail behavior via the empirical rank-frequency distribution of the strategy PnL (based on AR(1) with autocorrelation $0.5$). The columns represent the strategy types.}
\label{fig:synthetic_tail_ar5}
\end{figure}

\paragraph{Learning the temporal and correlation patterns.}
Figures \ref{fig:synthetic_corr} and \ref{fig:synthetic_autocorr} show the correlation  and auto-correlation patterns of market data (Figures \ref{fig:synthetic_corr_1} and 
\ref{fig:synthetic_autocorr_1}) and simulated data from {\TGAN}-Raw (Figures \ref{fig:synthetic_corr_2} and 
\ref{fig:synthetic_autocorr_2}), {\TGAN}-Static (Figures \ref{fig:synthetic_corr_3} and 
\ref{fig:synthetic_autocorr_3}), {\TGAN} (Figures \ref{fig:synthetic_corr_4} and \ref{fig:synthetic_autocorr_4}), and WGAN (Figures \ref{fig:synthetic_corr_wgan} and \ref{fig:synthetic_autocorr_wgan}).

\begin{figure}[!htbp]
    \centering
\subfigure[Market data.]
{ \label{fig:synthetic_corr_1}
\includegraphics[width=0.17\textwidth, trim=3cm 3cm 3cm  3cm, clip]{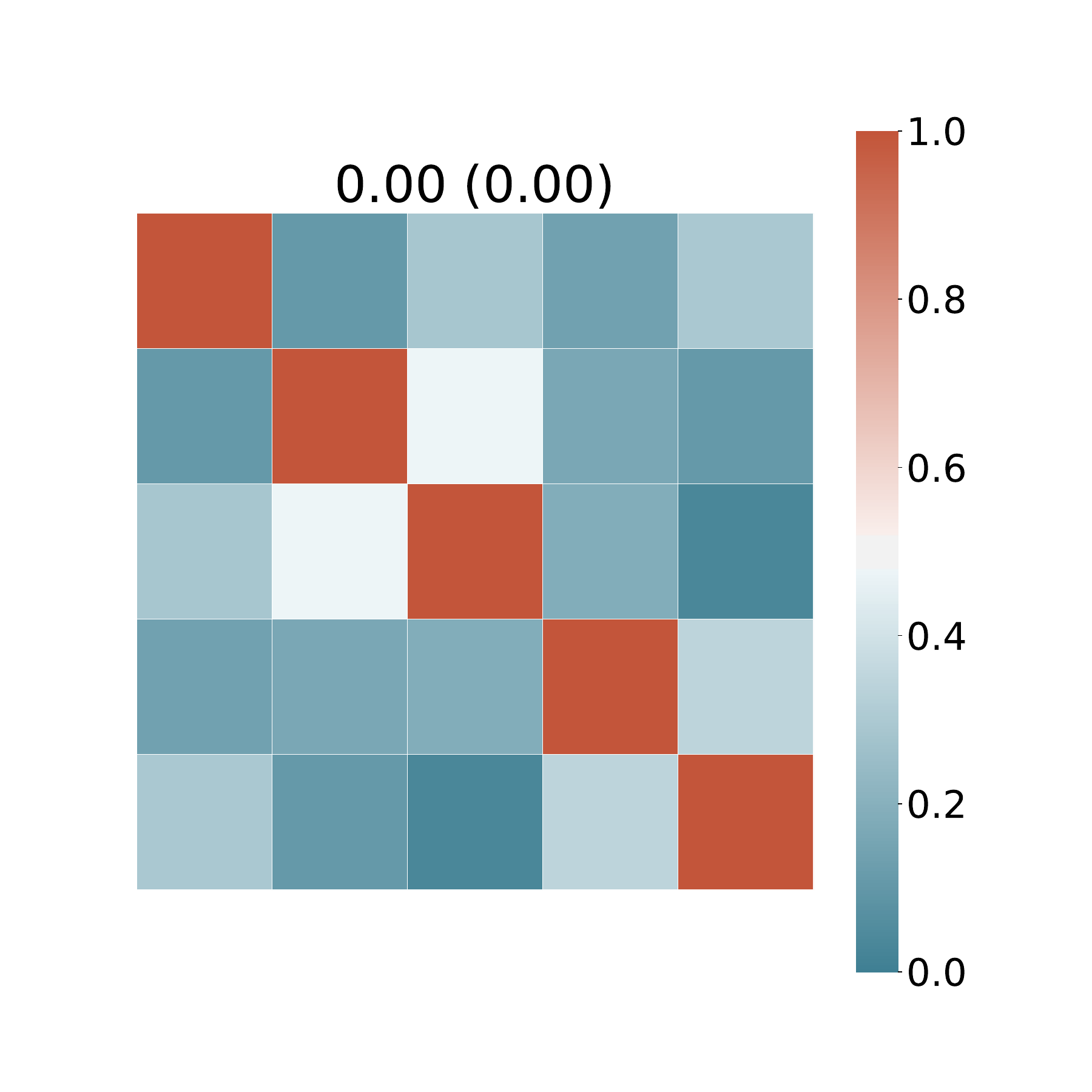}
}
\subfigure[{\TGAN}-Raw.]
%GAN trained with static single-asset portfolio]
{ \label{fig:synthetic_corr_2}
\includegraphics[width=0.17\textwidth, trim=3cm 3cm 3cm  3cm,clip]{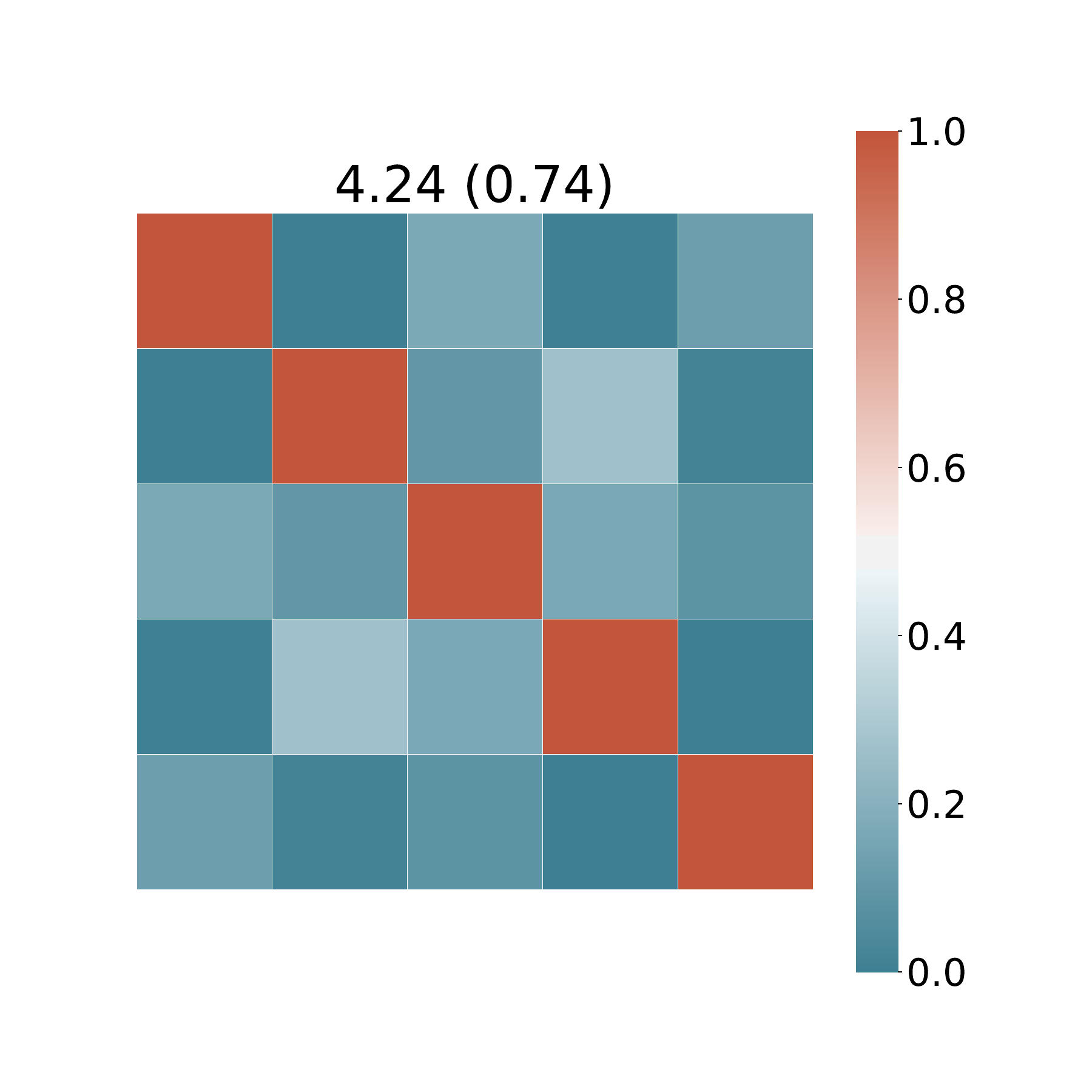}
}
\subfigure[{\TGAN}-Static.]
%GAN trained with static multi-asset portfolio]
{ \label{fig:synthetic_corr_3}
\includegraphics[width=0.17\textwidth, trim=3cm 3cm 3cm  3cm,clip]{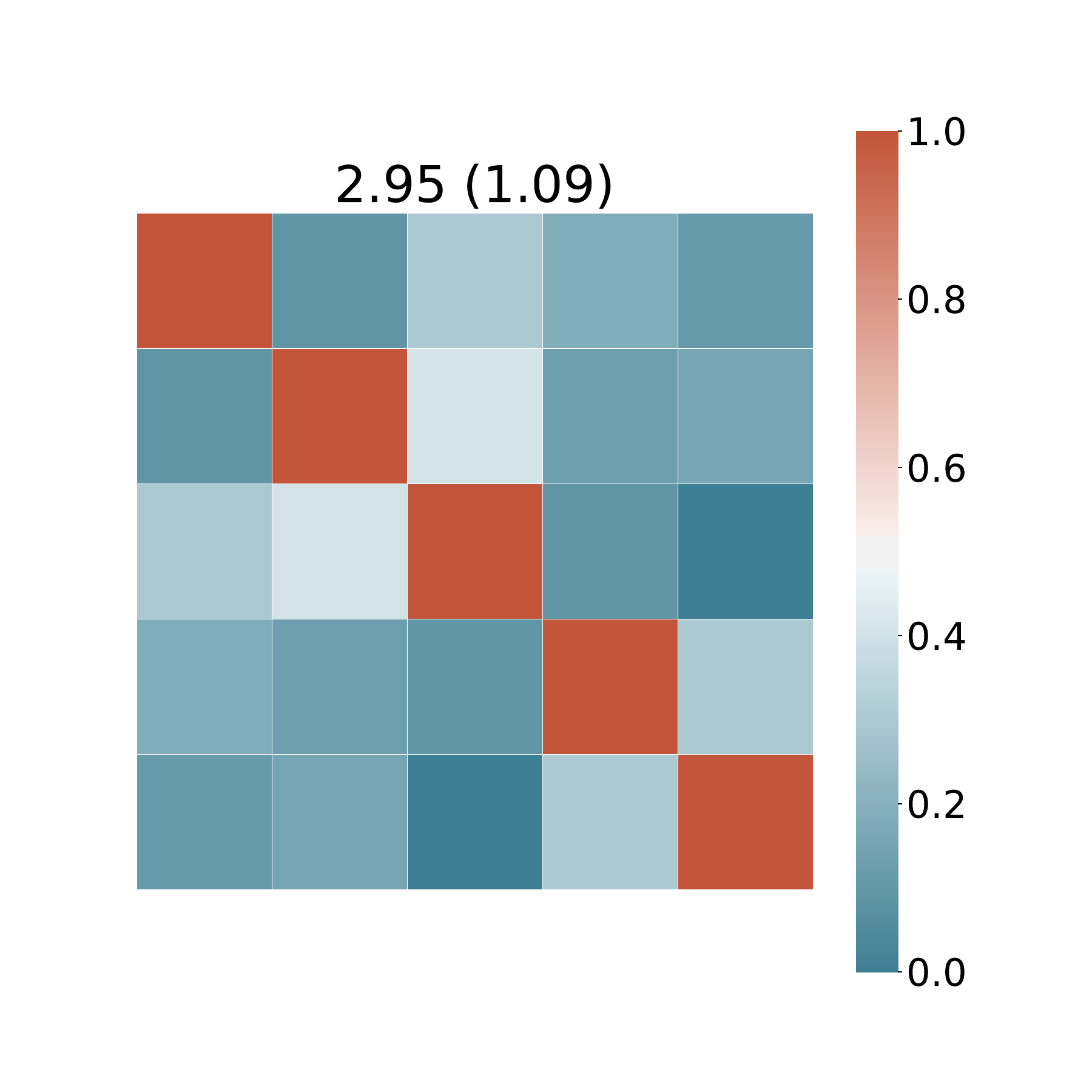}
}
\subfigure[{\TGAN}.]
%GAN trained with static multi-asset portfolio and dynamic strategies]
{ \label{fig:synthetic_corr_4}
\includegraphics[width=0.17\textwidth, trim=3cm 3cm 3cm  3cm,clip]{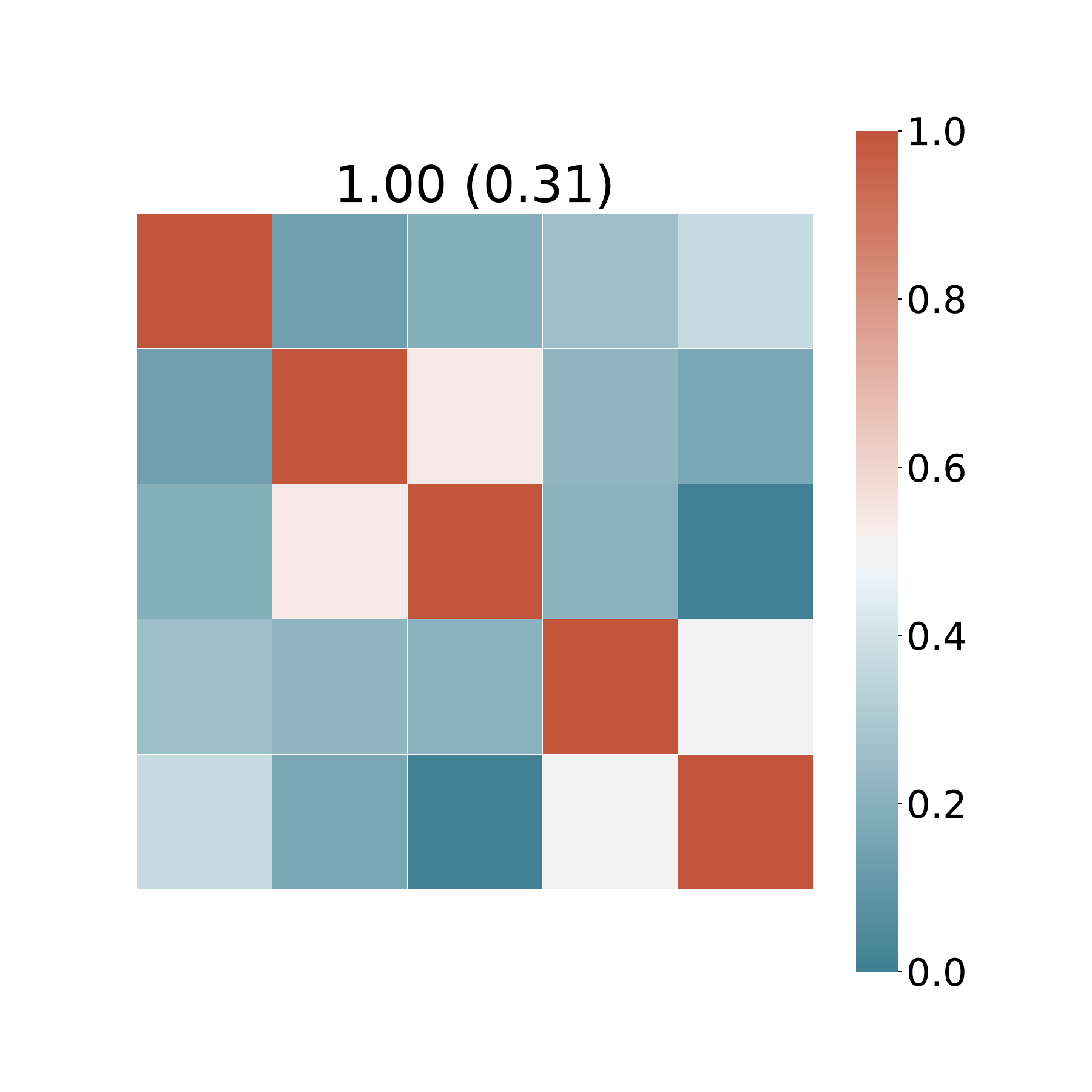}
} 
\subfigure[WGAN.]
{ \label{fig:synthetic_corr_wgan}
\includegraphics[width=0.17\textwidth, trim=3cm 3cm 3cm  3cm,clip]{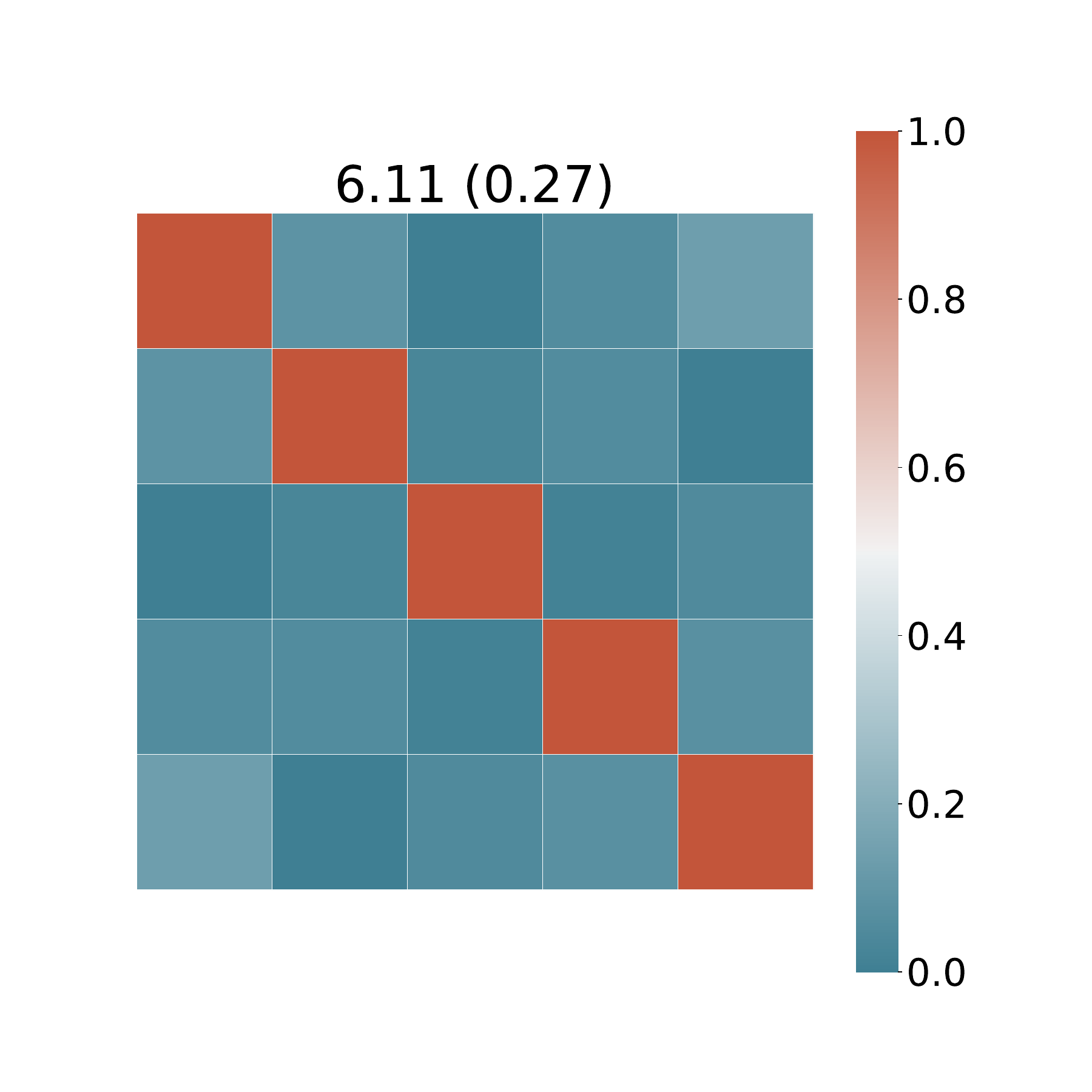}
} 
\caption{\label{fig:synthetic_corr} Correlations {of the price increments} from different trained GAN models: (1) {\TGAN-Raw}, (2) {\TGAN-Static}, (3) {\TGAN}, and (4) WGAN.
 The numbers at the top of each plot denote the mean and standard deviation (in parentheses) of the
sum of the absolute element-wise difference between the correlation matrices, computed with 10,000 training samples and 10,000 generated samples.}
\end{figure}

\begin{figure}[!htbp]
    \centering
\subfigure[Market Data.]
{ \label{fig:synthetic_autocorr_1}
\includegraphics[width=0.17\textwidth, trim=1.3cm 2cm 2cm  2cm,clip]{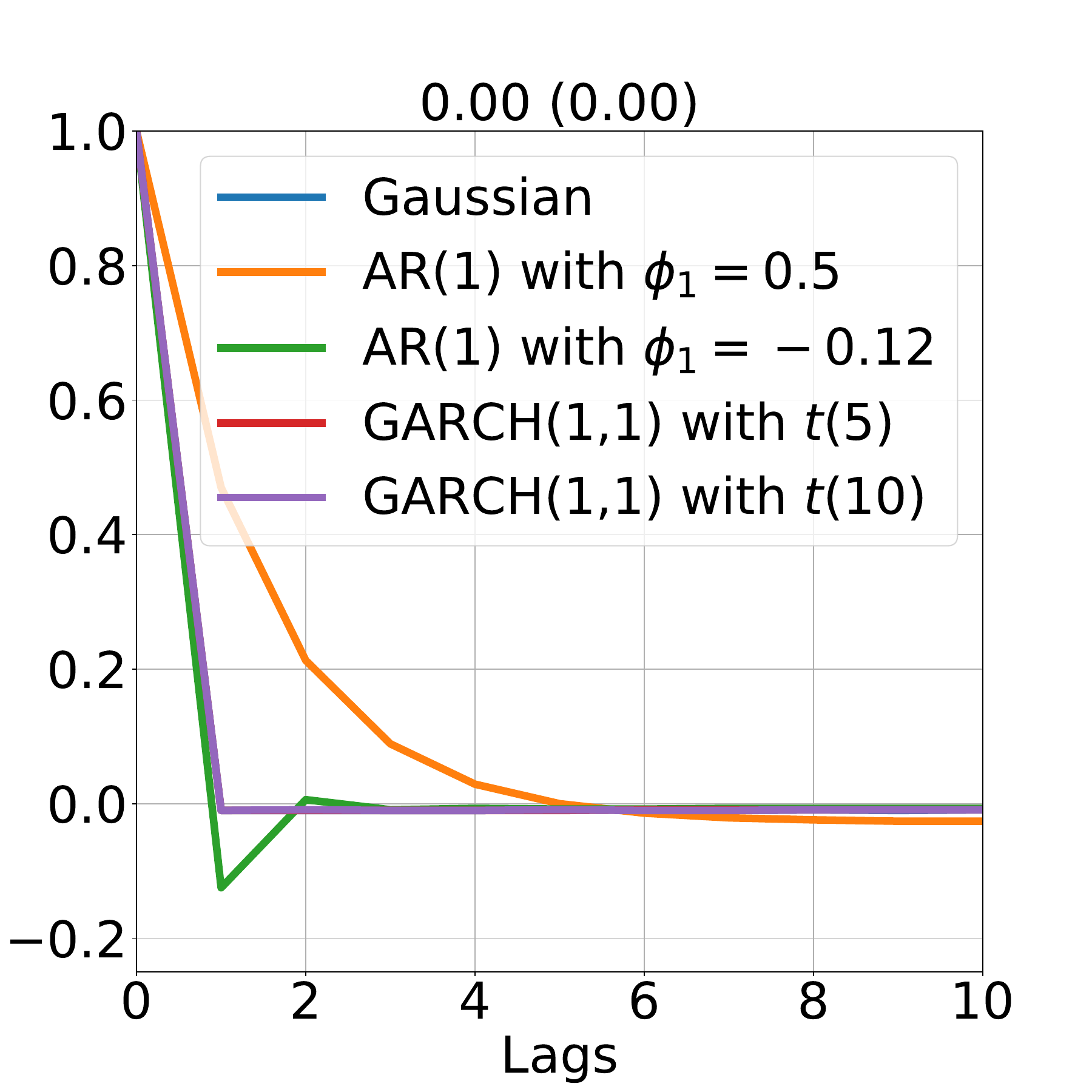}
}
\subfigure[{\TGAN-Raw}.]
%GAN trained with static single-asset portfolio]
{ \label{fig:synthetic_autocorr_2}
\includegraphics[width=0.17\textwidth, trim=1.3cm 2cm 2cm  2cm,clip]{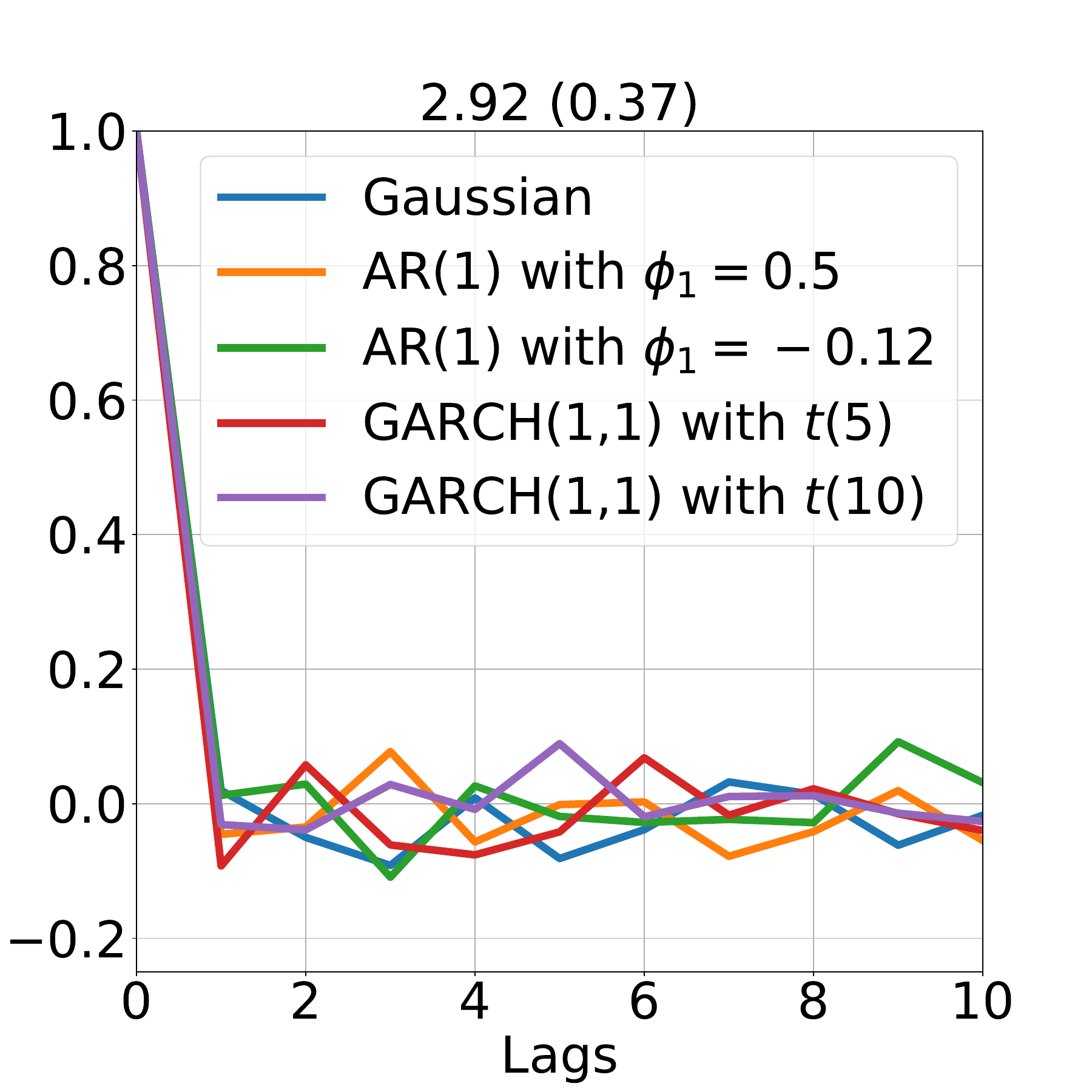}
}
\subfigure[{\TGAN-Static}.]
%GAN trained with static multi-asset portfolio]
{ \label{fig:synthetic_autocorr_3}
\includegraphics[width=0.17\textwidth, trim=1.3cm 2cm 2cm  2cm,clip]{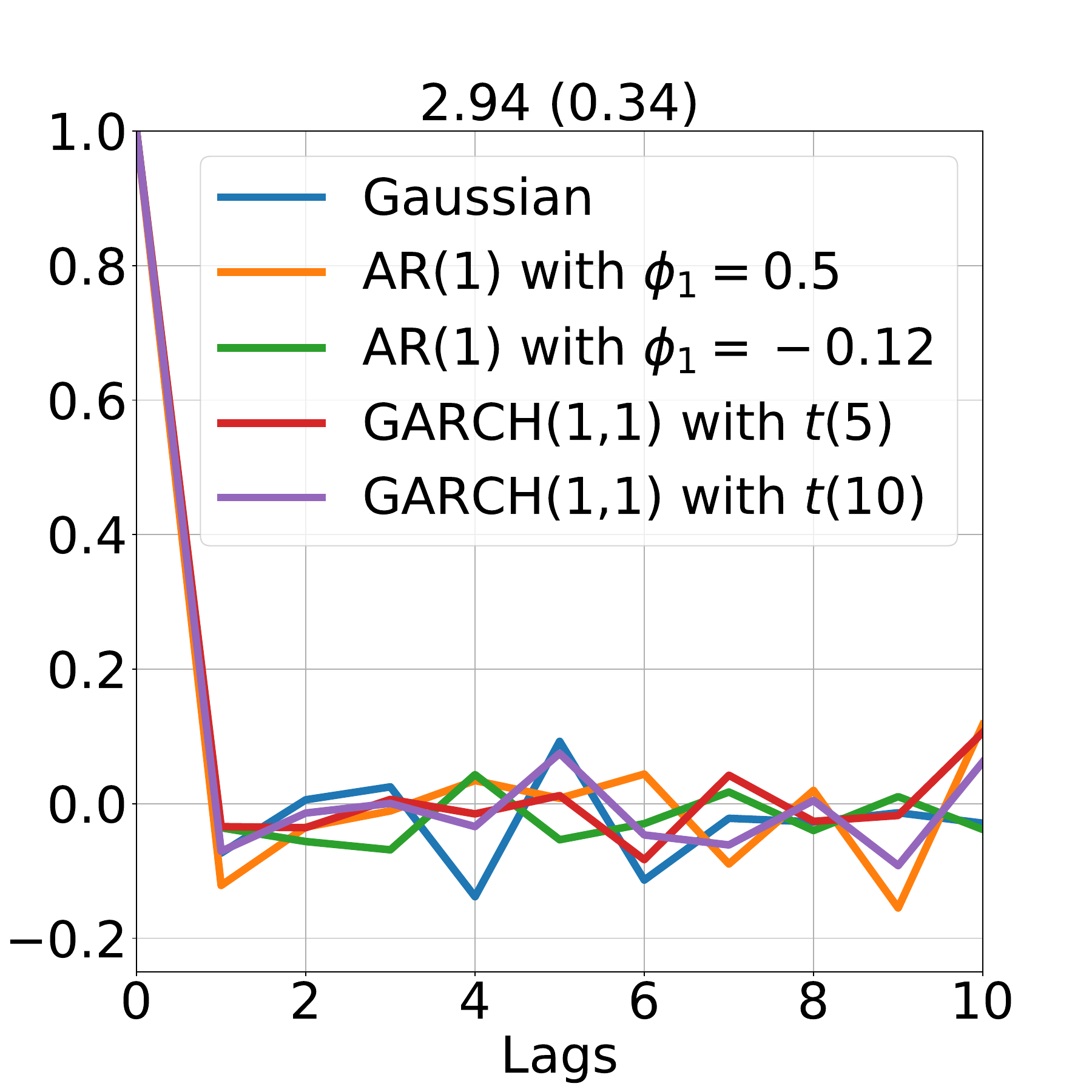}
}
\subfigure[{\TGAN}.] 
{ \label{fig:synthetic_autocorr_4}
\includegraphics[width=0.17\textwidth, trim=1.3cm 2cm 2cm  2cm,clip]{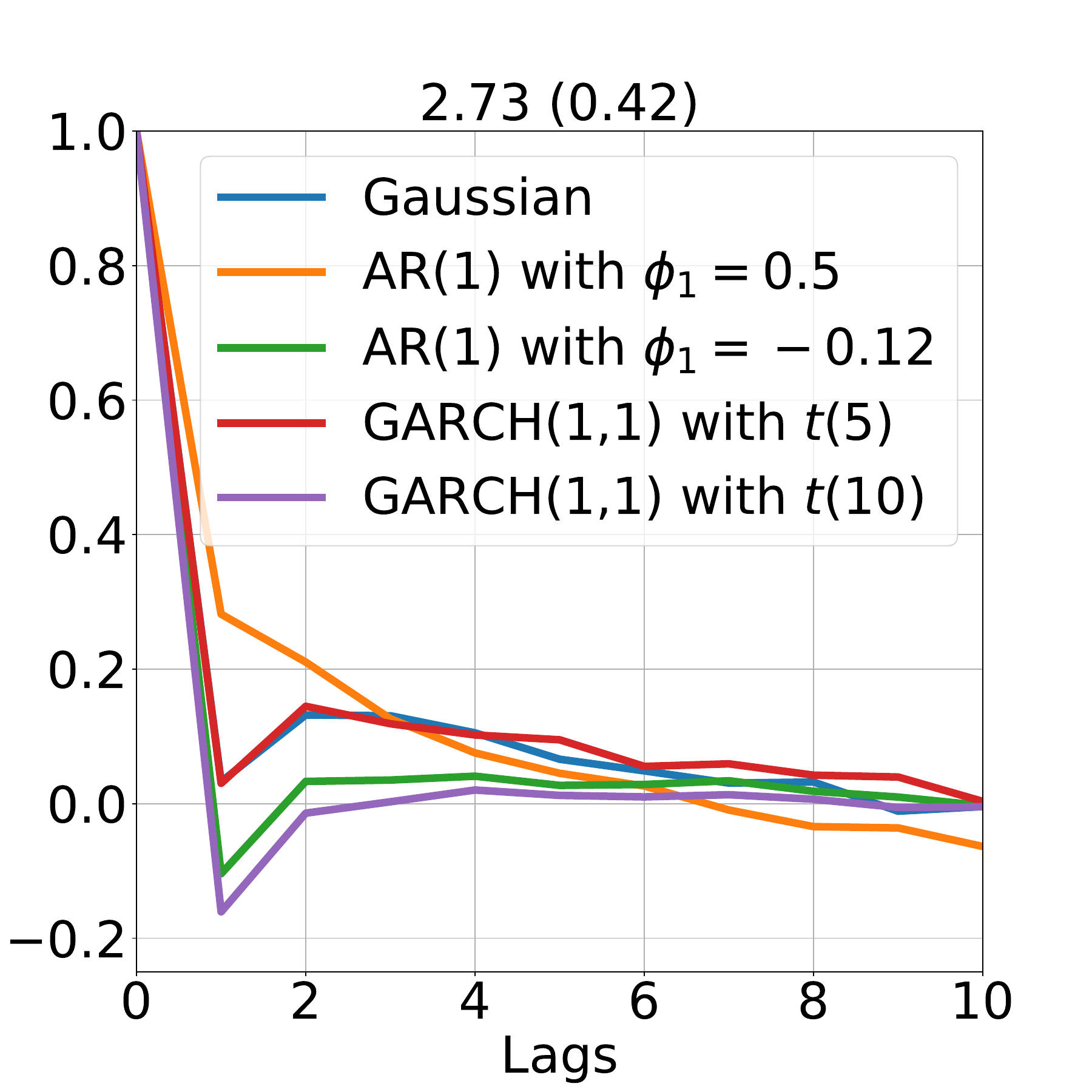}
}
\subfigure[WGAN.]
{ \label{fig:synthetic_autocorr_wgan}
\includegraphics[width=0.17\textwidth, trim=1.3cm 2cm 2cm  2cm,clip]{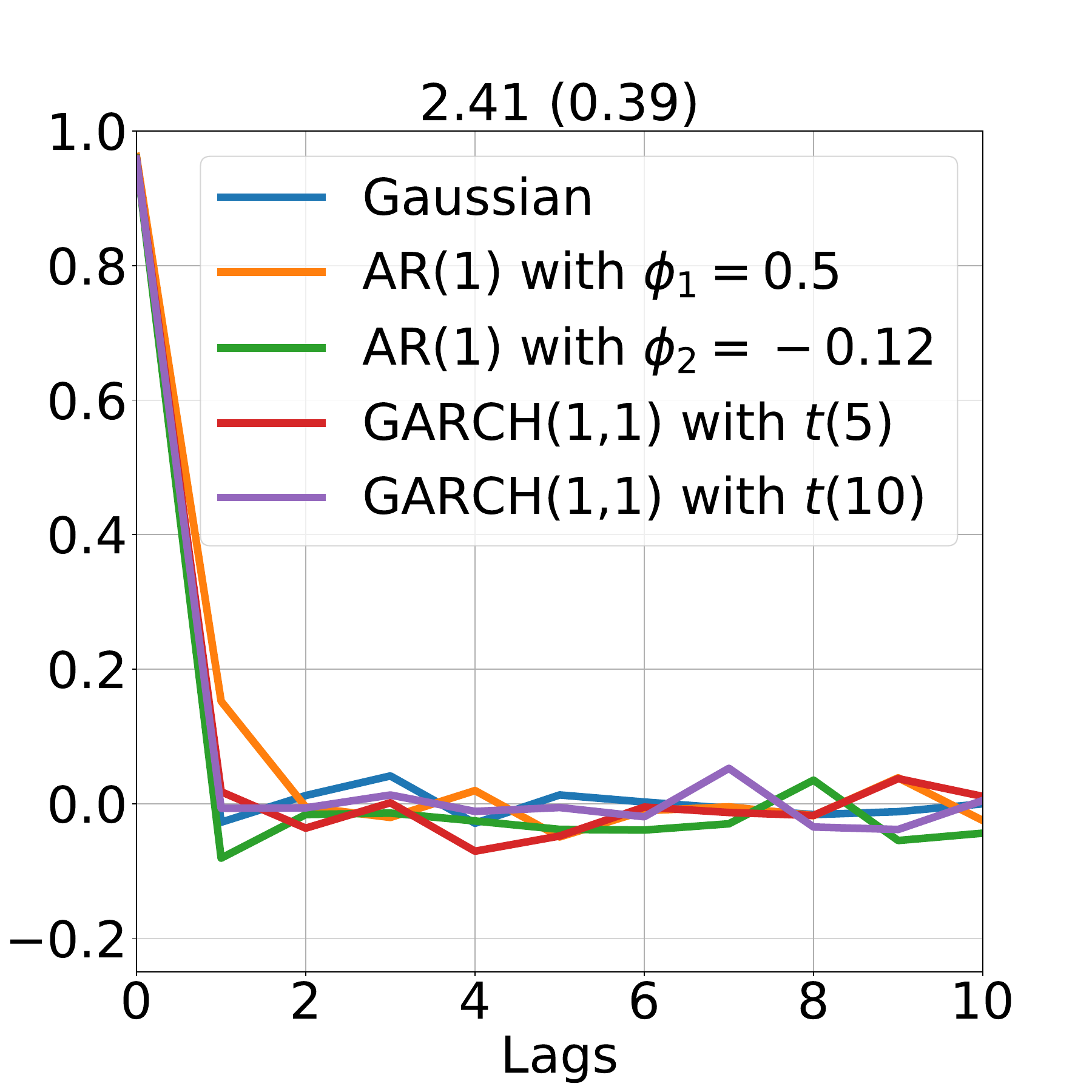}
}
\caption{\label{fig:synthetic_autocorr} Auto-correlations {of the price increments} from different trained GAN models: (1) {\TGAN-Raw}, (2) {\TGAN-Static}, (3) {\TGAN}, and (4) WGAN. The numbers at the top of each plot denote the mean and standard deviation (in parentheses) of the sum of the absolute difference between the auto-correlation coefficients computed with 10,000 training samples and 10,000 generated samples.}
\end{figure}

Figures \ref{fig:synthetic_corr} and \ref{fig:synthetic_autocorr}  demonstrate that the auto-correlation  and cross-correlations  returns are best reproduced by {\TGAN}, trained on multi-asset dynamic portfolios. On the contrary, {\TGAN-Raw} and WGAN, trained on raw returns, have the lowest accuracy in this respect. This illustrates the importance of training the algorithm on benchmark strategies instead of only raw returns.

\paragraph{Statistical significance.} Table \ref{tab:statistical_tests} summarizes the statistical test results for Historical Simulation Method, {\TGAN}-Raw, {\TGAN}-Static, {\TGAN}, and WGAN. Table \ref{tab:statistical_tests} suggests that {\TGAN} achieves the lowest average rejection rate of the null hypothesis described in Section \ref{sec:performance_evaluation_criteria}. 
In other words, scenarios generated by {\TGAN}} yield more accurate VaR and ES values for benchmark strategies compared to those of other generators.

\begin{table}[H]
    \centering
    \resizebox{0.85\textwidth}{!}{\begin{tabular}{lccccc}
\toprule
                 &   HSM     & {\TGAN}-Raw & {\TGAN}-Static & {\TGAN} & WGAN \\\midrule
Coverage Test (\%) &      17.9    &     53.6   &    22.9   &      17.1    &  44.9 \\
Score-based Test (\%)     &         0.00     &    21.3   &   15.4   &     0.00      &   11.4 \\
\bottomrule
\end{tabular}
}
    \caption{Average rejection rate of the null hypothesis in two tests across strategies. We use sample size $1,000$ and repeat the above experiment 100 times on testing data.} 
    \label{tab:statistical_tests}
\end{table}

\FloatBarrier
\subsection{Multiple risk levels}\label{sec:risk_levels}
{
VaR and ES are special cases of {\it spectral risk measures} \citep{kusuoka2001law}, defined as a weighted average of quantiles:
\begin{eqnarray}\label{eq:phi}
    \rho^\phi(X,\mathbb{P}) = \int_{[0,1]}{\rm VaR}_\alpha(X,\mathbb{P})\,\phi({\rm d}\alpha),
\end{eqnarray}
where the {\it spectrum} $\phi$ is a probability measure on $[0,1]$.
Theorem 5.2 in \citet{FZ2016} shows that, if $\phi$ has finite support $\{\alpha_1,...,\alpha_M\}$ then $({\rm VaR}_{\alpha_i},...,{\rm VaR}_{\alpha_M}, \rho^\phi)$ are jointly elicitable. %We refer the reader to a class of score functions in \citet{FZ2016}. 
This enables us to train {\TGAN} with multiple quantile levels $\{\alpha_m\}_{m=1}^M$ simultaneously. The theoretical developments in Section \ref{sec:framework}  generalize to this case.}

To verify that {\TGAN} is effective for not only one particular risk level, we evaluate the previously   trained model {\TGAN}(5\%)  at some different quantile levels.  Here {\TGAN}$(a\%)$ represents {\TGAN} model trained at risk level $a$. As shown in the second column of Table \ref{tab:multi_levels}, the performance of {\TGAN}(5\%) is comparable to the baseline estimate SE(1000). We also train the model at a different level 10\% (see Column {\TGAN}(10\%)). 
% We observe that {\TGAN}(10\%) is slightly worse than {\TGAN}(5\%) in estimating the tail risks at levels 1\% and 5\%, but better in 10\%. 
We observe that {\TGAN}(10\%) is slightly worse than {\TGAN}(5\%) in terms of generating scenarios {to match} the tail risks at levels 1\% and 5\%, but better in 10\%. 
Furthermore, we investigate the performance of {\TGAN} trained with multiple risk levels. In particular, the model {\TGAN}(1\%\&5\%) is trained with the spectral risk measure defined in \eqref{eq:phi}. The results suggest that, in general, including multiple levels can further improve the simulation accuracy.

\begin{table}[H]
    \centering
    \resizebox{0.85\textwidth}{!}{\begin{tabular}{lcccc}
\toprule
OOS Error  & SE(1000)   & {\TGAN}(5\%)  & {\TGAN}(10\%) & {\TGAN}(1\%\&5\%) \\\midrule
\multirow{2}{*}{$\alpha=1\%$} & 4.3 & 6.4 & 7.0 & 5.9 \\
          &  (3.0) & (2.7) & (3.1) & (2.6) \\\midrule
\multirow{2}{*}{$\alpha=5\%$} & 3.0 & 4.6 & 4.8 & 4.2 \\
          &  (2.2) & (1.6) & (1.6) & (1.8) \\\midrule
\multirow{2}{*}{$\alpha=10\%$} & 2.9 & 3.7 & 3.5 & 3.5 \\
          &  (2.1) & (1.5) & (1.5) & (1.7) \\\bottomrule
\end{tabular}}
    \caption{Mean and standard deviation (in parentheses) of relative errors for various risk levels. The columns represent the models trained for certain risks. The rows represent the out-of-sample performance for different risk levels.} 
    \label{tab:multi_levels}
\end{table}

\subsection{Generalization error} \label{sec:onlyG}
If the training and test errors closely follow one another, it is said that a learning algorithm has good generalization performance, see  \citet{arora2017generalization}. 
Generalization error quantifies the ability of machine learning models to capture certain inherent properties from the data or the ground-truth model. In general, machine learning models with a good generalization performance are meant  to learn ``underlying rules'' associated with the data generation process, rather than only memorizing the training data, so that they are able to extrapolate learned rules from the training data to new unseen data. Thereby, the generalization error of a generator $G$ can be measured as the difference between the empirical divergence of the training data $d(\mathbb{P}_r^{(n)}, \mathbb{P}_G^{(n)})$ and the ground-truth divergence $d(\mathbb{P}_{r}, \mathbb{P}_{G})$.

To  quantify the generalization capability, we use the notion of generalization proposed in \citet{arora2017generalization}. 
For a fixed sample size $n$, the generalization error of $\mathbb{P}_G$ is defined as 
\begin{eqnarray}\label{eq:generalizaton_error_eps}
\left| d(\mathbb{P}_{r}^{(n)}, \mathbb{P}_G^{(n)})-d(\mathbb{P}_{r}, \mathbb{P}_{G})\right|,
\end{eqnarray}
where $\mathbb{P}_r^{(n)}$ is the empirical distribution of $\mathbb{P}_{r}$ with $n$ samples, i.e., the distribution of the training data, and $\mathbb{P}_G^{(n)}$ is the empirical distribution of $\mathbb{P}_{G}$ with $n$ samples drawn after the generator $G$ is trained.
 A small generalization error under definition \eqref{eq:generalizaton_error_eps} implies that GANs with good generalization property should have consistent performances with the empirical distributions (i.e., $\mathbb{P}_r^{(n)}$ and  $\mathbb{P}_G^{(n)}$) and with the true distributions  (i.e., $\mathbb{P}_r$ and  $\mathbb{P}_G$). We consider two choices for the divergence function: $d_q$ based on quantile divergence, and $d_s$ based on our score function we use. 
 The quantile divergence between two distributions $P$ and $Q$ is defined as  \citet{ostrovski2018autoregressive}
\begin{eqnarray*}\label{eq:div2_gen}
q(P, Q):=\int_{0}^{1}\left[\int_{F_{P}^{-1}(\tau)}^{F_{Q}^{-1}(\tau)}\left(F_{P}(x)-\tau\right) \dd x\right] \dd \tau, \nonumber
\end{eqnarray*}
where $F_{P}$ (resp. $F_{Q}$) is the CDF of $P$ (resp. $Q$). Motivated by this definition, we define the {\it local divergence}, which focuses on the tails of  loss distributions for the benchmark strategies:
\begin{eqnarray} \label{eq:div3_gen}
d_{q}(\mathbb{P}_{r}, \mathbb{P}_{G}):= \frac{1}{K} \sum_{k=1}^K \int_{0}^{\alpha}\left[\int_{F_{\Pi^k,\mathbb{P}_{r}}^{-1}(\tau)}^{F_{\Pi^k,\mathbb{P}_{G}}^{-1}(\tau)}\left(F_{\Pi^k,\mathbb{P}_{r}}(x)-\tau\right) \dd x\right] \dd \tau,
\end{eqnarray}
where $F_{\Pi^k,\mathbb{P}_{r}}$ (resp. $F_{\Pi^k,\mathbb{P}_{G}}$) is the CDF of $\Pi^k(\mathbf{p})$ with $\mathbf{p} \sim \mathbb{P}_r$ (resp. $\Pi^k(\mathbf{q})$ with $\mathbf{q} \sim \mathbb{P}_G$). 
Recall that the score function used in \eqref{newminimax_distribution} can also be constructed 
as a ``divergence''  between two distributions in terms of their respective VaR and ES values
\begin{equation}\label{eq:div1_gen}
\begin{aligned}
d_{s} \left({\mathbb{P}}_{r}, {\mathbb{P}}_{G}\right) := \frac{1}{K} \sum_{k=1}^{K} \mathbb{E}_{\mathbf{p}\sim \mathbb{P}_r} & \Bigg[ S_{\alpha}\big({\rm VaR}_{\alpha}\left( \Pi^k,\mathbb{P}_{G}\right), {\rm ES}_{\alpha}\left( \Pi^k,\mathbb{P}_{G}\right), {\Pi^k(\mathbf{p})}\big) \\
&-  S_{\alpha}\big({\rm VaR}_{\alpha}\left( \Pi^k,\mathbb{P}_{r}\right), {\rm ES}_{\alpha}\left( \Pi^k,\mathbb{P}_{r}\right), {\Pi^k(\mathbf{p})}\big) \Bigg].
\end{aligned}
\end{equation}

\paragraph{Comparison with supervised learning}
To illustrate the generalization capabilities of {\TGAN}, we compare it with a supervised learning benchmark using the same loss function.

 Given the optimization problem \eqref{eq:generator_supervised}-\eqref{eq:distriminator_constrain}, a natural idea is to evaluate the generator   using   empirical VaR and ES estimates from the output scenarios. To this end, we consider the following optimization 
 % problem: 
\begin{eqnarray}\label{eq:generator_supervised2}
\min_{G\in \mathcal{G}} \frac{1}{Kn}\,  \sum_{k=1}^K\sum_{j=1}^n S_{\alpha}\Big(\Big({\mbox{VaR}_{\alpha}}\big(\Pi^k, \mathbb{P}_G^{(n)}\big),{\mbox{ES}_{\alpha}}\big(\Pi^k,\mathbb{P}_G^{(n)}\big)\Big),\,\,\Pi^k(\mathbf{p}_j)\Big),
\end{eqnarray}
where $\mathbb{P}_G^{(n)}$ is the empirical measure of $n$ samples drawn from $\mathbb{P}_G$, and $\mathbf{p}_j$ are samples under the measure $\mathbb{P}_r$ ($j=1,2,\ldots,n$).
%$\widehat{\mbox{VaR}_{\alpha}}\big(\Pi^k(\mathbf{q}_i), i \in [n]\big)$ and $\widehat{\mbox{ES}_{\alpha}}\big(\Pi^k(\mathbf{q}_i), i \in [n]\big)$ are empirical $\alpha$-VaR and $\alpha$-ES estimated with $n$ independent samples from the generator $G$. 
The optimization problem \eqref{eq:generator_supervised2} is a supervised learning problem. 
Compared with {\TGAN}, this setting has   several disadvantages. The first issue is the bottleneck in statistical accuracy. When using $\mathbb{P}_r^{(n)}$ as the guidance for supervised learning, as indicated in \eqref{eq:generator_supervised2}, it is not possible for the $\alpha$-VaR and $\alpha$-ES values of the simulated price scenarios $\mathbb{P}_G$ to improve on the sampling error of the empirical $\alpha$-VaR and $\alpha$-ES values estimated with the $n$ samples.  In particular, ES is very sensitive to tail events, and the empirical estimate of ES may not be stable even with 10,000 samples. The second issue concerns the limited ability in generalization. A generator constructed via supervised learning tends to mimic the exact patterns in the input financial scenarios $\mathbb{P}_r^{(n)}$, instead of generating new scenarios that are equally realistic compared to the input financial scenarios under the evaluation of the score function. 

To compare {\TGAN} with   a generator trained by supervised learning according to \eqref{eq:generator_supervised2}, we use the sorting procedure in Section \ref{app:neural_sorting} and use $\left( x^k_{(\floor{\alpha n}{})}, \frac{1}{\floor{\alpha n}{}} \sum_{i=1}^{\floor{\alpha n}{}} x^k_{(i)}\right)$ to estimate the values of $\alpha$ -VaR and $\alpha$ -ES, where $x^k_{(n)} \geq \ldots \geq x^k_{(2)} \geq x^k_{(1)}$ are the PnL sorted from $\mathbf{x}^k$ via the differentiable neural sorting architecture.  We train the generator on synthetic / real price scenarios,  with both multi-asset portfolio and dynamic strategies. The setting is similar to  {\TGAN} (described in Table \ref{tab:configuration}), except that there is no discriminator. 

\begin{figure}[!htbp]
\centering
\includegraphics[width=.6\textwidth, trim=1.2cm 0cm 3cm  3cm,clip]{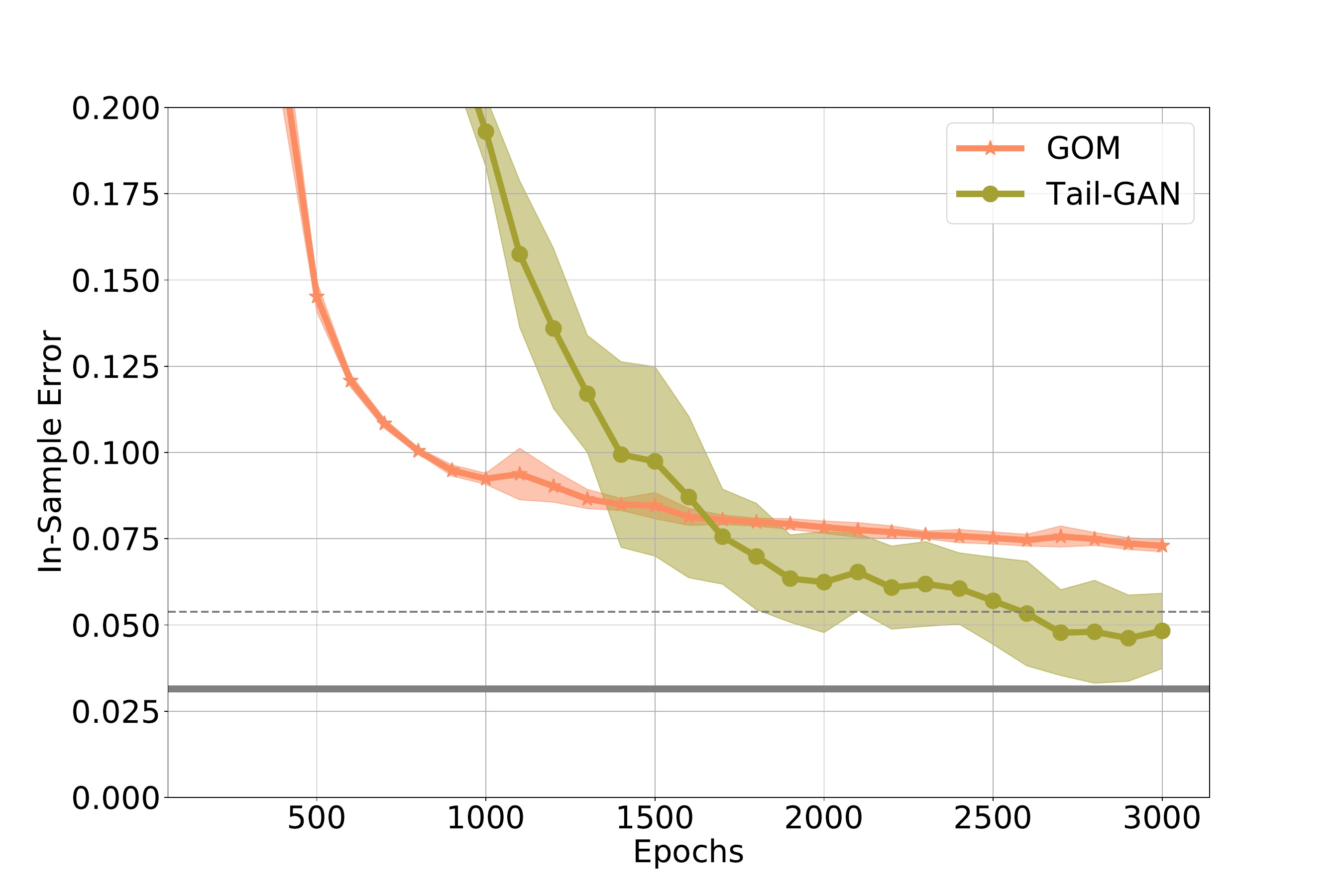}
\vspace{-3mm}
\caption{Training performance of supervised learning and {\TGAN}, as a function of the number of iterations in training. Grey horizontal line: average simulation error {${\rm SE}(1000)$}.  Dotted line: average simulation error plus one standard deviation. Each experiment is repeated five times with different random seeds. The performance is visualized with mean (solid lines) and standard deviation (shaded areas).}
\label{fig:onlyG_validation_performance} 
\end{figure}

Figure \ref{fig:onlyG_validation_performance} reports the convergence of in-sample errors, and Table \ref{tab:gom_oos_error} summarizes the out-of-sample errors of supervised learning and {\TGAN}. 
From Table \ref{tab:gom_oos_error}, we observe that the relative error of {\TGAN} is 4.6\%, which is a 30\% reduction compared to the relative error of around 7.2\% for supervised learning. Compared to \eqref{eq:generator_supervised2}, the advantage of using neural networks to learn the VaR and ES values, as designed in {\TGAN}, is that it memorizes information in previous iterations during the training procedure, and therefore the statistical bottleneck with $n$ samples can be  overcome when the number of iterations increases. Therefore, we conclude that {\TGAN} outperforms supervised learning in terms of simulation accuracy, demonstrating the importance of the discriminator.  

\begin{table}[!htbp]
    \centering
    \resizebox{0.4\textwidth}{!}{\begin{tabular}{lcc}
\toprule
             & {\TGAN} & Supervised learning\\\midrule
Out of sample error (\%)    & 4.6 &  7.2  \\
           & (1.6) & (0.2) \\\bottomrule
\end{tabular}}
    \caption{Mean and standard deviation (in parentheses) of relative errors for out-of-sample tests. Each experiment is repeated five times with different random seeds.} 
    \label{tab:gom_oos_error}
\end{table}

Table \ref{tab:gene_error} provides the generalization errors, under both $d_q$ and $d_s$, for {\TGAN} and supervised learning.  We observe that under both criteria, the generalization error of supervised learning is twice that of {\TGAN}, implying that {\TGAN} has better generalization power, in addition to higher performance accuracy.

\begin{table}[!htbp]
    \centering
    \resizebox{0.4\textwidth}{!}{\begin{tabular}{ccc}
\toprule
Error metric  &  {\TGAN}  & Supervised learning \\\midrule
\multirow{2}{*}{$d_{q}$}    &   0.214      &  0.581  \\
& (0.178)  &  (0.420)  \\\midrule
\multirow{2}{*}{$d_{s}$}   & 0.017    & 0.032 \\
& (0.014)       & (0.026)     \\
\bottomrule
\end{tabular}}
    \caption{Mean (in percentage) and standard deviation (in parentheses) of generalization errors under both divergence functions (see their mathematical formulations in \eqref{eq:div3_gen} and \eqref{eq:div1_gen}). Results are averaged over 10 repeated experiments (synthetic data sets).} 
    \label{tab:gene_error}
\end{table}

\FloatBarrier
\subsection{Scalability}\label{sec:scale}
In practice many applications require simulation of scenarios for a large number of assets.
We show that using {\it eigenportfolios} \citet{avellaneda2010}, defined as 
 $L^1$-normalized principal components of the sample correlation matrix of returns, in the training phase is an effective way of extracting information on cross-asset dependence for high-dimensional data sets.
 The resulting  eigenportfolios are uncorrelated by construction and their loss distributions provide a set of useful features for training, whose sizescales {\it linearly} with the dimension of the data set. 
This idea mqkes {\TGAN} scalable to high-dimnensional data sets.

We train {\TGAN} with eigenportfolios of 20 assets, and compare its performance with {\TGAN} trained on 50 randomly generated portfolios. {\TGAN} with the eigenportfolios shows dominating performance, which is also comparable to simulation error (with the same number of samples).  The detailed steps of the eigenportfolio construction are deferred to Appendix \ref{app:eigenportfolios}.

\paragraph{Data.} To showcase the scalability of {\TGAN}, we simulate the price scenarios of 20 financial assets for a given correlation matrix $\mathbf{\rho}$, with different temporal patterns and tail behaviors in return distributions. Among these 20 financial assets, five of them have IID  Gaussian returns, another five follow AR(1) models,  another five follow GARCH(1, 1) with Gaussian noise, and the rest follow GARCH(1, 1) with heavy-tailed noise.  Other settings are the same as in Section \ref{sec:validation_highdimension}.

\begin{figure}[H]
\centering
\includegraphics[width=.55\textwidth, trim=1.5cm 0cm 3cm  3cm,clip]{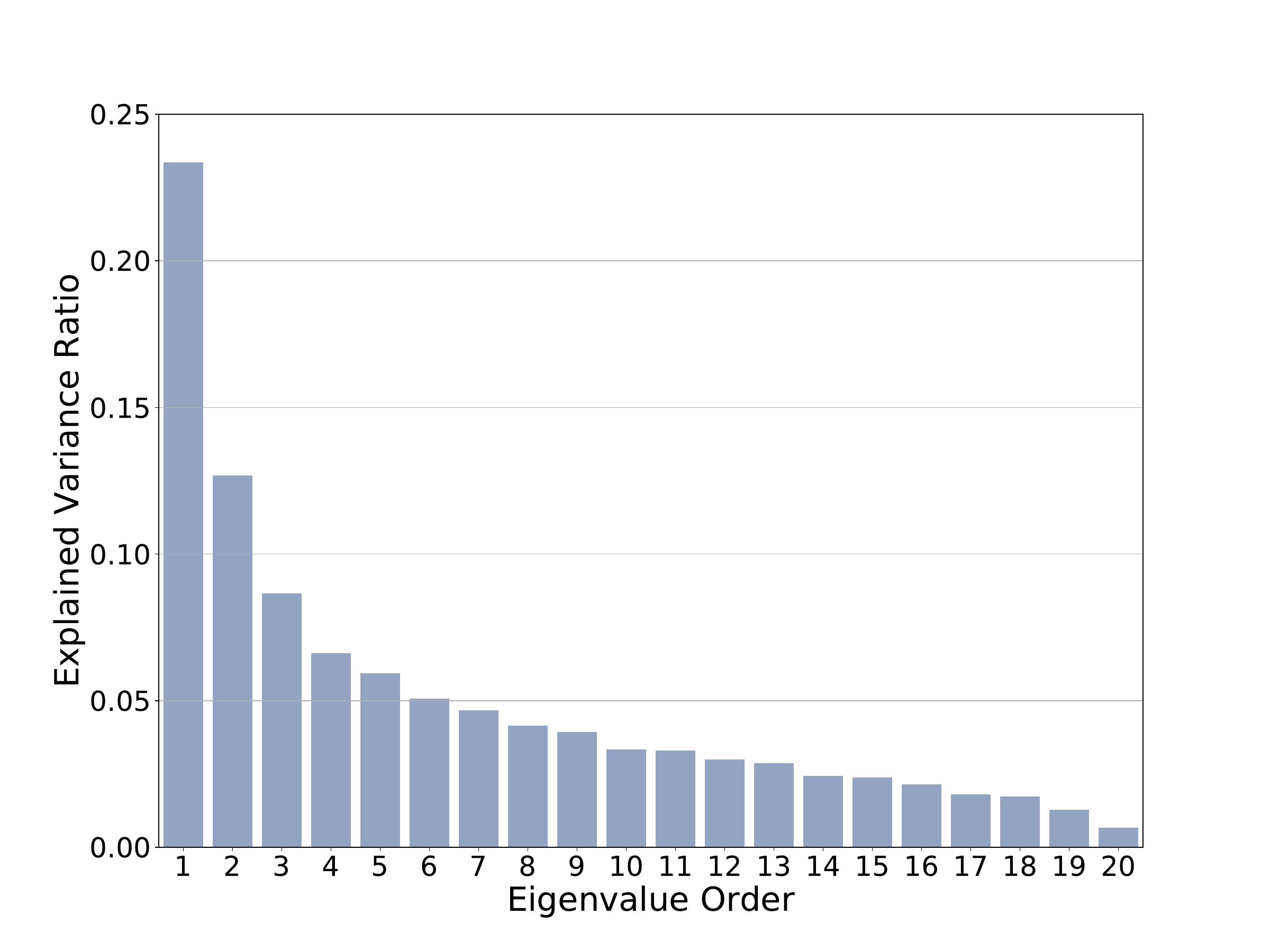}
\vspace{-3mm} 
\caption{Explained variance ratios of eigenvalues.}
\label{fig:EVR}
\end{figure}

\paragraph{Results.}
Figure \ref{fig:EVR} shows the percentage of explained variance of the principal components.
% i.e. $\frac{\mathbf{\Lambda}_{i,i}}{\sum_{m=1}^M \mathbf{\Lambda}_{m,m}}$. 
We observe that the first principal component accounts for more than 23\% of the total variation across the 20 asset returns. 

To identify and demonstrate the advantages of the eigenportfolios, we compare the following two {\TGAN} architectures 
\begin{enumerate}[label={(\arabic*)}]
	\item {\TGAN}(Rand): GAN trained with 50 multi-asset portfolios and dynamic strategies, 
	\item {\TGAN}(Eig): GAN trained with 20 multi-asset eigenportfolios and dynamic strategies.
\end{enumerate}
The weights of static portfolios in {\TGAN}(Rand) are randomly generated such that the  absolute values of the weights sum up to one. The out-of-sample test consists of $K=90$ strategies, including 50 convex combinations of eigenportfolios (with weights randomly generated), 20 mean-reversion strategies, and 20 trend-following strategies.

\begin{figure}[!htbp]
\centering
\includegraphics[width=.6\textwidth, trim=1.5cm 0cm 3cm  3cm,clip]{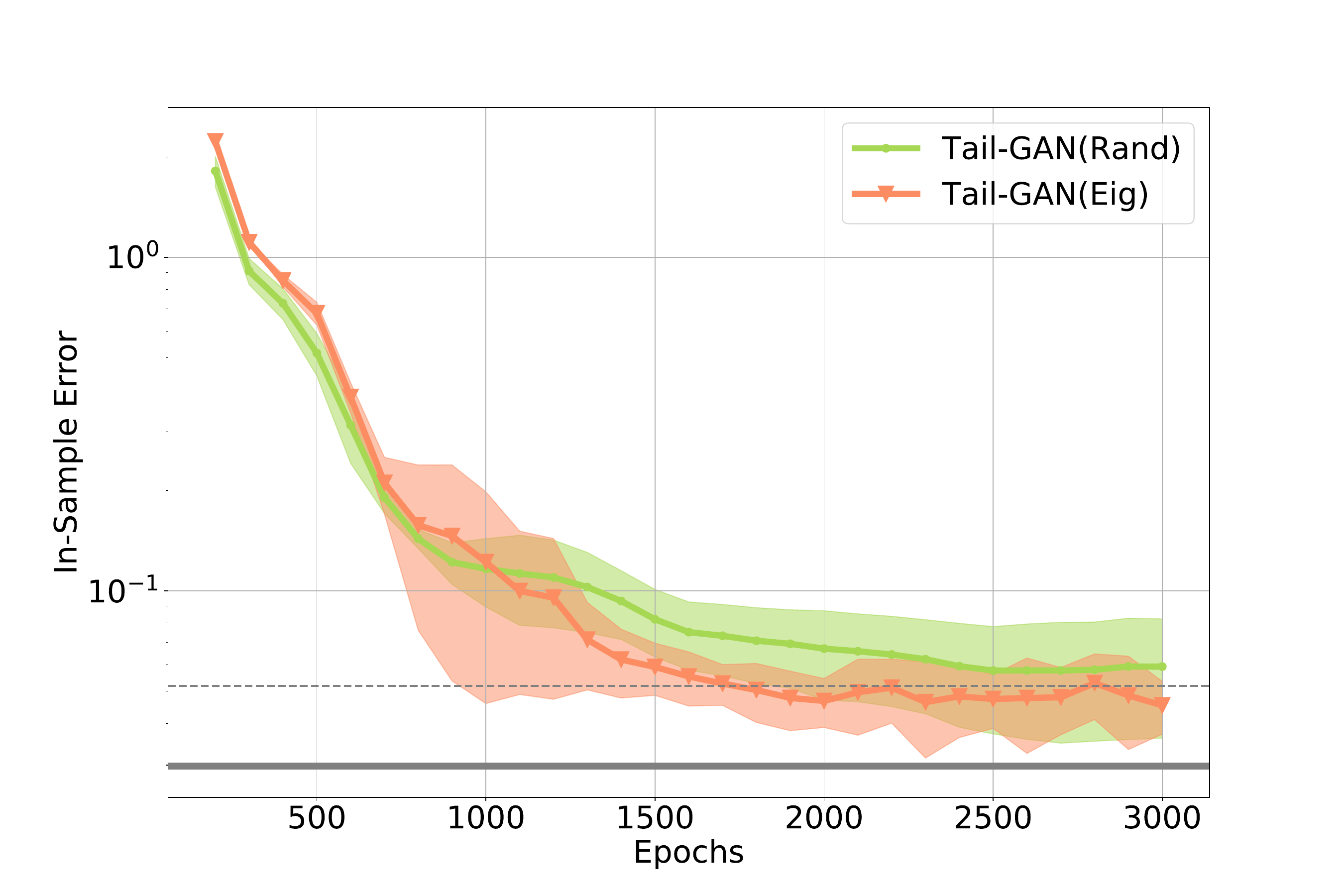}
\caption{Training performance on 50 random portfolios vs 20 eigenportfolios as described in Section \ref{sec:scale}: mean of relative error RE(1000) and standard deviation (shaded areas). Grey horizontal line: average simulation error. Dotted line: average simulation error plus one standard deviation. Each experiment is repeated five times with different random seeds. The performance is reported with mean (solid lines) and standard deviation (shaded areas).} 
\label{fig:syn_eig}
\end{figure}

\paragraph{Performance accuracy.} 
Figure \ref{fig:syn_eig} reports the convergence of in-sample errors. Table \ref{tab:syn_eig_error} summarizes the out-of-sample errors and shows that {\TGAN}(Eig) achieves better performance than {\TGAN}(Rand) with fewer number of portfolios.

\begin{table}[!htbp]
    \centering
    \resizebox{0.66\textwidth}{!}{\begin{tabular}{lccc}
\toprule
            & HSM &  {\TGAN}(Rand) & {\TGAN}(Eig) \\\midrule
OOS Error (\%)  & 3.5  &  10.4 & 6.9  \\
          & (2.3) &  (1.8) & (1.5) \\\bottomrule
\end{tabular}}
    \caption{Mean and standard deviation (in parentheses) for relative errors in out-of-sample tests. Each experiment is repeated five times with different random seeds.   }  
    \label{tab:syn_eig_error}
\end{table}
\FloatBarrier

%%%%%%%%%%%%%%%%%%%%%%%%%%%%%%%%%%%%%%%%%%%%%%%%%%%%%%%%%%%%%%%%%%%%%%%%%%%%%%%%%%%%
\section{Application to simulation of intraday market scenarios} \label{sec:market_simulation}

%\paragraph{Setup for intraday price  scenarios.} 
We train  {\TGAN} on intraday high-frequency Nasdaq ITCH data  for  the following five stocks: AAPL, AMZN, GOOG, JPM, QQQ from the  LOBSTER\footnote{https://lobsterdata.com/}  database for the time interval 10:00AM-3:30PM, from 2019-11-01 to 2019-12-06.    

The mid-prices (average of the best bid and ask prices) of these assets are sampled at a $\Delta=9$-second frequency, with $T=100$ for each price series representing a financial scenario during a 15-minute interval. We sample the 15-minute paths every one minute, leading to an overlap of 14 minutes between two adjacent paths\footnote{ {\TGAN} are also trained on  market data with no time overlap and the conclusions are similar.}. The architecture and configurations are the same as those reported in Table \ref{tab:configuration} in Appendix \ref{app:configuration}, except that the training period here is from 2019-11-01 to 2019-11-30, and the testing period is % within 
the first week of 2019-12. Thus, the size of the training data is $N = 6300$. 
Table \ref{tab:real_estimates} reports the 5\%-VaR and 5\%-ES values of several strategies calculated with the market data of AAPL, AMZN, GOOG, JPM, and QQQ.

\begin{table}[H]
    \centering
    \resizebox{0.68\textwidth}{!}{\begin{tabular}{lcccccc}
\toprule
& \multicolumn{2}{c}{Static buy-and-hold} & \multicolumn{2}{c}{Mean-reversion} & \multicolumn{2}{c}{Trend-following} \\
 \cmidrule(lr){2-3}  \cmidrule(lr){4-5}\cmidrule(lr){6-7}
  & VaR                   & ES                   & VaR                   & ES                    & VaR                    & ES                    \\\midrule
AAPL & -0.351                & -0.548               & -0.295                & -0.479                & -0.316                 & -0.485                \\
AMZN & -0.460                 & -0.720               & -0.398                & -0.639                & -0.399                 & -0.628                \\
GOOG & -0.316                & -0.481               & -0.272                & -0.426                & -0.273                 & -0.419                \\
JPM & -0.331                & -0.480                & -0.275                & -0.419                & -0.290                  & -0.427                \\
QQQ & -0.254                & -0.384               & -0.202                & -0.321                & -0.210 & -0.328 \\\bottomrule           
\end{tabular}

% \begin{tabular}{l|cc|cc|cc|cc}
%   & \multicolumn{2}{c}{Single-Asset   Portfolio} & \multicolumn{2}{c}{Multi-Asset   Portfolio} & \multicolumn{2}{c}{Mean-Reversion   Strategy} & \multicolumn{2}{c}{Trend-Following   Strategy} \\
%   & VaR                   & ES                   & VaR                  & ES                   & VaR                   & ES                    & VaR                    & ES                    \\\hline
% 1 & -0.351                & -0.548               & -0.253               & -0.382               & -0.295                & -0.479                & -0.316                 & -0.485                \\\hline
% 2 & -0.460                 & -0.720                & -0.121               & -0.177               & -0.398                & -0.639                & -0.399                 & -0.628                \\\hline
% 3 & -0.316                & -0.481               & -0.147               & -0.222               & -0.272                & -0.426                & -0.273                 & -0.419                \\\hline
% 4 & -0.331                & -0.480                & -0.209               & -0.316               & -0.275                & -0.419                & -0.290                  & -0.427                \\\hline
% 5 & -0.254                & -0.384               & -0.303               & -0.464               & -0.202                & -0.321                & -0.210 & -0.328 \\\hline             
% \end{tabular}
}
    \caption{Empirical VaR and ES values for trading strategies evaluated on the training data.}
    \label{tab:real_estimates}
\end{table}

\begin{figure}[H]
\centering
\label{fig:financial_is_all}
\includegraphics[width=.6\textwidth, trim=2.5cm 0cm 3cm  3cm,clip]{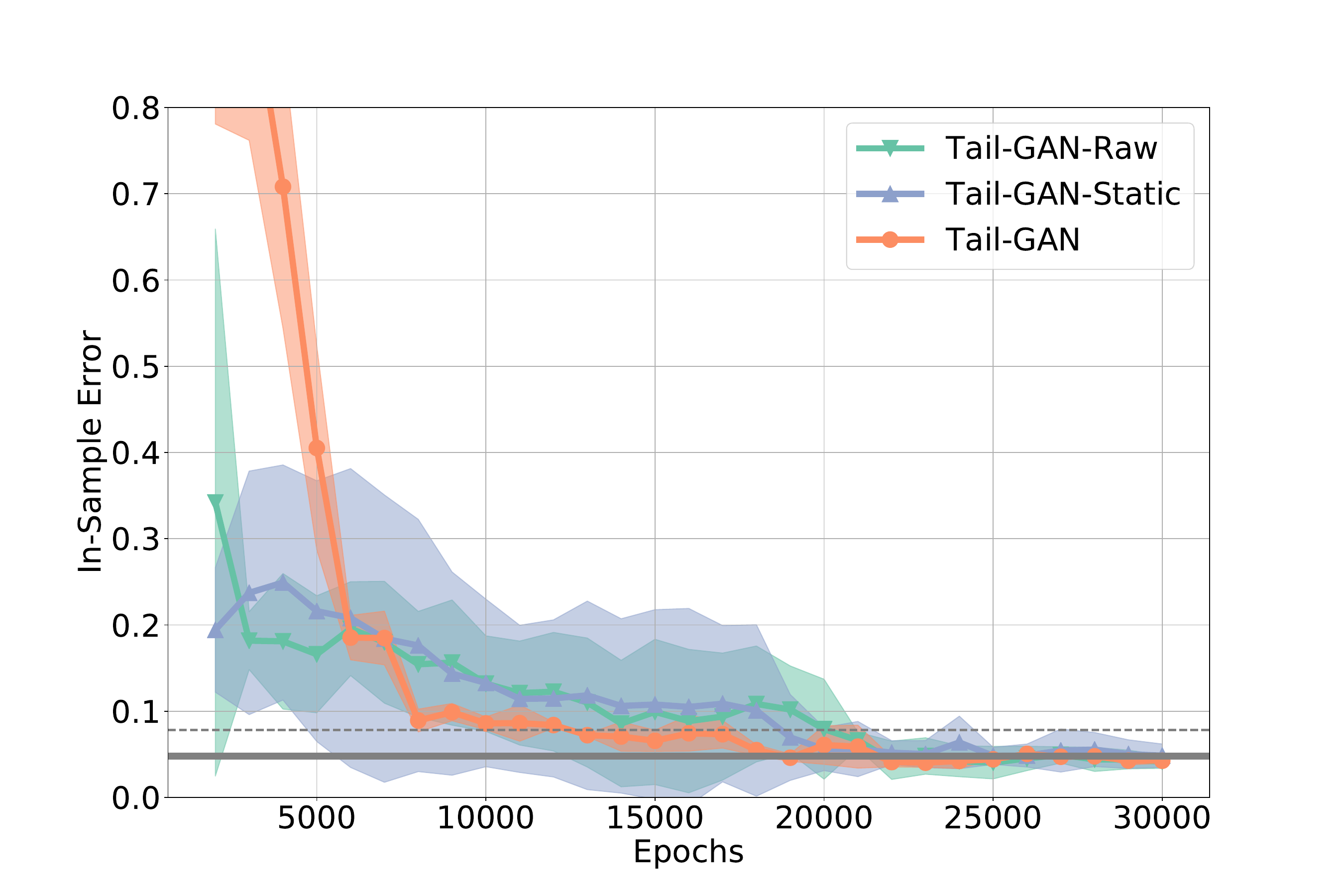} 
\vspace{-3mm}
\caption{Training performance: relative error ${\rm RE}(1000)$ with 1000 samples. Grey horizontal line: average simulation error {${\rm SE}(1000)$}. Dotted line: average simulation error plus one standard deviation. Each experiment is repeated  5 times with different random seeds. The performance is visualized with mean (solid lines) and standard deviation (shaded areas).}
\label{fig:financial_performance}
\end{figure}

\paragraph{Performance accuracy.}
Figure \ref{fig:financial_performance} reports the convergence of in-sample errors and Table \ref{tab:real_error} summarizes the out-of-sample errors. 

\begin{table}[H]
\centering
\vspace{-5mm}
\resizebox{0.95\textwidth}{!}{\begin{tabular}{lcccccc}
\toprule
            & ``Oracle'' & HSM  & {\TGAN-Raw} & {\TGAN-Static} & {\TGAN} & WGAN \\\midrule
Out of sample     &  2.4 & 10.4 & 112.8 & 75.8 & 10.1 & 26.9 \\
    Error (\%)      &  (1.6) & (3.6)  & (7.8) & (8.0) & (1.1) & (1.7) \\\bottomrule
\end{tabular}}
\caption{Mean and standard deviation (in parentheses) for relative errors in out-of-sample tests. ``Oracle'' represents the sampling error of the testing data. Each experiment is repeated five times with different random seeds.}
\label{tab:real_error}
\vspace{-2mm}
\end{table}

We draw the following conclusions from the results of Figure \ref{fig:financial_performance} and Table \ref{tab:real_error}.

\begin{itemize}
\item For the evaluation criterion ${\rm RE}(1000)$ based on in-sample data (see Figure \ref{fig:financial_performance}), all three GAN generators, {\TGAN}-Raw, {\TGAN}-Static and {\TGAN}, converge within 20,000 epochs and reach in-sample errors smaller than 5\%. 

\item For the evaluation criterion ${\rm RE}(1000)$, with both static portfolios and dynamic strategies based on out-of-sample data (Table \ref{tab:real_error}), only {\TGAN} converges to an error of 10.1\%, whereas the other  two \textsc{Tail-GAN} variants fail to capture the temporal information in the input price scenarios. 
% As expected, the lowest error is attained by the Oracle    
\item The HSM method comes close with an error of 10.4\% and WGAN reaches an error of 26.9\%. As expected, all methods attain higher errors than the sampling error of the testing data (denoted by ``oracle'' in Table \ref{tab:real_error}).
\end{itemize}

\begin{figure}[!htbp]
\centering
\includegraphics[width=1.0\textwidth, trim=5cm 0mm 0cm 0cm,clip]{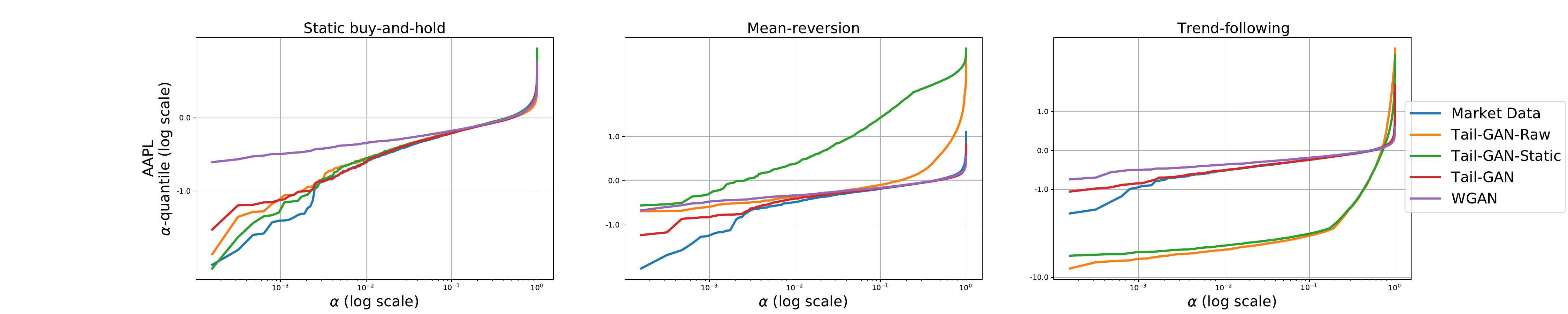}
\caption{Tail behavior via the empirical rank-frequency distribution of the strategy PnL (based on AAPL). The columns represent the strategy types.}
\label{fig:real_tail_aapl}
\end{figure}

To study the tail behavior of the intraday scenarios, we implement the same rank-frequency analysis as in Section \ref{sec:validation_highdimension}. For the AAPL stock, we draw the following conclusions from Figure \ref{fig:real_tail_aapl}.  
\begin{itemize}
\itemsep0em
\item All three {\TGAN} generators are able to capture the tail properties of static single-asset portfolio for quantile levels above 1\%. 
\item For the PnL distribution of the dynamic strategies, only {\TGAN} is able to generate scenarios with comparable (tail) PnL distribution.
That is, only scenarios sampled from {\TGAN} can correctly describe the risks of the trend-following and the mean-reversion strategies. 
\item {\TGAN}-Raw and {\TGAN}-Static underestimate the risk of loss from the mean-reversion strategy at the $\alpha=5\%$ quantile level, and overestimate the risk of loss from the trend-following strategy at the $\alpha=5\%$ quantile level.
\item While WGAN can effectively generate scenarios that align with the bulk of PnL distributions (e.g. above 10\%-quantile), it  fails to accurately capture the tails, usually resulting in underestimation of risks.

\end{itemize}
Note that some of the blue curves corresponding to the market data (almost) coincide with the red curves corresponding to {\TGAN}, indicating a promising performance of {\TGAN} to capture the tail risk of various trading strategies. See Figure \ref{fig:real_tail} in Appendix \ref{sec:supply_results} for the results for other assets.
% \FloatBarrier

\begin{figure}[!htbp]
\centering
\subfigure[Market data.]
{ \label{fig:financial_corr_1}
\includegraphics[width=0.17\textwidth, trim=3cm 3cm 3cm  3cm,clip]{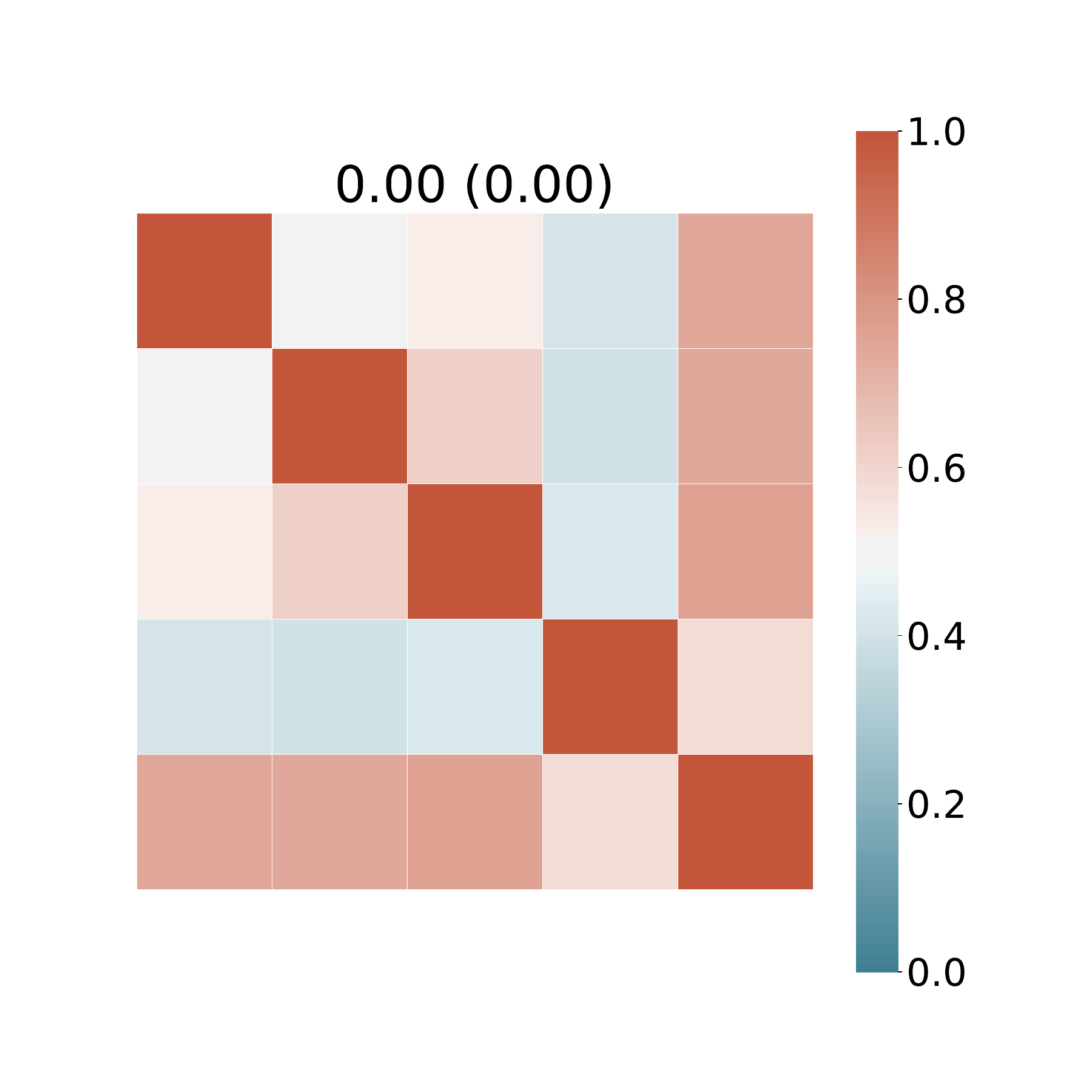}
}
\subfigure[{\TGAN}-Raw.]
%GAN trained with static single-asset portfolio]
{ \label{fig:financial_corr_2}
\includegraphics[width=0.17\textwidth, trim=3cm 3cm 3cm  3cm,clip]{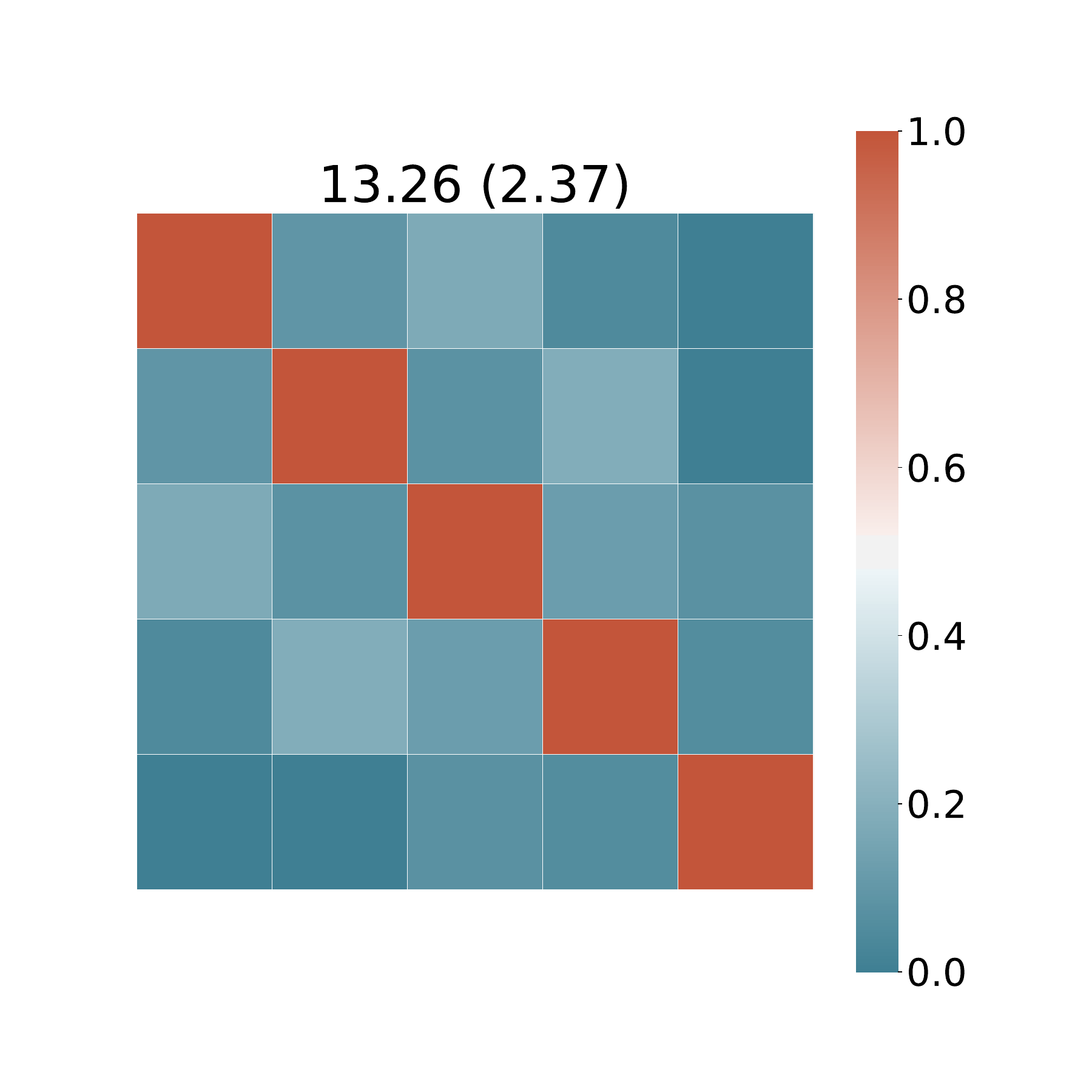}
}
\subfigure[{\TGAN}-Static.]
%GAN trained with static multi-asset portfolio]
{ \label{fig:financial_corr_3}
\includegraphics[width=0.17\textwidth, trim=3cm 3cm 3cm  3cm,clip]{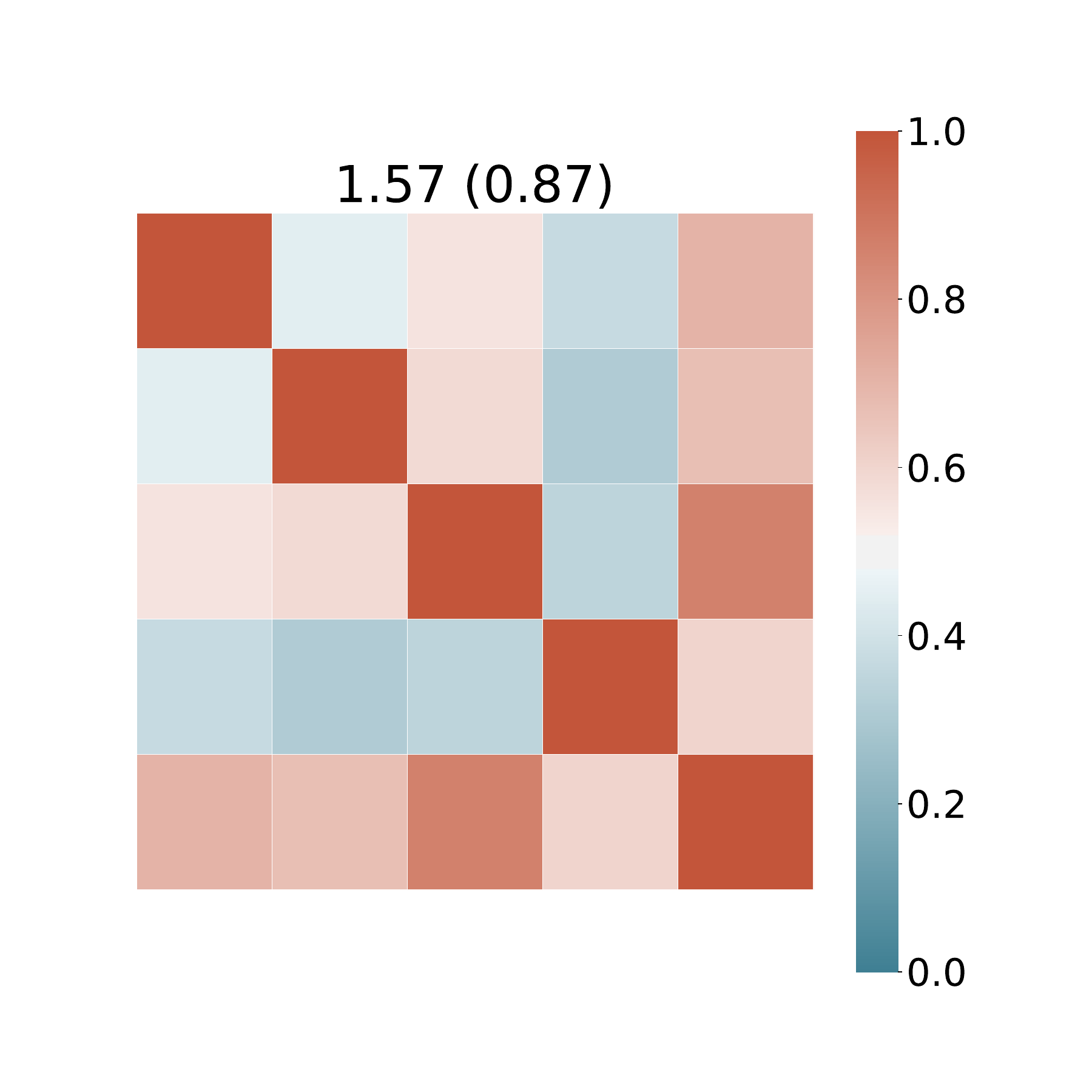}
}
\subfigure[{\TGAN}.]
%GAN trained with static multi-asset portfolio and dynamic strategies]
{ \label{fig:financial_corr_4}
\includegraphics[width=0.17\textwidth, trim=3cm 3cm 3cm  3cm,clip]{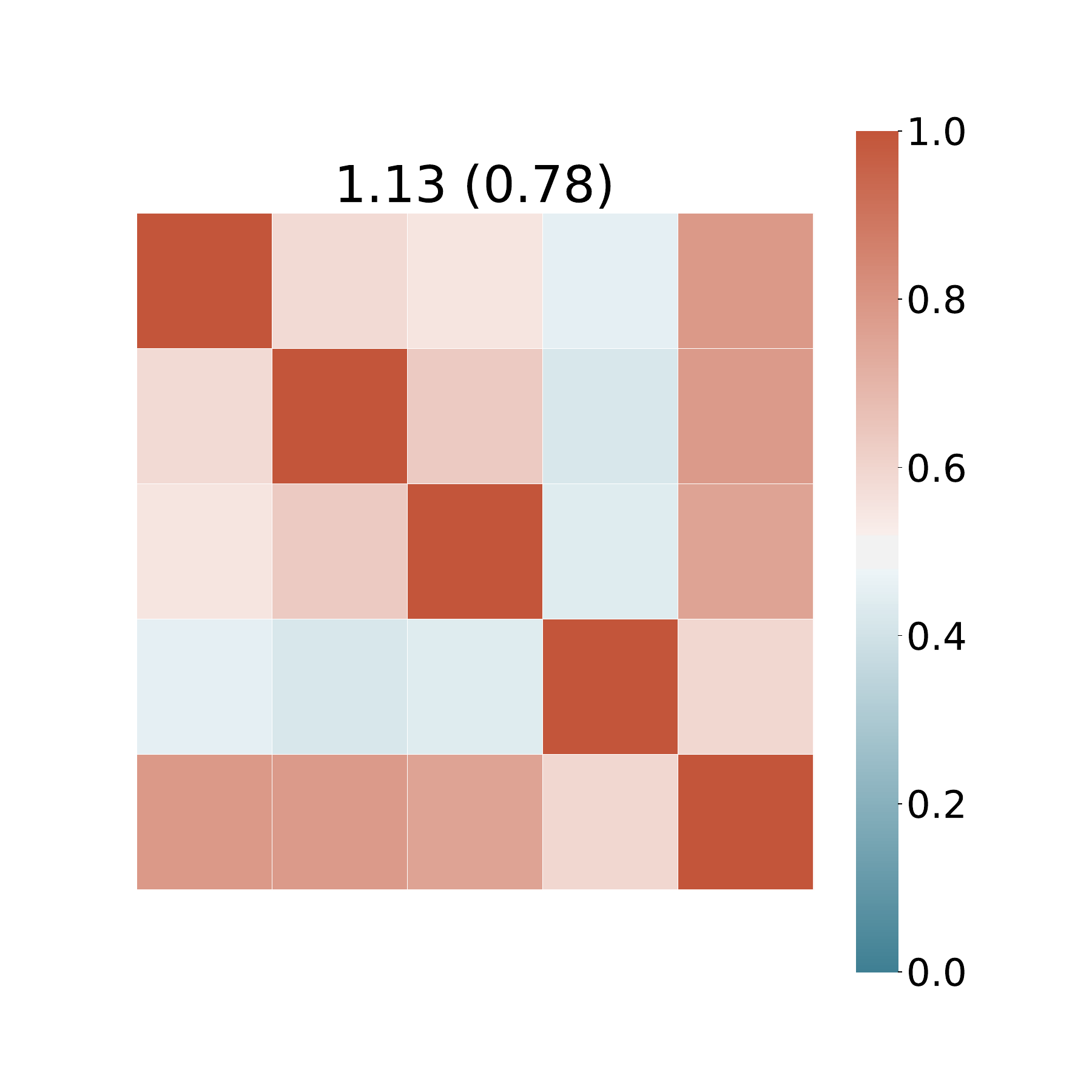}
}
\subfigure[{WGAN}.]
{ \label{fig:financial_corr_wgan}
\includegraphics[width=0.17\textwidth, trim=3cm 3cm 3cm  3cm,clip]{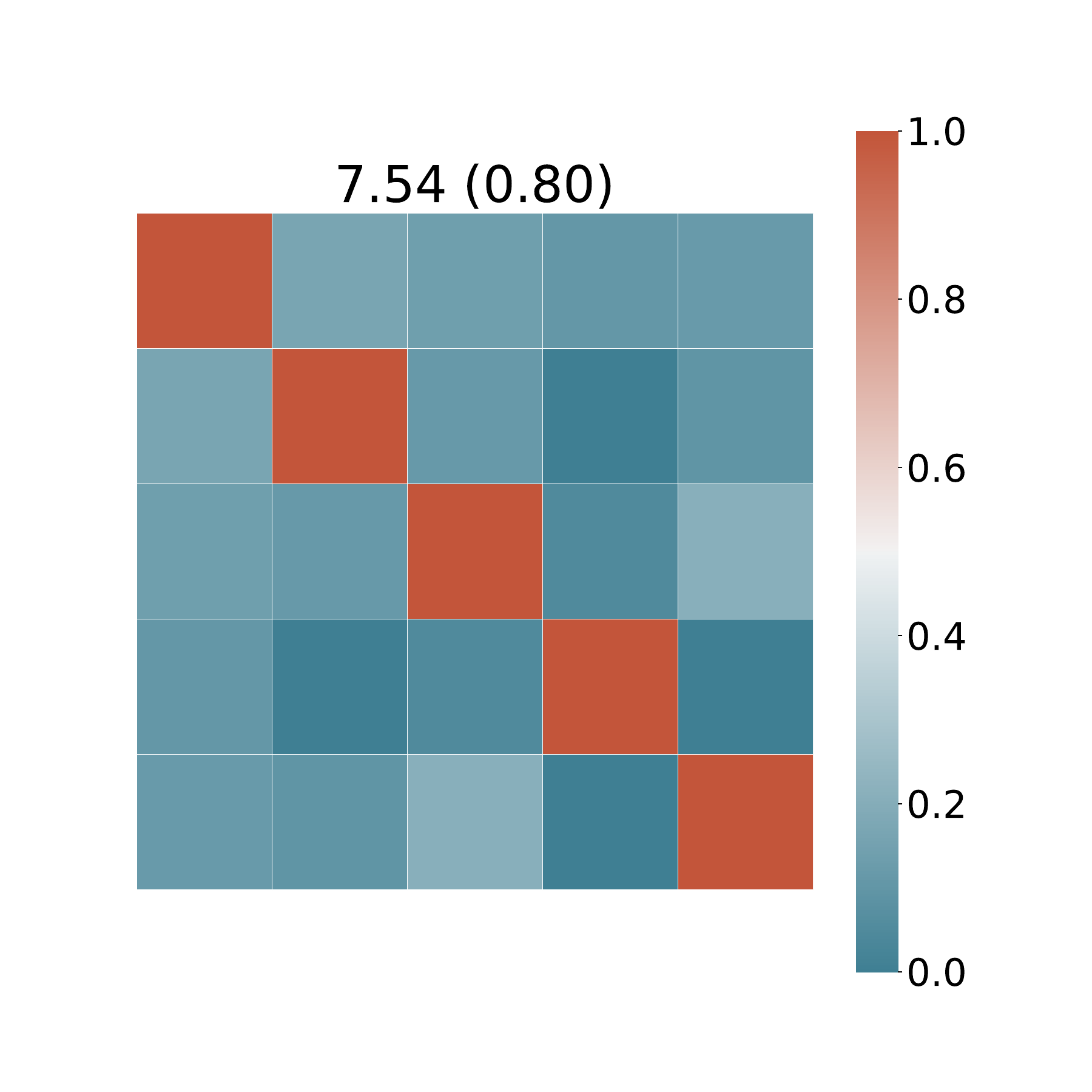}
}
\caption{\label{fig:financial_corr} Cross-asset correlations of the price increments in the market data and from different trained GAN models (1) {\TGAN}-Raw, (2) {\TGAN}-Static, (3) {\TGAN}, and (4) WGAN. Numbers on the top: mean and standard deviation (in parentheses) of the sum of the absolute difference between the correlation coefficients computed with all training samples and 1,000 generated samples.}
\end{figure}

\begin{figure}[!htbp]
\vspace{-3mm}
\centering
\subfigure[Market data.]
{ \label{fig:financial_autocorr_1}
\includegraphics[width=0.17\textwidth, trim=1.3cm 2cm 2cm  2cm,clip]{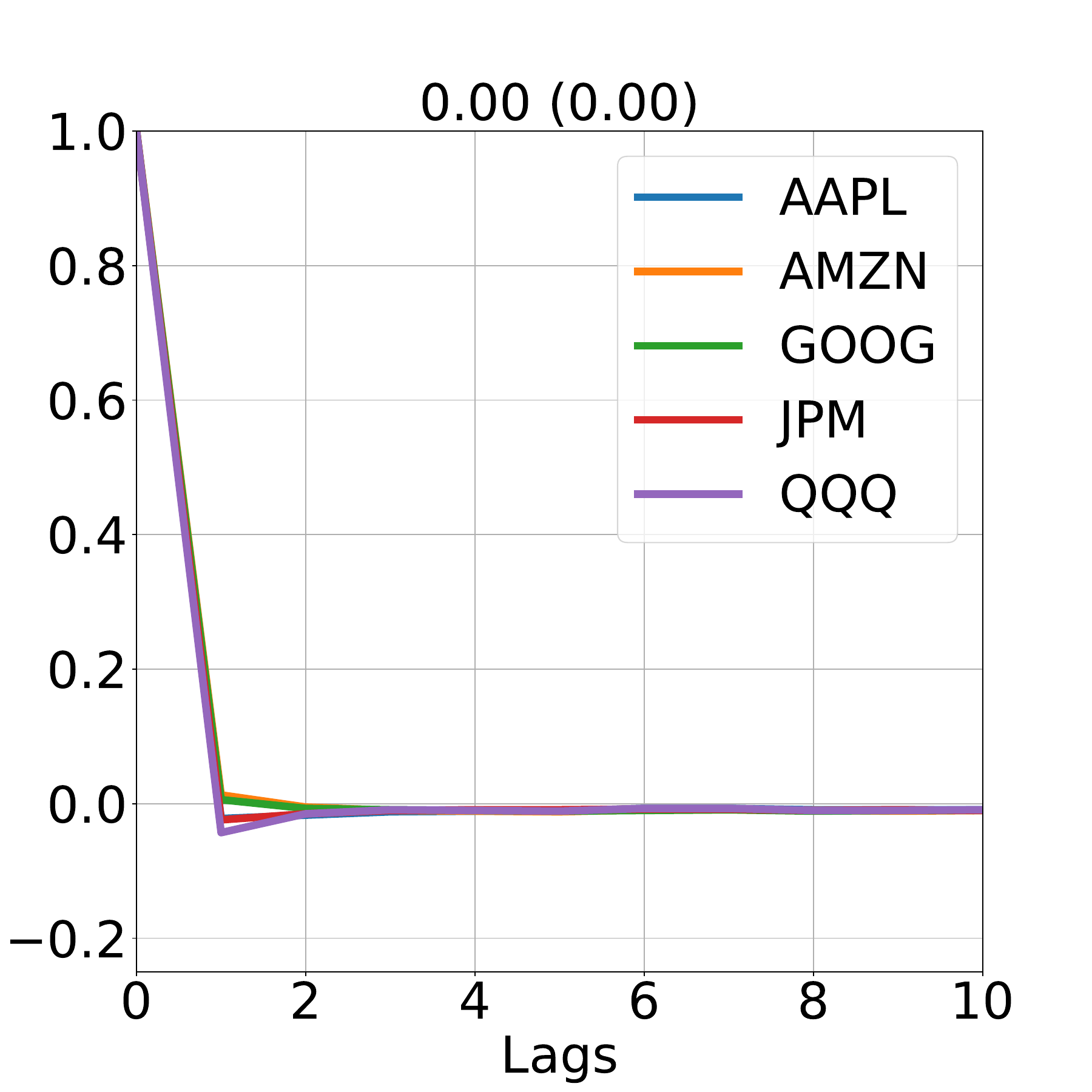}
}
\subfigure[{\TGAN}-Raw.]
%GAN trained with static single-asset portfolio.]
{ \label{fig:financial_autocorr_2}
\includegraphics[width=0.17\textwidth, trim=1.3cm 2cm 2cm  2cm,clip]{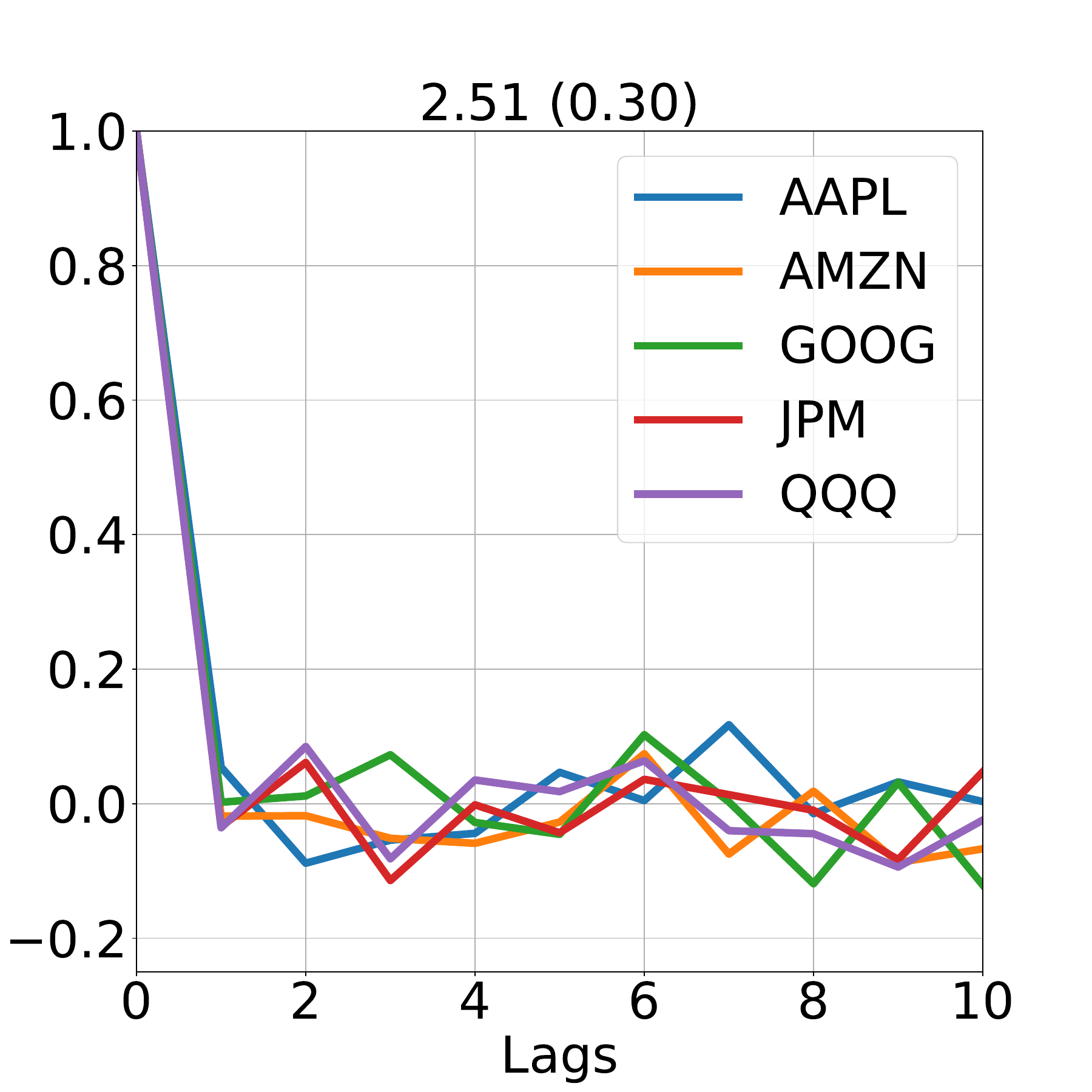}
}
\subfigure[{\TGAN}-Static.]
%GAN trained with static multi-asset portfolio.]
{ \label{fig:financial_autocorr_3}
\includegraphics[width=0.17\textwidth, trim=1.3cm 2cm 2cm  2cm,clip]{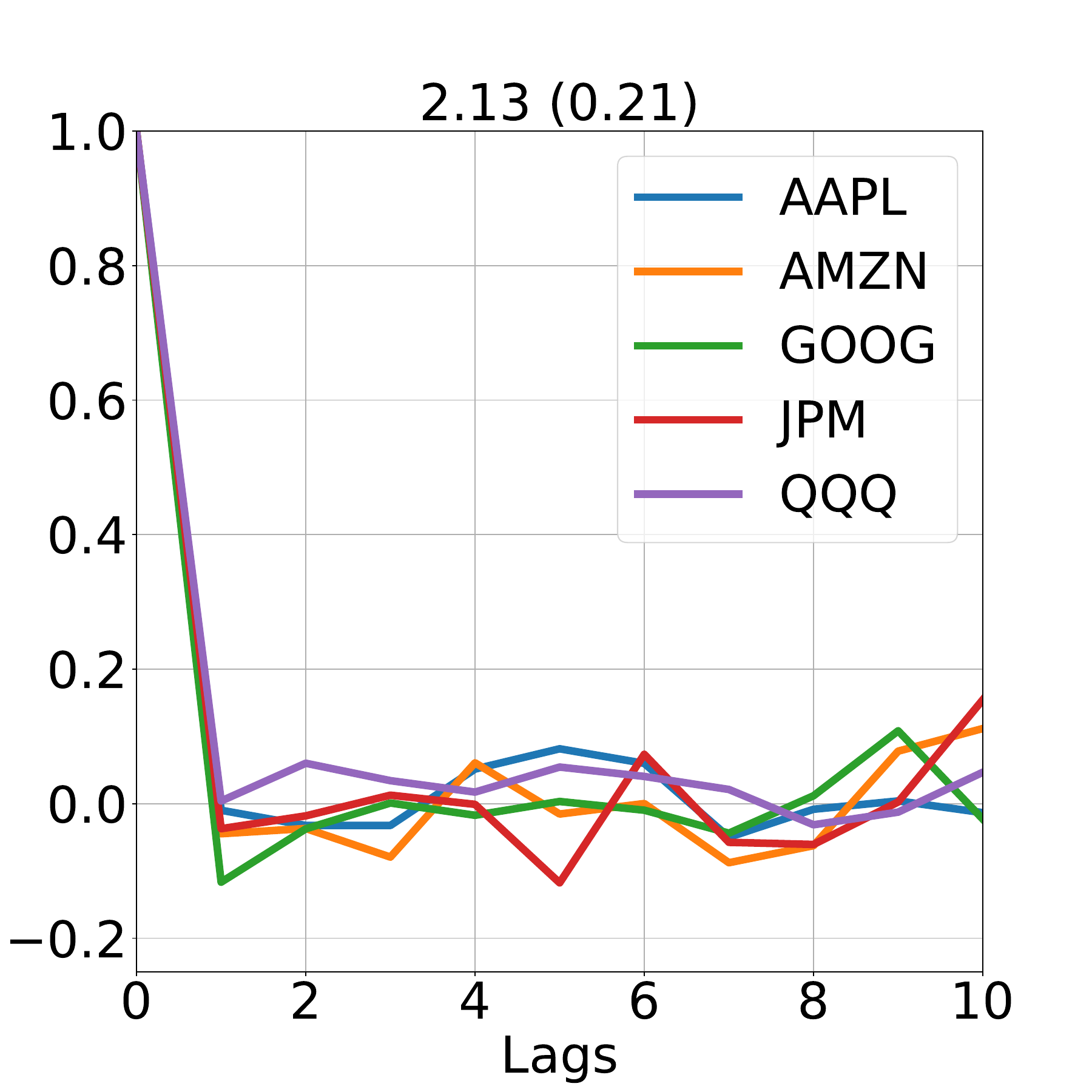}
}
\subfigure[{\TGAN}.]
%GAN trained with static multi-asset portfolio and dynamic strategies.]
{ \label{fig:financial_autocorr_4} 
\includegraphics[width=0.17\textwidth, trim=1.3cm 2cm 2cm  2cm,clip]{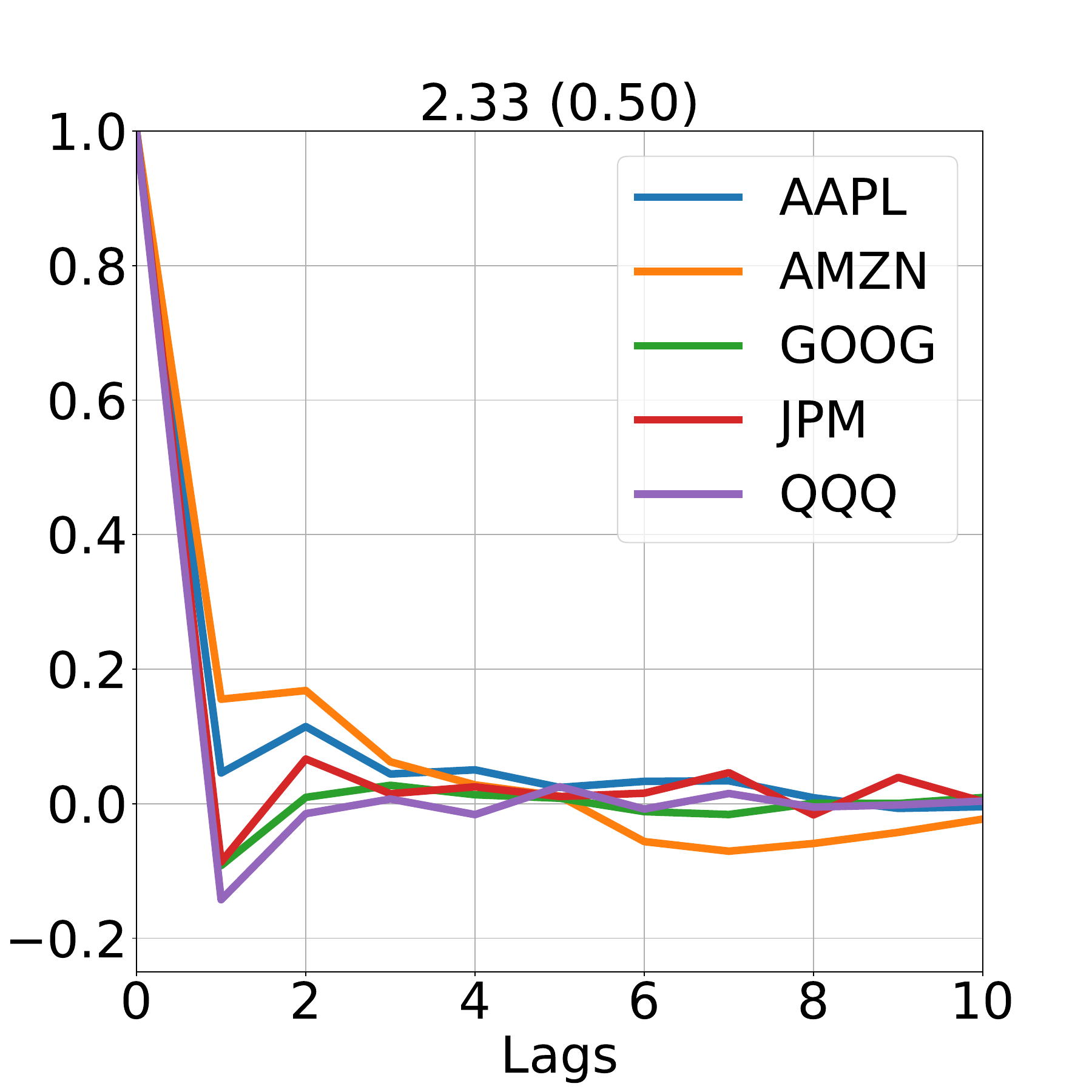}
}
\subfigure[WGAN.]
%GAN trained with static multi-asset portfolio and dynamic strategies.]
{ \label{fig:financial_autocorr_wgan} 
\includegraphics[width=0.17\textwidth, trim=1.3cm 2cm 2cm  2cm,clip]{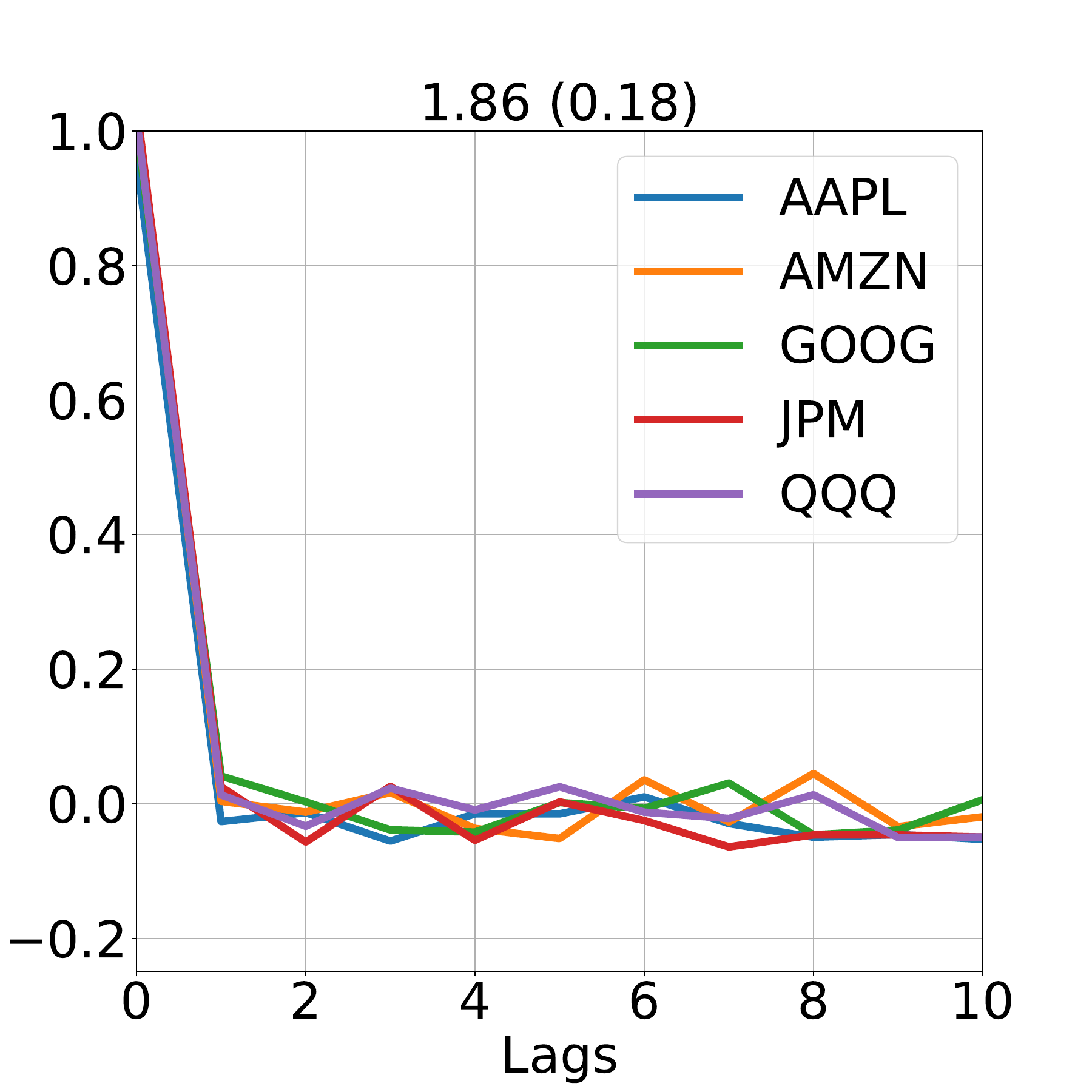}
}
\caption{\label{fig:financial_autocorr} 
Auto-correlations {of the price increments} from different trained GAN models: (1) {\TGAN}-Raw, (2) {\TGAN}-Static, (3) {\TGAN}, and (4) WGAN. Numbers on the top: mean and standard deviation (in parentheses) of the sum of the absolute element-wise difference between auto-correlation coefficients computed with all training samples and 1,000 generated samples.}
\end{figure}

\paragraph{Learning temporal and cross-correlation patterns.}
Figures \ref{fig:financial_corr} and \ref{fig:financial_autocorr} present the in-sample correlation and auto-correlation patterns of 
the market data (Figures \ref{fig:financial_corr_1} and \ref{fig:financial_autocorr_1}),
and simulated data from {\TGAN}-Raw  (Figures \ref{fig:financial_corr_2} and \ref{fig:financial_autocorr_2}), 
{\TGAN}-Static  (Figures \ref{fig:financial_corr_3} and \ref{fig:financial_autocorr_3}), {\TGAN} (Figures \ref{fig:financial_corr_4} and \ref{fig:financial_autocorr_4}) and WGAN (Figures \ref{fig:financial_corr_wgan} and \ref{fig:financial_autocorr_wgan}).

As shown in Figures \ref{fig:financial_corr} and \ref{fig:financial_autocorr}, {\TGAN} trained on dynamic strategies learns the information on cross-asset correlations more accurately than {\TGAN}-Raw and WGAN, which are trained on raw returns.

%\RX{Discussion: Figure \ref{fig:real_eig}.}
\begin{figure}[!htbp]
\vspace{-5mm}
\centering
\includegraphics[width=.6\textwidth, trim=1.3cm 0.2cm 3.8cm 3.65cm, clip]{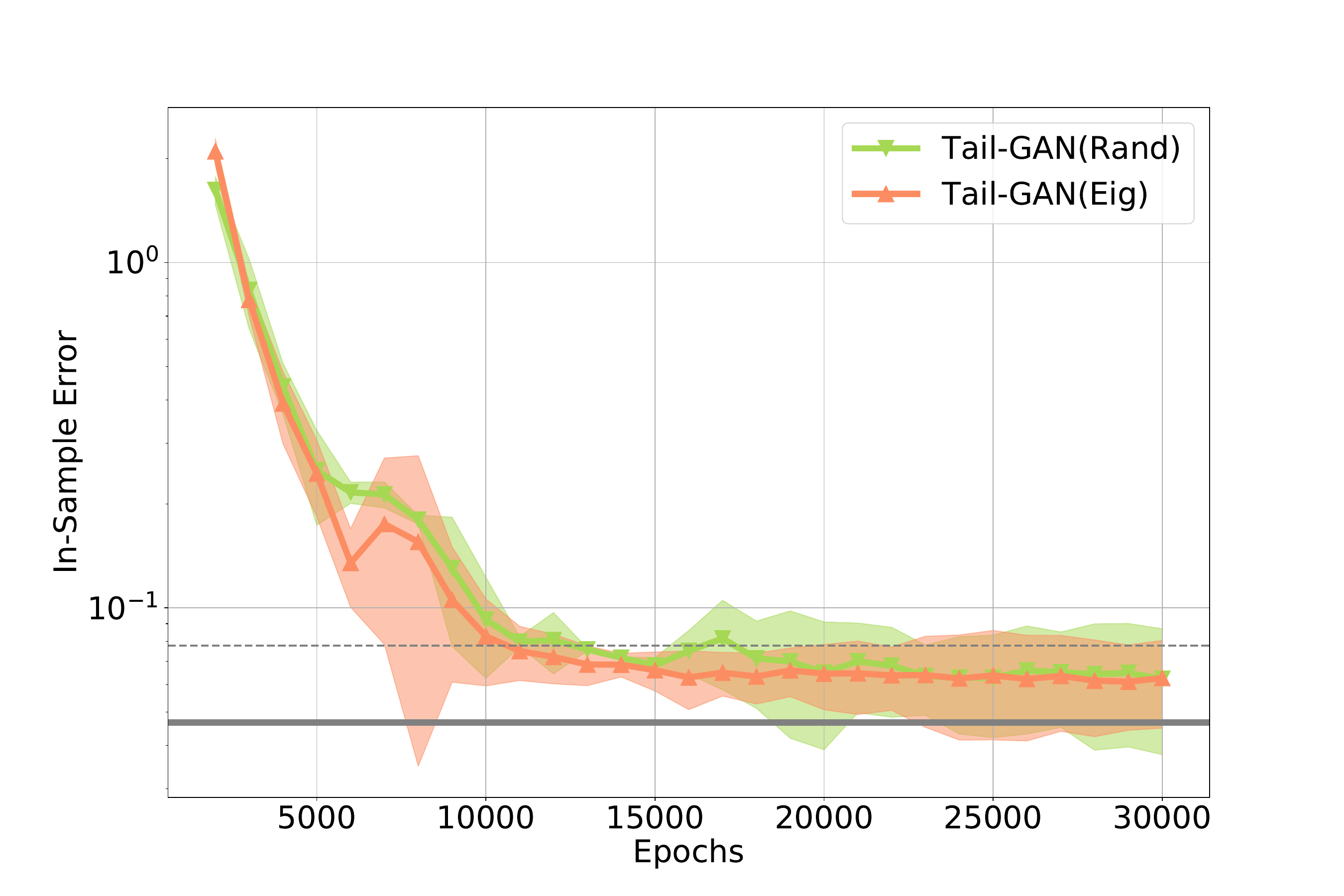}
\vspace{-2mm} 
\caption{Training performance on 50 random portfolios vs 20 eigenportfolios, as  in Section \ref{sec:scale}: mean of relative error ${\rm RE}(1000)$ and standard deviation (shaded areas). Grey horizontal line: average simulation error. Dotted line: average simulation error plus one standard deviation. Each experiment is repeated 5 times with different random seeds.}
\label{fig:real_eig}
\vspace{-2mm}
\end{figure}

\FloatBarrier
\paragraph{Scalability.} To test the scalability property of {\TGAN} on realistic scenarios, we conduct a similar experiment as in Section \ref{sec:scale}. The stocks considered here include the top 20 stocks in the S\&P500 index. The training period is between 2019-11-01 and 2019-11-30. 

Using eigenportfolios improves the convergence of in-sample errors, shown in Figure \ref{fig:real_eig}, and decreases out-of-sample errors, reported in Table \ref{tab:real_eig_error}, with fewer training portfolios.

\begin{table}[!htbp]
\vspace{-2mm}
\centering
\resizebox{0.75\textwidth}{!}{\begin{tabular}{lcccc}
\toprule
            & ``Oracle'' & HSM &  {\TGAN}(Rand) & {\TGAN}(Eig) \\\midrule
Out of sample  & 2.2 & 25.9  &  31.0 & 25.6  \\
  Error (\%)        & (1.7) & (5.1) &  (1.0) & (1.0) \\\bottomrule
\end{tabular}

% \begin{tabular}{l|ccc}
%             & HSM &  {\TGAN}(Rand) & {\TGAN}(Eig) \\\hline
% Out-Of-Sample Error (\%)  & 25.9  &  31.0 & 25.6  \\
%           & (5.1) &  (1.0) & (1.0) \\\hline
% \end{tabular}
}
\caption{Mean and standard deviation (in parentheses) for relative errors in out-of-sample tests.  ``Oracle'' represents the sampling error of the testing data. Each experiment is repeated five times with different random seeds.}  
\label{tab:real_eig_error}
\end{table}

The codes associated with the experiments can be found at: \url{https://github.com/chaozhang-ox/Tail-GAN}.

\section{Conclusion}
We have introduced a novel data-driven methodology for the accurate simulation of ``tail risk'' scenarios for high-dimensional multi-asset portfolios. 
Through detailed numerical experiments, we have illustrated the adequate performance of the algorithms compared to other generative models; in particular, we have demonstrated that   {\TGAN}  correctly captures the tail risks for a broad class of trading strategies in and out of sample.

Our proposed  framework lends itself to  various generalizations which are worthy of exploring. One important extension is to use data other than price histories as inputs; for example, joint   information from prices and the limit order book  (\cite{cont2023limit}) may be used to  enable the generation of high-frequency financial scenarios which lead to realistic profit/loss distributions for commonly used high-frequency trading strategies.

\FloatBarrier

\bibliographystyle{plainnat}
\bibliography{GAN.bib}
\newpage

%%%%%%%%%%%%%%%%%%%%%%%%%%%%%%%%%%%%%%%%%%%%%%%%%%%%%%%%%%%%%%%%%%%%%%%%%%%%%%%%%%%%%%%%%%%%%%%%%%%%%%%%%%%%%%%%%%%%%%%%%%
\appendix

For better clarity in explaining the theoretical results, we introduce the pushforward mapping. Given measurable spaces  $(X_{1},\Sigma _{1})$ and $(X_{2},\Sigma _{2})$, a measurable mapping $ \Phi\colon X_{1}\to X_{2}$  and a measure  $\mu \colon \Sigma _{1}\to [0,+\infty ]$, the pushforward of $\mu$  is defined to be the measure $\Phi\#\mu$ given by, for any $B \in \Sigma_2$,
\begin{eqnarray}\label{eq:push_forward_mapping}
\Phi\#\mu(B) = \mu\Big(\Phi^{-1}(B)\Big).
\end{eqnarray} 

For example, $\Pi^k\#\mathbb{P}_r$ denotes the distribution of $\Pi^k(\mathbf{p})$ under $\mathbb{P}_r$.

\section{Equivalence between bi-level optimization and max-min game}\label{app:equivalence}
Here, we provide the details underlying Theorem \ref{thm:equivalent_formulation_informal}, which establishes the equivalence between the bi-level optimization problem and the corresponding max-min game.

To start, ideally, the discriminator $\Dbar$ takes strategy PnL distributions as inputs, and outputs two values for each of the $K$ strategies, aiming to provide the correct $({\rm VaR}_{\alpha},{\rm ES}_{\alpha})$.  Mathematically, this amounts to 
\begin{eqnarray}\label{eq:theoretical_discriminator}
\Dbar^* \in \arg \min_{\Dbar} \frac{1}{K}\sum_{k=1}^{K} \mathbb{E}_{\mathbf{p}\sim\mathbb{P}_r}\left[S_{\alpha}\left( \overbrace{ \Dbar(\underbrace{\Pi^k \# \mathbb{P}_r}_{\text{strategy PnL distribution}})}^{\text{VaR and ES prediction  from } \Dbar};\mathbf{p}\right)\right].
\end{eqnarray}

\paragraph{Bi-level optimization problem.} We first start with a theoretical version of the objective function to introduce some insights  and then provide the practical sample-based version for training. Given two classes of functions $\overline{\mathcal{G}}:=\{\Gbar \,:\, \mathbb{R}^{N_z}\rightarrow \Omega \}$ and $\overline{\mathcal{D}}:=\{\Dbar \,:\, \mathcal{P}(\mathbb{R}) \rightarrow \mathbb{R}^{2} \}$, our goal is to find a generator $\Gbar^*\in \overline{\mathcal{G}}$ and  a discriminator $\Dbar^*\in \overline{\mathcal{D}}$ via the following bi-level (or constrained) optimization problem
\begin{eqnarray}\label{eq:generator_supervised}
\Gbar^* \in \arg \min_{\Gbar\in \overline{\mathcal{G}}} \frac{1}{K}\,\sum_{k=1}^K \mathbb{E}_{\mathbf{p}\sim \mathbb{P}_r} \left[S_{\alpha}\Big(\Dbar^*\big(\Pi^k\#\mathbb{P}_{\Gbar}\big),\,\,\Pi^k(\mathbf{p})\Big)\right],
\end{eqnarray}
where  $\mathbb{P}_{\Gbar}\in \mathcal{P}(\Omega)$ is the distribution of the samples from $\Gbar$ and 
\begin{eqnarray}\label{eq:distriminator_constrain}
\Dbar^* \in \arg \min_{\Dbar\in \overline{\mathcal{D}}} \frac{1}{K}\,\sum_{k=1}^K  \mathbb{E}_{\mathbf{p}\sim \mathbb{P}_r}\left[S_{\alpha}\Big(\Dbar\big(\Pi^k \# \mathbb{P}_{r} \big),\,\,\Pi^k(\mathbf{p})\Big)\right].
\end{eqnarray}

In the bi-level optimization problem \eqref{eq:generator_supervised}-\eqref{eq:distriminator_constrain},
the discriminator $\Dbar^*$ aims to map  the PnL distribution  to the associated $\alpha$-VaR and $\alpha$-ES values. Given the definition of the score function and the joint elicitability property of VaR and ES, we have $\Dbar^*(\cdot) := ({\rm VaR}_{\alpha},{\rm ES}_{\alpha})(\cdot)$ according to  \eqref{M_statistics}. Assume $\Dbar^*$ solves  \eqref{eq:distriminator_constrain}, the generator  $\Gbar^* \in \overline{\mathcal{G}}$ in \eqref{eq:generator_supervised} aims to  map the noise input to a price scenario that has  {\it consistent} ${\rm VaR}$ and  ${\rm ES}$ values of the strategy PnLs applied to $\mathbb{P}_r$.

\paragraph{From bi-level optimization to max-min game.}
In practice, constrained optimization problems are difficult to solve, and one can instead relax the constraint by applying the Lagrangian relaxation method with a dual parameter $\lambda>0$, leading to a max-min game between two neural networks $\Dbar$ and $\Gbar$,
\begin{eqnarray}\label{newminimax_distribution} 
\max_{\Dbar \in \overline{\mathcal{D}}_0}\min_{\Gbar\in \overline{\mathcal{G}}}
\frac{1}{K}\, \sum_{k=1}^K\Bigg[ \mathbb{E}_{\mathbf{p}\sim \mathbb{P}_r}\left[\,\,
S_{\alpha}\Big(\Dbar\big(\Pi^k\#\mathbb{P}_{\Gbar}\big),\Pi^k(\mathbf{p})\Big)\right] - \lambda \,\,  \mathbb{E}_{\mathbf{p}\sim \mathbb{P}_r}\left[S_{\alpha}\Big(\Dbar(\Pi^k\#\mathbb{P}_r),\,\,\Pi^k(\mathbf{p})\Big)\right]\,\,\Bigg],
\end{eqnarray}
where   
\begin{eqnarray}\label{eq:smaller_D}
   \overline{\mathcal{D}}_0:= \Big\{\Dbar\,:& \mathcal{P}(\mathbb{R}) \rightarrow \mathbb{R}^{2} \,\,\textit{ and }\,\, \exists\mu \in \mathcal{P}(\Omega) \text{ with a finite first moment }\,\, \text{s.t.}\nonumber\\
  & \Dbar(\Pi^k\#\mu) = ({\rm VaR}_\alpha(\Pi^k,\mathbb{P}_r),{\rm ES}_\alpha(\Pi^k,\mathbb{P}_r)), k=1,2,\cdots,K  \,\Big\}
\end{eqnarray}
is a smaller set of discriminators such that $ \overline{\mathcal{D}}_0\subseteq \mathcal{D}$. The set \eqref{eq:smaller_D} of discriminators may be described as the set of maps $\overline{\mathcal{D}}\,: \mathcal{P}(\mathbb{R}) \rightarrow \mathbb{R}^{2}$ which 
    can match the target VaR and ES values at least for {\it some} probability measure $\mu$ on market scenarios.
    This is a feasibility constraint, which can be viewed as a ``non-degeneracy'' or expressibility requirement on the map $\Dbar$ and excludes trivial cases such as constant maps etc.

\begin{theorem}[Equivalence of the Formulations]\label{thm:equivalent_formulation} 
Set $N_z = M\times T$. Assume that $\mathbb{P}_z$ has a finite first moment and is absolutely continuous with respect to the Lebesgue measure.
   Then the max-min game \eqref{newminimax_distribution} with $ \overline{\mathcal{D}}_0$  is equivalent to the bi-level optimization problem \eqref{eq:generator_supervised}-\eqref{eq:distriminator_constrain} for any $\lambda>0$.
\end{theorem}

The proof of Theorem \ref{thm:equivalent_formulation} is deferred to Appendix \ref{proof4}.

\section{Technical proofs}\label{sec:proofs}
\subsection{Proof of Proposition \ref{thm:optimization}}
\label{proof1}
{\it Proof of Proposition \ref{thm:optimization}.}
First we check the elicitability condition for $H_1(v)$ and $H_2(e)$ on region $\mathcal{B}$.  When $H_2(e)=\frac{\alpha}{2}e^2$, we have $H_2^{\prime}(e) = \alpha\,e$ and $H_2^{\prime \prime}(e) = \alpha$. For any $(v,e)\in \mathcal{B}$, this amounts to 
\begin{eqnarray}
\frac{\partial R_{\alpha}(v,e)}{\partial v} = e -W_{\alpha}v \geq 0, \nonumber
\end{eqnarray}
where $R_{\alpha}(v, e)$ is defined in \eqref{eq:r_function}.

Recall the score function $S_{\alpha}(v,e,x)$ defined in \eqref{eq:quant_score}, %assume $X$ follows PDF $f(x)$ and CDF $F(x)$, 
and  $s_{\alpha}(v,e)$ defined in \eqref{eq:expected_score}. Then
\begin{eqnarray}
s_{\alpha}(v,e)
&= & -(\mu(X\leq v)-\alpha)\frac{W_{\alpha}}{2}v^2+\frac{W_{\alpha}}{2}\int_{-\infty}^{v}x^2\mu(\dd x)+\mu(X \leq v)v\,e \nonumber\\
%&&-e\int_{-\infty}^v x\mathbb{P}(X\in \dd x)+\alpha e (e-v)-\alpha e+\text{const.}
&&-e\int_{-\infty}^v x\mu(\dd x)+\alpha e \left(\frac{e}{2}-v\right)+\text{const.} \nonumber
\end{eqnarray}
Therefore,
\begin{eqnarray*}
\frac{\partial s_{\alpha}}{\partial v}(v,e) &=& \Big(\mu(X \leq v)-\alpha\Big)\left(-W_{\alpha}v+ e\right),\\
\frac{\partial s_{\alpha}}{\partial e}(v,e)&=&  \mu(X \leq v)v-\int_{-\infty}^v x\mu(\dd x)+\alpha(e-v).
\end{eqnarray*}
And hence
\begin{eqnarray*}
\frac{\partial^2 s_{\alpha}}{\partial v^2}(v,e)&=&\frac{\mu(\dd v)}{\dd v}\left(-W_{\alpha}v+e \right)-W_{\alpha}(\mu(X \leq v)-\alpha),\label{h_vv}\\
\frac{\partial^2 s_{\alpha}}{\partial e^2}(v,e)&=&\alpha,\quad
\frac{\partial^2 s_{\alpha}}{\partial e\partial v}(v,e)= \mu(X \leq v)-\alpha.
\end{eqnarray*}
Since $\frac{\mu(X \in \dd v)}{\dd v}\geq 0$ and $-W_{\alpha}v+e>0$ hold on region $\mathcal{B}$, we have
\[
\frac{\partial^2 s_{\alpha}}{\partial v^2}(v,e) \geq -W_{\alpha}(\mu(X \leq v)-\alpha), \textit{ on } \mathcal{B}.
\]
Therefore $\frac{\partial^2 s_{\alpha}}{\partial v^2}(v,e)\ge 0$ holds  since  $v\leq {\rm VaR}_{\alpha}(\mu)$ on region $\mathcal{B}$. Next when $(v,e)\in \mathcal{B}$,
\begin{eqnarray}
&&\frac{\partial^2 s_{\alpha}}{\partial v^2} \frac{\partial^2 s_{\alpha}}{\partial e^2 } - \left(\frac{\partial^2 s_{\alpha}}{\partial v \partial e}\right)^2=\alpha\,\frac{\mu(X\in \dd v)}{\dd v}\left(-W_{\alpha}v+e \right)-\alpha W_{\alpha}(\mu(X \leq v)-\alpha) - \left(\mu(X \leq v)-\alpha \right)^2\nonumber\\
&\ge& (\alpha-\mu(X \leq v))(\alpha W_{\alpha}-\alpha+\mu(X \leq v))\label{eq:inq1}\\
&\ge&  (\alpha-\mu(X \leq v))\mu(X \leq v) \ge 0.\label{eq:inq2}
\end{eqnarray}
Note that \eqref{eq:inq1} holds since $-W_{\alpha}v+e\ge 0$,  and \eqref{eq:inq2} holds since $W_{\alpha} \ge 1$ and $\mu(X \leq v) \leq \alpha$ on $\mathcal{B}$. Therefore $\nabla^2\,s_{\alpha}$ is positive semi-definite on the region $\mathcal{B}$. 

In addition, when condition \eqref{eq:addition_assumption} holds, we show that $s_{\alpha}(v,e)$ is positive semi-definite on $\widetilde{\mathcal{B}}$.  

Denote $\widetilde{\mathcal{B}}^1 = \widetilde{\mathcal{B}}\cap \big\{(v,e)\in \mathbb{R}^2\,|\,v\leq {\rm VaR}_{\alpha}(\mu)\big\}$ and $\widetilde{\mathcal{B}}^2 = \widetilde{\mathcal{B}}\cap \{(v,e)\in \mathbb{R}^2|v > {\rm VaR}_{\alpha}(\mu)\}$. Then $\widetilde{\mathcal{B}}^1 \cup \widetilde{\mathcal{B}}^2=\widetilde{\mathcal{B}}$ and $\widetilde{\mathcal{B}}^1 \cap \widetilde{\mathcal{B}}^2=\emptyset$. The positive semi-definite property for $ \nabla^2 s_{\alpha}$ on $\widetilde{\mathcal{B}}^1$ follows a similar proof as above. 

We only need to show that $\nabla^2 s_{\alpha}$ is positive semi-definite on $\widetilde{\mathcal{B}}^2$. 
In this case, we have
\begin{eqnarray}
\frac{\partial^2 s_{\alpha}}{\partial v^2}(v,e)&=&\frac{\mu(\dd v)}{\dd v}\big(-W_{\alpha}v+e \big)-W_{\alpha}(\mu(X \leq v)-\alpha)\nonumber\\
&\geq& \delta_{\alpha}\, z_{\alpha}-W_{\alpha}(\mu(X \leq v)-\alpha) \ge 0, 
\label{eq:second_order}
\end{eqnarray}
which holds since  $\frac{\delta_{\alpha}\, z_{\alpha}}{W_{\alpha}}+\alpha \ge \beta_{\alpha}+\alpha \ge  \mu(X \leq v)$ on $\widetilde{\mathcal{B}}$. In addition,
\begin{eqnarray}
&&\frac{\partial^2 s_{\alpha}}{\partial v^2} \frac{\partial^2 s_{\alpha}}{\partial e^2 } - \left(\frac{\partial^2 s_{\alpha}}{\partial v \partial e}\right)^2=\alpha\,\frac{\mu(\dd v)}{\dd v}\left(-W_{\alpha}v+e \right)-\alpha W_{\alpha}(\mu(X \leq v)-\alpha) - \left(\mu(X \leq v)-\alpha \right)^2\nonumber\\
&\ge& \alpha\delta_{\alpha}\, z_{\alpha}+ (\mu(X \leq v)-\alpha)(-\alpha W_{\alpha}+\alpha-\mu(X \leq v))\label{eq:second_order2}\\
&\ge& \alpha\delta_{\alpha}\, z_{\alpha} - \beta_{\alpha}(\alpha W_{\alpha}+\beta_{\alpha})\label{eq:second_order3}\ \ge 0\label{eq:second_order4}.
\end{eqnarray}
Here \eqref{eq:second_order2} holds since $\frac{\mu( \dd v)}{\dd v} \ge \delta_{\alpha}$ and $z_{\alpha} \ge \left(-W_{\alpha}v+e \right)$. \eqref{eq:second_order3} holds since $\mu(X \leq v) \in (\alpha,\alpha +\beta_{\alpha}]$ on $\widetilde{\mathcal{B}}^2$.
To show \eqref{eq:second_order4}, it suffices to show
\begin{eqnarray}
\alpha\delta_{\alpha}\, z_{\alpha} - \frac{\delta_{\alpha}\, z_{\alpha}}{2W_{\alpha}}\left(\alpha W_{\alpha}+\frac{\delta_{\alpha}\, z_{\alpha}}{2W_{\alpha}}\right)  \ge 0, \label{eq:second_order5}
\end{eqnarray}
since $\beta_{\alpha}\leq \frac{\delta_{\alpha}\, z_{\alpha}}{2W_{\alpha}}$. 
Finally,  \eqref{eq:second_order5} holds since $W_{\alpha}>\frac{1}{\sqrt{\alpha}}$, $\delta_{\alpha}\in (0,1)$, and  $z_{\alpha}\in \left(0,\frac{1}{2}-\alpha\right)$.  \hfill \qedsymbol{}\\

\subsection{Proof of Theorem \ref{thm:universial_VaR_ES}}
\label{proof2}

\noindent {\it Proof of Theorem \ref{thm:universial_VaR_ES}.} 
 \underline{Step 1.} Consider the optimal transport problem in the semi-discrete setting: the source measure $\mathbb{P}_z$ is continuous  and the target measure $P_n$ is discrete. Under Assumption \ref{ass:distribution}, we can write $\mathbb{P}_{z}({\rm d}x) = m(x){\rm d}x$ for some probability density $m$. $P_n$ is discrete and we can write $P_n=\sum_{i=1}^n \nu_i \delta_{y_i}$ for some $\{y_i\}_{i=1}^n\subset \Omega$, $\nu_j\ge 0$ and $\sum_{j=1}^n \nu_j=1$. In this semi-discrete setting, the Monge's problem is defined as
\begin{eqnarray}\label{eq:Monge}
    \inf_{\Phi} \int \frac{1}{2}\|x-\Phi(x)\|^2 m(x){\rm d}x \quad \text{ s.t. } \int_{\Phi^{-1}(y_j)}{\rm d} \mathbb{P}_z = \nu_j, \,\, j=1,2,\cdots,n.
\end{eqnarray}
In this case, the transport map assigns each point $x\in\Omega$ to one of these $y_j$. Moreover, by taking advantage of the discreteness of the measure $\nu$, one sees that the dual Kantorovich problem in the semi-discrete case is maximizing the following functional:
 \begin{eqnarray}\label{eq:kirly_F}
        \mathcal{F}(\psi) =\mathcal{F}(\psi_1,\cdots,\psi_n) =\int \inf_j \Big( \frac{1}{2}\|x-y_j\|^2-\psi_j\Big)m(x) {\rm d}x +\sum_{j=1}^n \psi_j \nu_j.
    \end{eqnarray}
The optimal Monge transport for \eqref{eq:Monge} may be characterized by the maximizer of $\mathcal{F}$. To see this, let us introduce the concept of power diagram. Given a finite set of points $\{y_j\}_{j=1}^n\subset \Omega$ and the scalars $\psi=\{\psi_j\}_{j=1}^n$, the power diagrams associated to the scalars $\psi$ and the points $\{y_j\}_{j=1}^n$ are the sets:
\begin{equation}
    S_j = \left\{x\in \Omega \,\,\,\,\left\vert \,\,\,\, \frac{1}{2}\|x-y_j\|^2-\psi_j\leq \frac{1}{2}\|x-y_k\|^2-\psi_k,\, \forall k \neq j\right\}\right..\nonumber
\end{equation}
By grouping the points according to the power diagrams $S_j$, we have from \eqref{eq:kirly_F} that
\begin{equation}
    \mathcal{F}(\psi) = \sum_{j=1}^n \Big[\int_{S_j}\left( \frac{1}{2}\|x-y_j\|^2 -\psi_j\right)m(x){\rm d}x +\psi_j\nu_j\Big].
\end{equation}

According to Theorem 4.2 in \citet{lu2020universal},
   the optimal transport plan $\Phi$ to solve the semi-discrete Monge's problem is given by 
    \begin{equation}
        \Phi(x) = \nabla \bar{\psi} (x),\nonumber
    \end{equation}
    where $\bar{\psi}(x) = \max_j \{x\cdot y_j +m_j\}$ for some $m_j\in \mathbb{R}$. Specifically, $\Phi(x) = y_j$ if $x\in S_j(x)$. Here $\psi = (\psi_1,\cdots,\psi_n)$ is an maximizer of $\mathcal{F}$ defined in
   \eqref{eq:kirly_F} and $\{S_j\}_{j=1}^n$ denotes the power diagrams associated to $\{y_j\}_{j=1}^n$ and $\psi$.

Proposition 4.1 in \citet{lu2020universal} guarantees that there exists a feed-forward neural network $G(\cdot;\gamma)$ with $L = \lceil \log n \rceil$ fully connected layers of equal width   $N=2^{L} $  and   ReLU activation such that $\bar{\psi}(\cdot) = G(\cdot;\gamma)$.

\underline{Step 2.} 
Denote $\mathbb{P}_r^{(n)}(\cdot):=\frac{1}{n}\sum_{i=1}^n {\bf 1}\{\cdot=\mathbf{p}_i\}$ as an empirical measure to approximate  $\mathbb{P}_r \in \mathcal{P}(\Omega)$  using $n$ i.i.d. samples $\{\mathbf{p}_i\}_{i=1}^n$.  Let $\{\Pi^k(\mathbf{p}_{[i]})\}_{i=1}^n$
be the order statistics of $\{\Pi^k(\mathbf{p}_i)\}_{i=1}^n$,
i.e., $\Pi^k(\mathbf{p}_{[1]})\leq \cdots\leq \Pi^k(\mathbf{p}_{[n]})$.

 Let $\hat{v}_{n,\alpha}^k$ and $\hat{e}_{n,\alpha}^k$ denote the estimates of VaR and ES at level $\alpha$ using the $n$ samples above. These quantities are defined as follows (\citet{serfling2009approximation}):
\begin{eqnarray*}
\hat{v}_{n,\alpha}^k &=& \Pi^k(\mathbf{p}_{[\lceil \alpha n\rceil]}), \text{ and }\\
\hat{e}_{n,\alpha}^k &=& \frac{1}{n(1-\alpha)}\sum_{i=1}^n \Pi^k(\mathbf{p}_{i}) {\bf 1}\{\Pi^k(\mathbf{p}_{i}) \le \hat{v}_{n,\alpha}^k \}.
\end{eqnarray*}
We first prove the result under the VaR criteria. According to \citet[Proposition 2]{kolla2019concentration},  with probability at least $\frac{1}{2}$ it holds that
\begin{equation}\label{eq:concentration_VaR}
  \left|\hat{v}_{n,\alpha}^k- {\rm VaR}_\alpha(\Pi^k,\mathbb{P}_r)\right|\leq \sqrt{\frac{\log(4)}{2nc}},
\end{equation}
where $c=c(\delta_k,\eta_k)$ is a constant that depends on $\delta_k$ and $\eta_k$, which are specified in Assumption {\bf A3}. Setting the RHS of \eqref{eq:concentration_VaR} as $\varepsilon$, we have $n = \mathcal{O}(\varepsilon^{-2})$. Under this choice of $n$, we have $\left|\hat{v}_{n,\alpha}^k-{\rm VaR}(\Pi^k,\mathbb{P}_r)\right|<\varepsilon$ holds with probability at least $\frac{1}{2}$. This implies that there must exist an empirical measure $\mathbb{P}_r^{(n)*}$ such that the corresponding  $\hat{v}_{n,\alpha}^k$ satisfies $\left|\hat{v}_{n,\alpha}^k-{\rm VaR}(\Pi^k,\mathbb{P}_r)\right|<\varepsilon$. $\mathbb{P}_r^{(n)*}$ will be the target (empirical) measure we input in Step 1. Therefore setting $n = \mathcal{O}(\varepsilon^{-2})$ leads to the fact that  $L = \lceil \log n \rceil = \mathcal{O}(\log(\varepsilon^{-2}))$ and  $N=2^{L} = \mathcal{O}(\varepsilon^{-2\log 2})$, which concludes the main result for the universal approximation under the VaR criteria.

We next prove the result under the ES criteria. Under Assumptions \ref{ass:strategy} and \ref{ass:distribution}, we have
\begin{equation}\label{eq:finite_moment_pi}
  \mathbb{E}_{\mathbb{P}_r}\left[ |\Pi^k(\mathbf{p})|^\beta \right]\leq (\ell_k) ^\beta   \mathbb{E}_{\mathbb{P}_r}\left[\|\mathbf{p}\|^\beta\right]<\infty.
\end{equation}
Take $n>\frac{16\log(8)}{(\eta_k\delta_k(1-\alpha))^2}$. Under \eqref{eq:finite_moment_pi} and Assumption {\bf A3}, with probability $\frac{1}{2}$ it holds that
\begin{eqnarray}
\label{eq:concentration_ES}
    \left|\hat{e}_{n,\alpha}^k-{\rm ES}(\Pi^k,\mathbb{P}_r)\right| &\leq& \frac{\left(5\Big(\mathbb{E}_{\mathbb{P}_r}[\|\Pi^k(\mathbf{p})\|^\beta]\Big)^{1/\beta} - {\rm VaR}(\Pi^k,\mathbb{P}_r)\right)}{(1-\alpha)}\left(\frac{1}{n}\right)^{1-\frac{1}{\beta}}\sqrt{\log(6)}\noindent + \frac{4}{\eta_k(1-\alpha)}\sqrt{\frac{\log(8)}{n}},\nonumber
\end{eqnarray}
where $\eta_k$ and $\delta_k$ are as defined in {\bf A3}. The result in \eqref{eq:concentration_ES} is a slight modification of \citet[Theorem 4.1]{prashanth2020concentration}. Setting the RHS of \eqref{eq:concentration_ES} as $\varepsilon$, we have $n = \mathcal{O}(\varepsilon^{-\frac{\beta}{\beta-1}})$. Under this choice of $n$, we have $\left|\hat{e}_{n,\alpha}^k-{\rm ES}(\Pi^k,\mathbb{P}_r)\right|<\varepsilon$ holds with probability at least $\frac{1}{2}$. This implies that there must exist an empirical measure $\mathbb{P}_r^{(n)*}$ such that $\left|\hat{e}_{n,\alpha}^k-{\rm ES}(\Pi^k,\mathbb{P}_r)\right|<\varepsilon$ holds. $\mathbb{P}_r^{(n)*}$ will be the target (empirical) measure we input in Step 1. 
Note that in this case, $L = \lceil \log n \rceil = \mathcal{O}(\log(\varepsilon^{-\frac{\beta}{\beta-1}}))$ and  $N=2^{L} =\mathcal{O}(\varepsilon^{-\frac{\beta}{\beta-1}\log 2})$, which concludes the main result for the universal approximation under the ES criterion. \hfill \qedsymbol{}\\

\subsection{Proof of Theorem \ref{thm:universial_general}}
\label{proof3}
\noindent {\it Proof of Theorem \ref{thm:universial_general}}. Step 1 is the same as Theorem \ref{thm:universial_VaR_ES}. It is sufficient to prove the corresponding Step 2.

\underline{Step 2.} 
Denote $\mathbb{P}_r^{(n)}(\cdot):=\frac{1}{n}\sum_{i=1}^n {\bf 1}\{\cdot=\mathbf{p}_i\}$ as an empirical measure to approximate  $\mathbb{P}_r \in \mathcal{P}(\Omega)$  using $n$ i.i.d. samples $\{\mathbf{p}_i\}_{i=1}^n$. Denote  $M_\beta:=\mathbb{E}_{\mathbb{P}_r}[\|\mathbf{p}\|^\beta] <\infty$.  From \citet[Theorem 3.1]{lei2020convergence} we have
  \begin{equation}\label{eq:expected_W1}
      \mathbb{E}\mathcal{W}_1 (\mu_n,\mu) \leq c_{\beta} M_\beta n^{-\frac{1}{(2\beta)\vee (M\times T)}\wedge(1-\frac{1}{\beta})} (\log n)^{\zeta_{\beta,M\times T}},
  \end{equation}
  where $c_{\beta}$ is a constant depending only on $\beta$ (not $M\times T$) %\CZ{$p$ is beta? and $d$?}, and 
  \begin{eqnarray*}
  \zeta_{\beta,M\times T} = 
      \begin{cases}
          2 & \quad  \textit{ if } M\times T= \beta = 2,\\
       1 & \quad  \textit{ if } ``M\times T\neq 2 \textit{ and } \beta=\frac{M\times T}{M\times T-1}\wedge 2" \textit{ or } ``\beta>M \times T=2",\\
      0 & \quad \textit{ otherwise}.
      \end{cases}
  \end{eqnarray*}
By  Kantorovich duality, we have
\begin{eqnarray}
   % \mathcal{W}_1(\mu,\mu_n)= {\frac {1}{K}}\sup _{\|f\|_{L}\leq K}\mathbb {E} _{x\sim \mu }[f(x)]-\mathbb {E} _{y\sim \mu_n }[f(y)]
    \mathcal{W}_1\Big(\Pi^k\#\mathbb{P}_r,\Pi^k\# \mathbb{P}_r^{(n)}\Big) &=& {\frac {1}{\ell}}\sup _{\|f\|_{L}\leq \ell}\mathbb {E} _{\mathbf{p}\sim \mathbb{P}_r }\Big[f(\Pi^k(\mathbf{p}))\Big]-\mathbb {E} _{\mathbf{q}\sim \mathbb{P}_r^{(n)} }\Big[f(\Pi^k(\mathbf{q}))\Big]\noindent\\
    &\leq& {\frac {1}{\ell}}\sup _{\|g\|_{L}\leq \ell\ell_k}\mathbb {E} _{\mathbf{p}\sim \mathbb{P}_r}\Big[g(\mathbf{p})\Big]-\mathbb {E} _{\mathbf{q}\sim \mathbb{P}_r^{(n)} }\Big[g(\mathbf{q})\Big]\label{result1}\\
    &\leq& \ell_k \mathcal{W}_1\Big(\mathbb{P}_r,\mathbb{P}_r^{(n)}\Big)\label{result2}
\end{eqnarray}
where $\|\cdot\|_L$ is the Lipschitz norm. \eqref{result1}  holds since
$f(\Pi^k(\cdot))$ is $\ell \ell_k$-Lipschitz when $f$ is $\ell$-Lipschitz and $\Pi^k$ is $\ell_k$-Lipschitz. \eqref{result2} holds by  Kantorovich  duality.

Taking expectation on \eqref{eq:risk_bound1} and
applying  \eqref{eq:expected_W1} and \eqref{result2}, we have
\begin{eqnarray}\label{eq:risk_measure_bound}
 \mathbb{E}  \Big|\rho(\Pi^k,\mathbb{P}_r)-\rho(\Pi^k,\mathbb{P}_r^{(n)})\Big| &\leq& L  \mathbb{E}
\Big(\mathcal{W}_1\Big(\Pi^k\#\mathbb{P}_r,\,\Pi^k\#\mathbb{P}_r^{(n)}\Big)\Big)^{\kappa}\\
 &\leq& L  
 \Big(\mathbb{E} \left[\mathcal{W}_1\Big(\Pi^k\#\mathbb{P}_r,\,\Pi^k\#\mathbb{P}_r^{(n)}\Big)\right]\Big)^{\kappa}\label{eq:Jensen}\\
 & \leq & L\Big(\ell_k c_{\beta} M_\beta n^{-\frac{1}{2\vee M\times T}\wedge(1-\frac{1}{\beta})} (\log n)^{\zeta_{\beta,M\times T}}\Big)^\kappa, \label{eq:last_eqn}
\end{eqnarray}
where \eqref{eq:Jensen} holds by Jensen's inequality since $\kappa \in (0, 1]$. \hfill 

\eqref{eq:risk_measure_bound} implies that there must exist an empirical measure $\mathbb{P}_r^{(n)*}$ such that 
% $\left|\hat{v}_{n,\alpha}^k-{\rm VaR}(\Pi^k(\mathbf{p});\mathbb{P}_r)\right|
$\Big|\rho(\Pi^k,\mathbb{P}_r)-\rho(\Pi^k,\mathbb{P}_r^{(n)}) \Big|<\varepsilon$ holds. This $\mathbb{P}_r^{(n)*}$ will be the target (empirical) measure we input in Step 1. 

It is easy to check that
\begin{itemize}
    \item $\frac{1}{2\vee (M\times T)}\wedge(1-\frac{1}{\beta}) = 1-\frac{1}{\beta}$ when $M= T=1$ and $1<\beta \leq 2$;
    \item $\frac{1}{2\vee (M\times T)}\wedge(1-\frac{1}{\beta}) = \frac{1}{2}$ when $M=T=1$ and $\beta \ge 2$;
    \item $\frac{1}{2\vee (M\times T)}\wedge(1-\frac{1}{\beta}) = \frac{1}{M\times T}$ when $M\times T \ge 2$ and $\frac{1}{M\times T} + \frac{1}{\beta}<1$;
     \item $\frac{1}{2\vee (M\times T)}\wedge(1-\frac{1}{\beta}) = 1-\frac{1}{\beta}$ when $M\times T \ge 2$ and $\frac{1}{M\times T} + \frac{1}{\beta}\ge 1$.
\end{itemize}
This concludes the universal approximation result under risk measures that are H\"older continuous.
\hfill \qedsymbol{}

\subsection{Proof of Theorem \ref{thm:equivalent_formulation}}
\label{proof4}

\begin{proof}[Proof of Theorem \ref{thm:equivalent_formulation}.]
For any $\Dbar \in \overline{\mathcal{D}}_0$,   there exists $\mu := \mu(\Dbar) \in \mathcal{P}(\Omega)$ with  finite first moment such that
\begin{eqnarray}\label{eq: reacheability_condition}
\Dbar(\Pi^k\#\mu) = \Big({\rm VaR}_\alpha(\Pi^k,\mathbb{P}_r),{\rm ES}_\alpha(\Pi^k,\mathbb{P}_r)\Big), \quad \forall k=1,2,\cdots,K.
\end{eqnarray}
Denote $\Sigma(\Dbar)$ as the set of all such $\mu\in \mathcal{P}(\Omega)$ with finite first moment that satisfies \eqref{eq: reacheability_condition}. Then given that both $\mu\in \Sigma(\Dbar)$ and $\mathbb{P}_z$ have finite first moments and that $\mathbb{P}_z$ is absolutely continuous with respect to the Lebesgue measure, we could find
a mapping $\overline{G} \in \overline{\mathcal{G}}$ such that $\overline{G}\# \mathbb{P}_{{z}} \in \Sigma(\Dbar)$ so that $ \mathbb{E}_{\mathbf{p}\sim \mathbb{P}_r}\left[\,\,
S_{\alpha}\Big(\Dbar\big(\Pi^k\#\mathbb{P}_{\Gbar}\big),\Pi^k(\mathbf{p})\Big)\right]$ is minimized (Theorem 7.1 in \citet{ambrosio2003existence}). That is,
\begin{eqnarray*}
  \min_{\Gbar\in \overline{\mathcal{G}}} \frac{1}{K}\sum_{k=1}^K \mathbb{E}_{\mathbf{p}\sim \mathbb{P}_r}\left[\,\,
S_{\alpha}\Big(\Dbar\big(\Pi^k\#\mathbb{P}_{\Gbar}\big),\Pi^k(\mathbf{p})\Big)\right] = \frac{1}{K}\sum_{k=1}^K \mathbb{E}_{\mathbf{p}\sim \mathbb{P}_r}\left[\,\,
S_{\alpha}\Big(\Big({\rm VaR}_\alpha(\Pi^k,\mathbb{P}_r),{\rm ES}_\alpha(\Pi^k,\mathbb{P}_r)\Big),\Pi^k(\mathbf{p})\Big)\right].
\end{eqnarray*}
In this case, for the maximization problem of $\Dbar$ over $\overline{\mathcal{D}}_0$,
\begin{eqnarray*}
    \eqref{newminimax_distribution} &=& \max_{\Dbar \in \overline{\mathcal{D}}_0}\
\frac{1}{K}\, \sum_{k=1}^K\Bigg[ \mathbb{E}_{\mathbf{p}\sim \mathbb{P}_r}\left[\,\,
S_{\alpha}\Big(\Big({\rm VaR}_\alpha(\Pi^k,\mathbb{P}_r),{\rm ES}_\alpha(\Pi^k,\mathbb{P}_r)\Big),\Pi^k(\mathbf{p})\Big)\right] - \lambda \,\,  \mathbb{E}_{\mathbf{p}\sim \mathbb{P}_r}\left[S_{\alpha}\Big(\Dbar(\Pi^k\#\mathbb{P}_r),\,\,\Pi^k(\mathbf{p})\Big)\right]\,\,\Bigg]\\
&=& -\lambda  \min_{\Dbar \in \overline{\mathcal{D}}_0}\frac{1}{K}\, \sum_{k=1}^K\mathbb{E}_{\mathbf{p}\sim \mathbb{P}_r}\left[S_{\alpha}\Big(\Dbar(\Pi^k\#\mathbb{P}_r),\,\,\Pi^k(\mathbf{p})\Big)\right].
\end{eqnarray*}
By the definition of $\overline{\mathcal{D}}_0$, we have
\begin{eqnarray*}
   && \min_{\Dbar \in \overline{\mathcal{D}}_0}\frac{1}{K}\, \sum_{k=1}^K \mathbb{E}_{\mathbf{p}\sim \mathbb{P}_r}\left[S_{\alpha}\Big(\Dbar(\Pi^k\#\mathbb{P}_r),\,\,\Pi^k(\mathbf{p})\Big)\right] \\
    &=& \frac{1}{K}\, \sum_{k=1}^K \mathbb{E}_{\mathbf{p}\sim \mathbb{P}_r}\left[S_{\alpha}\Big(\Big({\rm VaR}_\alpha(\Pi^k,\mathbb{P}_r),{\rm ES}_\alpha(\Pi^k,\mathbb{P}_r)\Big),\,\,\Pi^k(\mathbf{p})\Big)\right]\\
    &=& \min_{\Dbar \in \overline{\mathcal{D}}}\frac{1}{K}\, \sum_{k=1}^K \mathbb{E}_{\mathbf{p}\sim \mathbb{P}_r}\left[S_{\alpha}\Big(\Dbar(\Pi^k\#\mathbb{P}_r),\,\,\Pi^k(\mathbf{p})\Big)\right], 
\end{eqnarray*}
which is equivalent to \eqref{eq:distriminator_constrain}. Denote this minimizer as $\Dbar^*$, plugging this into the optimization problem for $\Gbar$ in the max-min game leads to the upper-level optimization problem \eqref{eq:generator_supervised}.
\end{proof}

%%%%%%%%%%%%%%%%%%%%%%%%%%%%%%%%%%%%%%%%%%%%%%%%%%%%%%%%%%%%%%%%%%%%%%%%%%%%%%%%%%%%%%%%%%%%%%%%%%%%%%%%%%%%%%%%%%%%%%%%%%

\section{Implementation details}
\label{app:implementation}

\subsection{Setup of parameters in the synthetic data set}\label{app:para_syn}

Mathematically, for any given time $t\in [0,T]$, we first sample $\mathbf{u}_{t} = (u_{1, t}, \ldots, u_{5, t})^\top \sim \mathcal{N}(0, \Sigma)$ with covariance matrix $\Sigma \in \mathbb{R}^{5 \times 5}$, $v_{1,t} \sim \chi^{2}(\nu_1)$ and $v_{2,t} \sim \chi^{2}(\nu_2)$. Here  $v_{1,t} $, $v_{2,t} $ are independent of $\mathbf{u}_{t}$. We then calculate the price increments according to the following equations
% \begin{eqnarray*}
% \Delta p_{1, t} = u_{1, t}; \qquad \quad 
% \Delta p_{2, t}  = \phi_1 \Delta p_{2, t-1} + u_{2, t};   \qquad \quad 
% \Delta p_{3, t}  = \phi_2 \Delta p_{3, t-1} + u_{3, t};  
% \end{eqnarray*}
% \vspace{-1cm}
% \begin{eqnarray*}\label{eq:syn_formula}
% \Delta p_{4, t} = \varepsilon_{4, t} = \sigma_{4, t} \eta_{1, t};  \quad  \qquad
% \Delta p_{5, t}= \varepsilon_{5, t} = \sigma_{5, t} \eta_{2, t},
% \end{eqnarray*}
\begin{equation*}\label{eq:syn_formula}
\begin{aligned}
\Delta p_{1, t} &= u_{1, t},\quad 
\Delta p_{2, t} = \phi_1 \Delta p_{2, t-1} + u_{2, t},  \quad
\Delta p_{3, t} = \phi_2 \Delta p_{3, t-1} + u_{3, t}, \\
\Delta p_{4, t} &= \varepsilon_{4, t} = \sigma_{4, t} \eta_{1, t}, \quad
\Delta p_{5, t} = \varepsilon_{5, t} = \sigma_{5, t} \eta_{2, t},
\end{aligned}
\end{equation*}
where $\sigma_{4, t}^{2}=\gamma_{4} + \kappa_{4} \varepsilon_{4, t-1}^{2} + \beta_{4} \sigma_{4, t-1}^{2}, \eta_{1, t} = \frac{u_{4, t}}{\sqrt{v_{1,t} / \nu_1}}, $ and  $\sigma_{5, t}^{2}=\gamma_{5} + \kappa_{5} \varepsilon_{5, t-1}^{2} + \beta_{5} \sigma_{5, t-1}^{2}, \eta_{2, t} = \frac{u_{5, t}}{\sqrt{v_{2,t} / \nu_2}}$.

 We set $T=100$ as the number of observations over one trading day. We first generate a correlation matrix $\rho$ with elements uniformly sampled from $[0,1]$. We then sample the annualized standard deviations $s$ with values between $0.3$ and $0.5$, and set $\Sigma_{ij} =\frac{s_{i}}{255\times T}  \frac{s_j}{255\times T} \rho_{ij} $ ($i,j=1,2,\ldots,5$); $\phi_1=0.5$ and $\phi_2=-0.15$; $\nu_1=5$ and $\nu_2=10$; $\kappa_4$ and $\kappa_5$ are sampled uniformly from [0.08, 0.12]; $\beta_4$ and $\beta_5$ are sampled uniformly from [0.825, 0.875]; and finally $\gamma_4$ and $\gamma_5$ are sampled uniformly from [0.03, 0.07]. We choose one quantile $\alpha=0.05$ for this experiment. 
 
Table \ref{tab:syn_estimates} reports the 5\%-VaR and 5\%-ES values of several strategies calculated with the synthetic financial scenarios designed above.

\begin{table}[H]
    \centering
    \resizebox{0.8\textwidth}{!}{\begin{tabular}{lcccccc} 
\toprule
& \multicolumn{2}{c}{Static  buy-and-hold} &   \multicolumn{2}{c}{Mean-reversion} &  \multicolumn{2}{c}{Trend-following} \\
 \cmidrule(lr){2-3}  \cmidrule(lr){4-5}\cmidrule(lr){6-7}
  & VaR & ES & VaR & ES & VaR & ES                    \\\midrule
Gaussian & -0.489                & -0.615                 & -0.432                & -0.553                & -0.409                 & -0.515                \\
AR(1) with $\phi_1=0.5$ & -0.876                & -1.100                & -0.850                 & -1.066                & -0.671                 & -0.829                \\
AR(1) with $\phi_2=-0.12$ & -0.461                & -0.581               & -0.399                & -0.513                & -0.387                 & -0.488                \\
GARCH(1,1) with $t(5)$ & -0.480                 & -0.603               & -0.420                 & -0.535                & -0.400                   & -0.501                \\
GARCH(1,1) with $t(10)$ & -0.403                & -0.507               & -0.354                & -0.453                & -0.328                 & -0.410   \\\bottomrule      
\end{tabular}

% \begin{tabular}{l|cc|cc|cc|cc}
%   & \multicolumn{2}{c}{Single-Asset   Portfolio} & \multicolumn{2}{c}{Multi-Asset   Portfolio} & \multicolumn{2}{c}{Mean-Reversion   Strategy} & \multicolumn{2}{c}{Trend-Following   Strategy} \\
%   & VaR                   & ES                   & VaR                  & ES                   & VaR                   & ES                    & VaR                    & ES                    \\\hline
% 1 & -0.489                & -0.615               & -0.392               & -0.492               & -0.432                & -0.553                & -0.409                 & -0.515                \\
% 2 & -0.876                & -1.100                 & -0.233               & -0.296               & -0.850                 & -1.066                & -0.671                 & -0.829                \\
% 3 & -0.461                & -0.581               & -0.238               & -0.297               & -0.399                & -0.513                & -0.387                 & -0.488                \\
% 4 & -0.480                 & -0.603               & -0.296               & -0.373               & -0.420                 & -0.535                & -0.400                   & -0.501                \\
% 5 & -0.403                & -0.507               & -0.356               & -0.450                & -0.354                & -0.453                & -0.328                 & -0.410   \\\hline             
% \end{tabular}
}
    \caption{Empirical VaR and ES values for trading strategies evaluated on the training data.} 
    \label{tab:syn_estimates}
\end{table}

\subsection{Setup of the configuration}\label{app:configuration}
\begin{table}[H]
    \centering
    \resizebox{0.95\textwidth}{!}{\begin{tabular}{ccc}
\toprule
                                  & Configuration & Values \\\midrule
\multirow{5}{*}{Discriminator}            & Architecture       & Fully-connected layers     \\
            & Activation       & Leaky ReLU     \\
            & Number of neurons in each layer    & (1000, 256, 128, 2)     \\   
            & Learning rate       & $10^{-7}$     \\
             & Dual parameter ($\lambda$)       & $1$     \\
            & Batch normalization       & No     \\\midrule

\multirow{6}{*}{Generator}            & Architecture       & Fully-connected layers     \\
            & Activation       & Leaky ReLU     \\
            & Number of neurons in each layer   & (1000, 128, 256, 512, 1024, $5 \times 100$)     \\ 
            & Learning rate       & $10^{-6}$     \\
            & Batch normalization       & Yes   \\\midrule
            
        \multirow{4}{*}{Strategies}             & Static portfolio with single asset& 5 \\
       & Static portfolio with multiple assets& 50 \\
             & Mean-reversion strategies & 5\\
                & Trend-following strategies & 5 \\\midrule
\multirow{5}{*}{Additional parameters}  
& Size of training data ($N$)& 50,000 \\
& Number of PnL samples ($N_{B}$)& 1,000 \\
& Noise dimension ($N_{z}$)& 1,000 \\
& Noise distribution & $t(5)$ \\
& $H_1, {H}_2$ & $H_1(v) = - 5 v^2$, ${H}_2(e) = \frac{\alpha}{2} e^2$ \\\bottomrule
\end{tabular}

}
    \caption{Network architecture configuration.  }
    \label{tab:configuration}
\end{table}

\paragraph{Discussion on the configuration.}
 \begin{itemize}     
        \item {\bf Choice of $\lambda$:}   Theorem \ref{thm:equivalent_formulation} suggests that {\TGAN} is effective as long as $\lambda > 0$. In our experiments, we set $\lambda=1$ and also tested values of 2, 10, and 100 to address the issue of hyper-parameter selection. We observed that $\lambda=2$ and $\lambda=10$ resulted in a similar performance to $\lambda=1$, while larger values such as $\lambda=100$ led to a worse performance similar to that of the supervised learning method. This may be due to the fact that larger $\lambda$  values could potentially harm the model's generalization power in practical settings.  
        \item  {\bf Choice of $S_\alpha$ ($H_1$ and $H_2$):} Proposition \ref{thm:optimization} demonstrates that choosing $H_1$ and $H_2$ as quadratic functions (as proposed in \cite{AS2014}) results in a positive semi-definite score function in a neighborhood region around the global minimum. This  evidence supports selecting quadratic functions for $H_1$ and $H_2$. 
        
        \item {\bf Neural network architecture:} Theorem \ref{thm:universial_VaR_ES} implies that a feed-forward neural network with fully connected layers of equal width and ReLU activation is capable of generating financial scenarios that are arbitrarily close to the scenarios sampled from the true distribution $\mathbb{P}_r$ under VaR and ES criteria. This sheds light on using a simple network architecture such as multi-layer perceptron (MLP) in the training of {\TGAN}.
        
       \qquad  While a more sophisticated neural network architecture may improve practical performance, our focus is not to compare different architectures, but rather to demonstrate the benefits of incorporating the essential component of tail risks of trading strategies into our {\TGAN} framework. Therefore, we choose to use a simple MLP, the same architecture used in Wasserstein GAN (\cite{arjovsky2017wasserstein}).  
    \end{itemize}

\subsection{Differentiable neural sorting}
\label{app:neural_sorting}
\begin{figure}[H] 
  \vspace{-3mm}
  \centering
  \includegraphics[width=.66\textwidth]{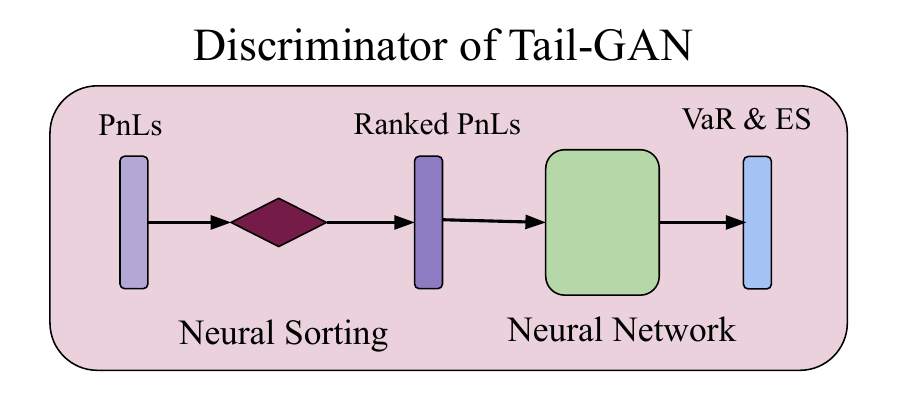}
  \vspace{-3mm}
  \caption{Architecture of the {\TGAN} discriminator.}
  \label{fig:discriminator}
\end{figure}

\vspace{-3mm}
The architecture of the {\TGAN} Discriminator has two key ingredients, as depicted in Figure \ref{fig:discriminator}. For the first ingredient, a differentiable sorting algorithm proposed by \citet{grover2019stochastic} is employed to rank the PnLs.  The second part adopts a standard neural network architecture, taking the ranked PnLs as the input, and providing the estimated $\alpha$-VaR and $\alpha$-ES values as the output.

 We follow the design in  \citet{grover2019stochastic} to include the differentiable sorting architecture, so that the input of the discriminator will be the ranked PnL's  (sorted in decreasing order). This design, based on the idea of using the \textsc{soft-max} operator to approximate the \textsc{arg-max} operator, enables  back-propagation of the gradient of the sorting function during the network training process.

Denote $\mathbf{x}^{k} = (x^k_1,x^k_2,\ldots,x^k_n)^{\top}$ as a real-valued vector of length $n$, representing the PnL samples of strategy $k$. Let $B({\mathbf{x}^{k}})$ denote the matrix
of absolute pairwise differences of the elements of $\mathbf{x}^{k}$, such that $B_{i,j}(\mathbf{x}^{k}) = |x_i^{k} -x_j^{k}|$. We then define the following permutation
matrix $\Gamma(\mathbf{x}^{k})$ following  \citet{grover2019stochastic,ogryczak2003minimizing} 
\begin{eqnarray*}
\Gamma_{i,j} (\mathbf{x}^{k})= 
\begin{cases}
1,\quad \mbox{ if } j=\arg \max ((n+1-2i)-B(\mathbf{x}^{k}){\bf 1}),\\
0,\quad \mbox{otherwise},
\end{cases}
\end{eqnarray*}
where $\bf 1$ is the all-ones vector. Then,  $\Gamma(\mathbf{x}^{k}) \mathbf{x}^{k}$  provides a ranked vector of $\mathbf{x}^{k}$ (\citet[Lemma 1]{ogryczak2003minimizing} and \citet[Corollary 3]{grover2019stochastic}). However, the \textsc{arg-max}  operator is {\it non-differentiable} which prohibits the direct usage of the permutation matrix for gradient computation. Instead, \citet{grover2019stochastic} propose to replace the \textsc{arg-max} operator with  \textsc{soft-max}, in order to obtain a continuous relaxation $\widehat{\Gamma}^{\tau}$ with a temperature parameter $\tau>0$. In particular, the $(i,j)$-th element of $\widehat{\Gamma}^{\tau}(\mathbf{x}^{k})$ is given by
\begin{eqnarray*}\label{eq:sorting_appx}
\widehat{\Gamma}^{\tau}_{i,j}(\mathbf{x}^{k}) = \frac{\exp\big(\big((n+1-2i)-B(\mathbf{x}^{k})_j{\bf 1}\big)/\tau\big)}{\sum_{l=1}^{n}\exp\big(\big((n+1-2i)-B(\mathbf{x}^{k})_l{\bf 1}\big)/\tau\big)},
\end{eqnarray*}
in which $B(\mathbf{x}^{k})_l$ is the $l$-th row of matrix $B(\mathbf{x}^{k})$. This relaxation is continuous everywhere and differentiable almost everywhere with respect to the elements of $\mathbf{x}^{k}$. In addition, \citet[Theorem 4]{grover2019stochastic} shows that $\widehat{\Gamma}^{\tau}_{i,j}(\mathbf{x}^{k})$ converges to $\Gamma_{i,j} (\mathbf{x}^{k})$
 almost surely when $x^k_1,\ldots,x^k_{n}$ are sampled IID from a distribution which is absolutely continuous with respect to the Lebesgue measure in $\mathbb{R}$.

Finally we could set in \eqref{eq:discriminator}:
$$\widetilde{\Gamma}(\mathbf{x}) =  \widehat{\Gamma}^{\tau}(\mathbf{x})\mathbf{x}.$$

\subsection{Construction of eigenportfolios}\label{app:eigenportfolios}
We construct eigenportfolios from the principal components of the sample correlation matrix $\hat{\mathbf{\rho}}$ of returns, ranked in decreasing order of eigenvalues: 
$\hat{\mathbf{\rho}} = \mathbf{Q} \mathbf{\Lambda} \mathbf{Q}^{-1}$ where $\mathbf{Q}$ is the orthogonal matrix with the $i$-th column being the eigenvector $\mathbf{q}_i \in \mathbb{R}^M$ of $\hat{\mathbf{\rho}}$, and $\mathbf{\Lambda}$ is the diagonal matrix whose diagonal elements are the corresponding eigenvalues, such that $\mathbf{\Lambda}_{1,1} \geq \mathbf{\Lambda}_{2,2} \geq \dots \geq \mathbf{\Lambda}_{M,M} \geq 0$.

Eigenportfolios are constructed from the principal components as 
follows. Denote $\mathbf{h}={\rm diag}{(\sigma_1, \dots, \sigma_M)}$, where $\sigma_i$ is the empirical standard deviation of asset $i$. For the $i$-th eigenvector $\mathbf{q}_i$, we consider its corresponding eigenportfolio
\begin{equation}
\frac{{(\mathbf{h}^{-1}\mathbf{q}_i)}^T\mathbf{p}}{\|\mathbf{h}^{-1}\mathbf{q}_i\|_1}, \nonumber
\end{equation}
where $\mathbf{p} \in \Omega$ is the price scenario, and $\|\mathbf{h}^{-1}\mathbf{q}_i\|_1$ is used to normalize the portfolio weights so that the absolute weights sum to unity. 

%%%%%%%%%%%%%%%%%%%%%%%%%%%%%%%%%%%%%%%%%%%%%%%%%%%%%%%%%%%%%%%%%%%%%%%%%%%%%%%%%%%%%%%%%%%%%%%%%%%%%%%%%%%%%%%%%%%%%%%%%%

\section{Additional numerical experiments}\label{sec:supply_results}\label{app:supply_results}
\begin{figure}[!htbp]
\centering
\includegraphics[width=0.8\textwidth, trim=3cm 0mm 4cm 4cm,clip]{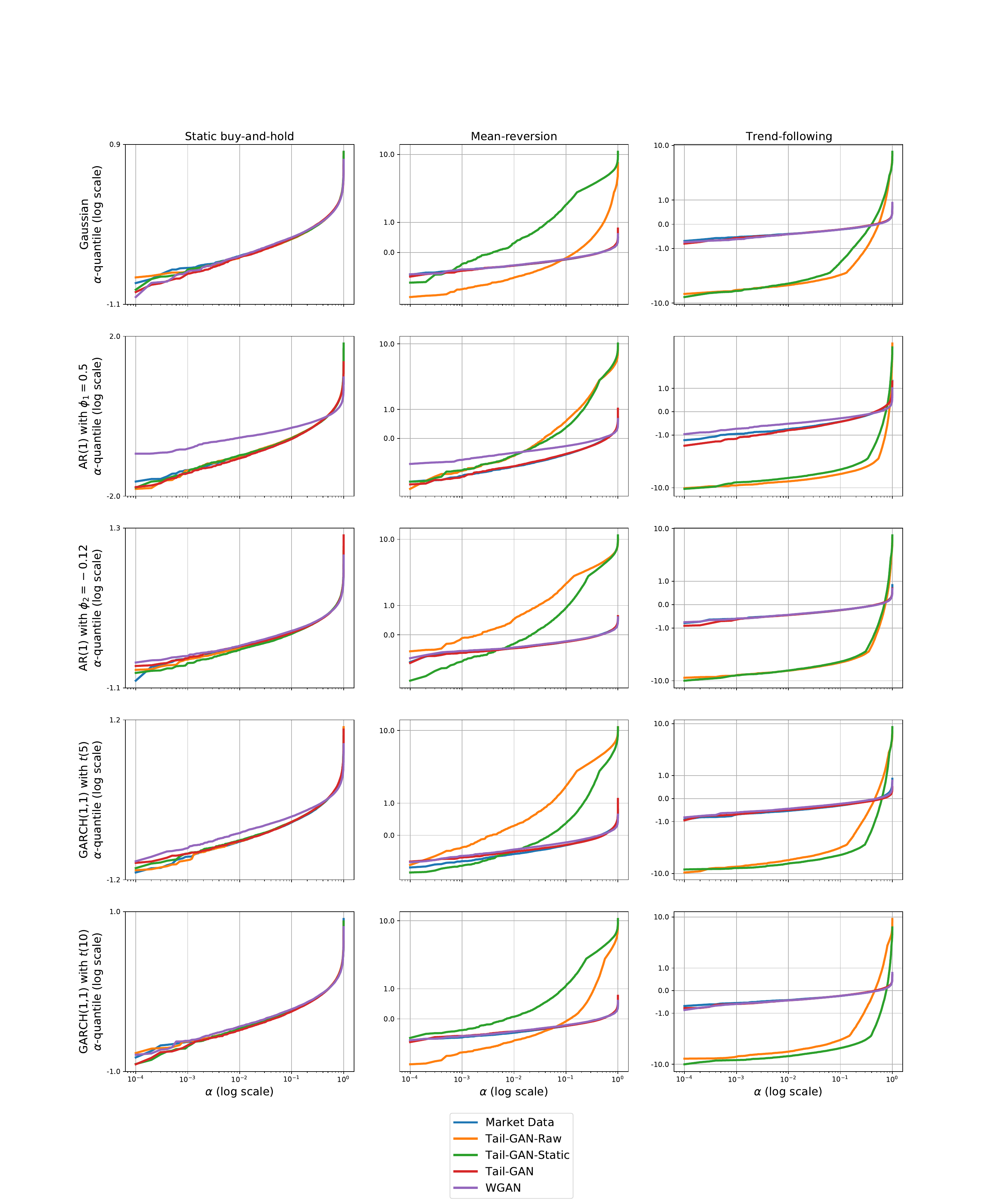}
\caption{Tail behavior via the empirical rank-frequency distribution of the strategy PnL. The rows index the various models used for generating the synthetic data, while the columns index the strategy types.}
\label{fig:synthetic_tail}
\end{figure}

\begin{figure}[!htbp]
\centering
\includegraphics[width=0.8\textwidth, trim=3cm 0mm 4cm 4cm,clip]{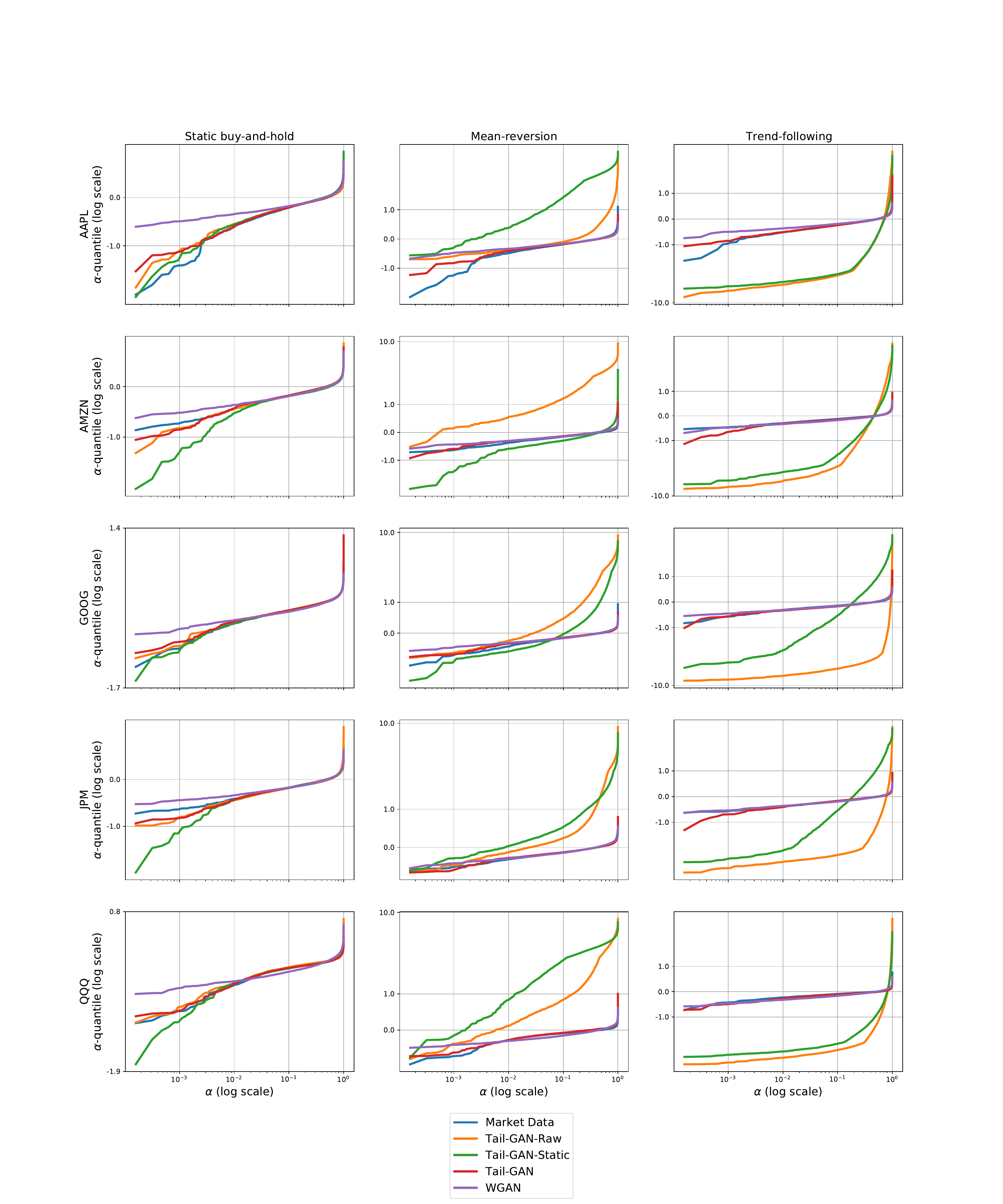}
\caption{Tail behavior via the empirical rank-frequency distribution of the strategy PnL. The rows index various stocks, while the columns index the strategy types.}
\label{fig:real_tail}
\end{figure}

\end{document}